\PassOptionsToPackage{numbers,sort&compress}{natbib}
\documentclass[preprintnumbers,amsmath,amssymb,prd, notitlepage,nofootinbib,twocolumn, superscriptaddress]{revtex4-2}

\usepackage{amsfonts,amssymb,amsmath}
\usepackage{gensymb}
\usepackage[usenames]{color}
\usepackage{graphicx}
\usepackage{xcolor}
\usepackage{multirow}
\usepackage[utf8]{inputenc}

\usepackage[colorlinks,bookmarks]{hyperref}
\definecolor{linkblue}{rgb}{0,0,0.8}
\definecolor{linkgreen}{rgb}{0,0.5,0}
\hypersetup{pdfpagemode=UseNone, pdfstartview=FitH, linkcolor=linkblue, %
            citecolor=linkgreen, urlcolor=linkblue}

\graphicspath{{./Images/}}
\usepackage[normalem]{ulem}

\newcommand{\greaterthanapprox}{\mathrel{\vcenter{
  \offinterlineskip\halign{\hfil$##$\cr
    >\cr\noalign{\kern2pt}\sim\cr\noalign{\kern-2pt}}}}}
    
    \newcommand{\lessthanapprox}{\mathrel{\vcenter{
  \offinterlineskip\halign{\hfil$##$\cr
    <\cr\noalign{\kern2pt}\sim\cr\noalign{\kern-2pt}}}}}

\newcommand{\vl}{\mathbf{L}} 

\newcommand{\lb}{\left(}
\newcommand{\rb}{\right)}

\newcommand{\lsb}{\left[}
\newcommand{\rsb}{\right]}
    
\newcommand{\ti}{\textit}

\newcommand{\be}{\begin{equation}}        
\newcommand{\ee}{\end{equation}}

\DeclareMathOperator{\Tr}{Tr}

\newcommand{\invMpc}{h\, {\rm Mpc}^{-1}\,}
\newcommand{\invMpcnoh}{ {\rm Mpc}^{-1}\,}

\begin{document}

\title{Avoiding baryonic feedback effects on neutrino mass measurements from CMB lensing}

\author{Fiona McCarthy}

\affiliation{Perimeter Institute for Theoretical Physics, Waterloo, Ontario, N2L 2Y5, Canada}
\affiliation{Department of Physics and Astronomy, University of Waterloo, Waterloo, Ontario, Canada, N2L 3G1}

\author{Simon Foreman}

\affiliation{Perimeter Institute for Theoretical Physics, Waterloo, Ontario, N2L 2Y5, Canada}
\affiliation{Dominion Radio Astrophysical Observatory, Herzberg Astronomy \& Astrophysics Research Centre,
National Research Council Canada, P.O. Box 248, Penticton, BC V2A 6J9, Canada}

\author{Alexander van Engelen}
\affiliation{School of Earth and Space Exploration, Arizona State University, Tempe, AZ 85287, USA}

\date{\today}

\begin{abstract}
A measurement of the sum of neutrino masses is one of the main applications of upcoming measurements of gravitational lensing of the cosmic microwave background (CMB). This measurement can be confounded by modelling uncertainties related to so-called ``baryonic effects" on the clustering of matter, arising from gas dynamics, star formation, and feedback from active galactic nuclei and supernovae. In particular, a wrong assumption about the form of baryonic effects on CMB lensing can bias a neutrino mass measurement by a significant fraction of the statistical uncertainty. In this paper, we investigate three methods for mitigating this bias: (1)~restricting the use of small-scale CMB lensing information when constraining neutrino mass; (2)~using an external tracer to remove the low-redshift contribution to a CMB lensing map; and (3)~marginalizing over a parametric model for baryonic effects on large-scale structure. We test these methods using Fisher matrix forecasts for experiments resembling the Simons Observatory and CMB-S4, using a variety of recent hydrodynamical simulations to represent the range of possible baryonic effects, and using cosmic shear measured by the Rubin Observatory's LSST as the tracer in method~(2). We find that a combination of (1) and (2), or (3) on its own, will be effective in reducing the bias induced by baryonic effects on a neutrino mass measurement to a negligible level, without a significant  increase in the associated statistical uncertainty. 
\end{abstract}
\maketitle

\section{Introduction}

With improving measurements of the Cosmic Microwave Background (CMB), a cosmological measurement of the sum of the neutrino masses is envisioned within the next decade.  Within the standard model of particle physics, the three neutrinos are massless particles; thus, the first measurements of neutrino flavour oscillations~\cite{PhysRevLett.81.1562,PhysRevLett.89.011301}, a process that only occurs if there exist mass differences between the species, were key developments in the search for beyond-standard-model physics. Neutrino oscillation experiments are sensitive to the difference in the squares of the masses of neutrinos, $\Delta m_{ij}^2\equiv m_i^2-m_j^2$; however, a cosmological neutrino detection will be sensitive to the sum of the neutrino masses $M_\nu\equiv\sum_{i=1}^3m_i$, and thus will be important in setting the overall scale of the neutrino masses. The current lower limit on $M_\nu$ (from neutrino oscillation experiments) is $M_\nu\greaterthanapprox60\,\text{meV}$ \cite{Patrignani:2016xqp}. Until now, cosmological experiments have only placed upper limits on $M_\nu$; the best is that of the  {\ti {Planck}} survey \cite{Aghanim:2018eyx}, which gives $M_\nu<120\,\text{meV}$.

Massive neutrinos have a well-understood effect on the matter power spectrum $P_m(k,z)$. After becoming non-relativistic when their temperature $T_\nu$ was comparable to their mass, they started contributing to $P_m(k,z)$; however, due to their small masses, they do not cluster on small scales, instead free-streaming, leading to a suppression of power on small scales. See \cite{Lesgourgues:2006nd,Lesgourgues:2012uu} for reviews on the cosmological effects of neutrinos.

{\ti{Planck}}'s upper bound on the neutrino mass was obtained from a joint analysis of the CMB temperature and polarisation maps, CMB lensing maps, and baryonic acoustic oscillation (BAO) measurements. As the neutrinos were still relativistic at the time of recombination when the CMB was released, the majority of a CMB survey's constraining power on $M_\nu$ comes from the CMB lensing information, which is sensitive to large-scale structure at all redshifts. In coming years, experiments such as the Simons Observatory (SO) \cite{Ade:2018sbj},  SPT-3G \cite{Benson:2014qhw},   and, further in the future, CMB-S4  \cite{Abazajian:2016yjj}, will make better measurements of the CMB lensing power spectrum, and have been forecast to measure the neutrino mass to   between 20 and 30 meV \cite{Ade:2018sbj,Allison:2015qca,Abazajian:2016yjj}.     

To be able to reach this level of constraint, the CMB lensing power spectrum must be well understood theoretically. In particular, the lensing power spectrum is a projection of the matter power spectrum $P_m(k,z)$ over all redshifts; however, there are certain effects that currently limit our understanding of $P_m(k,z)$, in particular effects due to baryonic processes (such as gas cooling and feedback from supernovae and active galactic nuclei) in the universe. Most predictions of $P_m(k,z)$ only account for gravitational forces, neglecting the complex baryonic interactions that we know exist; our current best method for understanding $P_m(k,z)$ including baryonic physics is performing large hydrodynamical simulations. Measurements of $P_m(k,z)$ from different simulations differ due to different numerical schemes and phenomenological implementations of baryonic processes that cannot be directly simulated at a given resolution. While the ``true" impact of baryonic effects on the matter power spectrum is not known, a general conclusion is that baryons contribute to a suppression of power on small scales (e.g.\ $0.1\invMpc \lesssim k \lesssim \mathcal{O}(\text{few}\times 10) \invMpc$ at $z=0$).  

Uncertainty due to baryonic feedback has been extensively studied in the context of cosmic shear surveys~\cite{White:2004kv,Zhan:2004wq,Jing:2005gm,Rudd:2007zx,Semboloni:2011fe,Natarajan:2014xba,Copeland:2019bho,Schneider:2019xpf}, as the scales affected by feedback directly correspond to the scales that current shear surveys are most sensitive to.  However, Ref.~\cite{Natarajan:2014xba} and recently Ref.~\cite{Chung:2019bsk} found that uncertainty from baryonic effects can also be important in the search for neutrino masses from CMB lensing, in spite of the higher redshifts and larger length scales involved. If we are to trust a measurement of $M_\nu$ from CMB lensing, it will be important to have an inference which is robust to these baryonic effects. Furthermore, the sensitivity of CMB lensing to baryonic effects implies that we could learn about the latter from observations of the former, an avenue explored for cosmic shear in Refs.~\cite{Harnois-Deraps:2014sva,Foreman:2016jzy,Huang:2020tpm,Yoon:2020bop}.

Ref.~\cite{Chung:2019bsk} explored the lensing power spectrum suppression and associated bias on neutrino mass from a suite of recent hydrodynamic simulations, finding that the range in possible biases is non-negligible comapared with expected statistical uncertainties. They found a significant scatter between different simulations,  comparable to the statistical uncertainty in the measurement.
In this paper, we consider various methods of mitigating this bias on the inference of $M_\nu$ from CMB lensing surveys similar to SO and CMB-S4. We describe various techniques to remove the sensitivity to the relevant baryonic processes and test, using Fisher forecasts, how these techniques will reduce the bias for  the series of simulations examined by Ref.~\cite{Chung:2019bsk}. The first mitigation method we consider is a simple scale cut, where the smallest scales of the CMB lensing convergence  (which are most sensitive to baryonic effects) are removed from the analysis. Second, we consider a method of using external cosmic shear measurements to estimate and subtract the low-$z$ contribution to the CMB lensing potential---as baryonic effects are relatively late-universe phenomena, we expect the high-$z$ portion of the CMB lensing potential to be less sensitive to them, while still retaining sensitivity to $M_\nu$. Finally, we ask whether marginalizing over the parameters of a general model for baryonic effects will reduce the bias while preserving the $M_\nu$ constraints.

We find that imposing a scale cut of $L_{\rm max}\sim1000$ on the lensing multipoles used for constraining $M_\nu$ can reduce the bias from baryonic effects by up to a factor of~2, with more aggressive cuts significantly increasing the statistical uncertainty. On the other hand, combining this scale cut with subtraction of a low-$z$ tracer, or marginalizing over a baryonic model (with or without a scale cut) will be much more effective in eliminating the bias, reducing it by at least a factor of 5 in the first case and 10 in the second case for all simulations we consider. The maximum residual bias associated with these simulations is $\sim$3 meV in either case, well below the level that would interfere with a high-significance detection of the minimum allowed neutrino mass sum. 

The paper is organized as follows. In Sec.~\ref{sec:lensing_baryons} we discuss the CMB lensing power spectrum and review the range of possible baryonic effects as represented by current hydrodynamical simulations. In Sec.~\ref{sec:bias_constraints} we present our Fisher forecast formalism for calculating the forecast constraint and biases on the inference of $M_\nu$ from these simulations. In Sec.~\ref{sec:Lmax_cutoff} we discuss the effects of a small angular-scale cutoff in $C_L^{\kappa\kappa}$. In Sec.~\ref{sec:subtraction_cosmicshear} we discuss the effect of ``subtracting'' a low-$z$ tracer to isolate the high-$z$ contribution to the CMB lensing map. In Sec.~\ref{sec:marginalisation_mead} we discuss the effect of marginalizing over parameters that describe the baryonic effects on the matter power spectrum. We discuss our results in Sec.~\ref{sec:discussion}.


\section{CMB Lensing and Baryons}\label{sec:lensing_baryons}

The CMB photons we detect have been gravitationally lensed by any matter they encounter along the paths they have travelled since their ``release" during recombination, at $z\sim1100$. Structures at any redshift after recombination can act as lenses, making CMB lensing a powerful probe of the evolution of the matter content of the universe.
We mainly quantify this information via the convergence power spectrum $C_L^{\kappa\kappa}$, which is a line-of-sight integral over the matter power spectrum $P_m(k,z)$ (see e.g.~\cite{Lewis:2006fu}), 
\be
C_L^{\kappa\kappa} = \int_0^{\chi_\mathrm{CMB}}   d\chi \frac{W_{\mathrm {CMB}}^\kappa(\chi)^2}{\chi^2}P_m\lb k=\frac{L+1/2}{\chi},z\rb\label{clkappakappa},
\ee
where $W_{\mathrm {CMB}}^\kappa(\chi)$ is the CMB lensing efficiency kernel
\be
W_{\mathrm{ CMB}}^\kappa(\chi) = \frac{3}{2}\Omega_m\lb \frac{H_0}{c}\rb^2\frac{\chi}{a(\chi)}\frac{\chi_{\rm CMB}-\chi}{\chi_{ \rm CMB}}\label{efficiency_CMB}
\ee
with $H_0$ the Hubble constant today, $c$ the speed of light, $\Omega_m$ the density of matter today, $\chi_{\rm CMB}$ the comoving distance to the surface of last scattering (at which the CMB was released), and $a(\chi)=\frac{1}{1+z(\chi)}$ and the scale factor at comoving distance $\chi$. Eq.~\eqref{clkappakappa} assumes the Limber approximation \cite{1953ApJ...117..134L,LoVerde:2008re}, which is valid in the small-scale, flat-sky limit; and the Born approximation, where the integral is taken over the photon's undeflected path, valid in the small-deflection limit~\cite{Pratten:2016dsm,Fabbian:2017wfp}.

To make an accurate inference of $M_\nu$ from a CMB lensing survey, we need to trust our theoretical model of the lensing convergence; i.e., we need to understand every component of Eq. \eqref{clkappakappa}. The cosmological ingredients that enter CMB lensing kernel $W^\kappa(\chi)$ are well understood. On the other hand, $P_m(k,z)$ is most commonly computed from linear gravitational perturbation theory on large scales, supplemented on small scales by non-linear extensions of gravitational perturbation theory, phenomenological models, $N$-body simulations, or emulators. Generally, these only account for gravitational interactions between the matter; i.e.\  they treat all matter as ``dark''.

However, about 15\%  of matter is not dark but baryonic, and has complex interactions with itself and with light. These interactions effect changes to how matter clusters on $\sim$Mpc and smaller scales: as examples, gas cooling and AGN feedback cause matter to condense and expand respectively (e.g.~\cite{Duffy:2010hf}). Our models currently lack a first-principles calculation of the power spectrum $P_m(k,z)$ incorporating these interactions, and they are typically neglected when considering CMB lensing surveys. However, some of the baryonic effects on $P_m(k,z)$---particularly the suppression of power on small scales---mimic the effects of  massive neutrinos, and neglecting these in the theoretical modelling of $C_L^{\kappa\kappa}$ can lead to significant biases on the neutrino mass inference~\cite{Natarajan:2014xba,Chung:2019bsk}. 

\begin{figure}[t]
\includegraphics[width=0.5\textwidth]{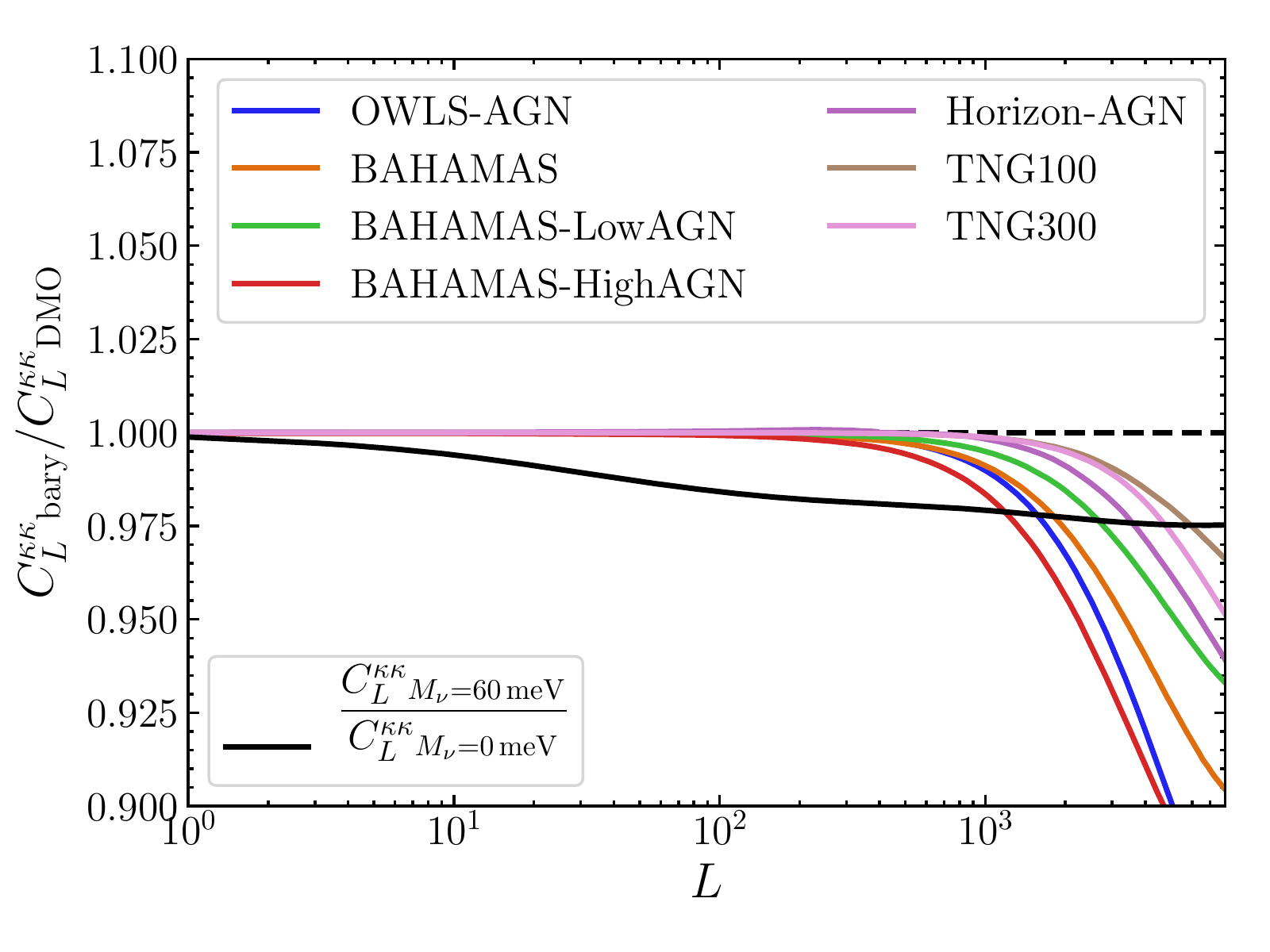}
\caption{The effect of baryons on the CMB lensing power spectrum $C_L^{\kappa\kappa}|_\mathrm{bary}$ in a selection of large hydrodynamical simulations (see Sec.~\ref{sec:sims}), as computed in Ref.~\cite{Chung:2019bsk}, shown as a ratio with the spectrum for dark matter only, $C_L^{\kappa\kappa}{}|_\mathrm{DMO}$. The effect of non-zero $M_\nu$ is also shown. The different scale dependences of baryonic and neutrino-mass effects indicate that it may be possible to distinguish between the two in lensing measurements.}\label{fig:baryons_clkk}
\end{figure}

In Fig.~\ref{fig:baryons_clkk}, we illustrate the suppression of the CMB lensing power spectrum by baryonic effects, as computed in Ref.~\cite{Chung:2019bsk}\footnote{These computations are available from \url{http://github.com/sjforeman/cmblensing_baryons}.} for a selection of recent hydrodynamical simulations (see Sec.~\ref{sec:sims}). At large scales, the power spectra coincide, while the baryonic suppression becomes relevant at $L\gtrsim1000$. We also show the effect of a non-zero neutrino mass on the lensing power spectrum, by plotting the ratio of the fiducial $M_\nu=60\,\text{meV}$ power spectrum to one where $M_\nu=0\,\text{meV}$ (with all other cosmological parameters unchanged). In this case, we see a power suppression with much milder scale-dependence than for baryonic effects. This difference indicates that it may be possible to disentangle the two types of suppression, motivating the methods we consider in this work.

\section{Neutrino mass: constraints and bias}\label{sec:bias_constraints}

\subsection{Forecasting the $1\sigma$ constraints}\label{sec:constraints}

We consider an analysis where $M_\nu$ is allowed to vary along with other cosmological parameters; as such we consider a parameter vector
\be
\vec \theta = \left(h, \Omega_bh^2,\Omega_ch^2,\tau, n_s,A_s,M_\nu\right)\label{parameter_vector}
\ee
with fiducial values $\{h=0.675, \Omega_bh^2=0.0222, \Omega_ch^2=0.1197, \tau=0.06, n_s=0.9655, A_s=2.2\times10^{-9}, M_\nu=0.06 \text{ eV}\}$ corresponding to the best-fit parameters of the {\ti{Planck}} analysis \cite{Aghanim:2018eyx} for the first six parameters, namely, the Hubble parameter in units of 100 km\,s$^{-1}$\,Mpc$^{-1}$, the physical baryon density, the physical cold dark matter density, the optical depth to recombination, the primordial scalar fluctuation slope and amplitude (with a pivot scale of 0.05 Mpc$^{-1}$). We take a fiducial value of 60 meV (the minimum allowed value) for the sum of the neutrino mass $M_
\nu$.  We compute the fiducial matter power spectrum with CAMB~\cite{Lewis:1999bs}, with the nonlinearities and treatment of neutrinos given by the extended halo model from Ref.~\cite{Mead:2016zqy}. 

As the cosmological parameters will also be constrained from the primary CMB, we include information from the primary CMB temperature and polarization as measured by the experiment we are forecasting for (see Sec.~\ref{sec:cmbexp}), as well as a prior from BAO measurements from DESI \cite{Aghamousa:2016zmz}, which improves the analysis by breaking the geometric degeneracy in the CMB.

To compute the information from the CMB and CMB lensing, we use the Fisher formalism, in which the Fisher matrix $F$ approximates the inverse covariance matrix of the parameters, with the diagonals of $F^{-1}$ giving the squares of the expected 1$\sigma$ uncertainties on each parameter (with all other parameters marginalized over). We calculate $F$ according to
\begin{widetext}
\be
F^{\rm CMB}_{ij}  = \sum_\ell \frac{2\ell+1}{2}f_{\rm sky}\Tr\lsb\frac{\partial C^{\rm CMB}_\ell}{\partial \theta^i}\lb C_\ell^{\rm CMB}\rb^{-1}\frac{\partial C^{\rm CMB}_\ell}{\partial \theta^j}\lb C_\ell^{\rm CMB}\rb^{-1}\rsb.
\label{eq:FCMB}
\ee
\end{widetext}
In Eq.~\eqref{eq:FCMB}, $f_{\rm sky}$ is the fraction of sky area which the surveys cover and $C_\ell^{\rm CMB}$ is the covariance matrix of the CMB:
\be
C_\ell^{\rm CMB} = \lb\begin{array}{c c c}
C_\ell^{TT}&C_\ell^{TE}&C_\ell^{T\kappa}\\
C_\ell^{TE}&C_\ell^{EE}&C_\ell^{E\kappa}\\
C_\ell^{T\kappa} & C_\ell^{E\kappa} & C_\ell^{\kappa\kappa}
\end{array}\rb\label{Cl_CMB}
\ee
where $C_\ell^{TT}$ is the power spectrum of the observed temperature anisotropies (including noise); $C_\ell^{EE}$ is the power spectrum of the observed $E$-mode polarisation anisotropies (also including noise); and $C_\ell^{TE}$ is their cross power spectrum. $C_\ell^{\kappa\kappa}$ includes the reconstruction noise for CMB lensing. Although the CMB we measure is lensed, we use the \ti{unlensed} primary CMB power spectra $C_\ell^{TT}, C_\ell^{EE}$ and $C_\ell^{TE}$ to avoid double-counting of the lensing information. A proper treatment including lensed CMB power spectra would involve including the covariances between the CMB power spectra induced by lensing, and also the covariances between the lensing convergence and the CMB power spectra~\cite{Peloton:2016kbw,Green:2016cjr}; neglecting the extra information on $M_\nu$ that comes from the lensed CMB power spectra makes our calculation conservative. $C_\ell^{T\kappa}$ and $C_\ell^{E\kappa}$, the cross power spectra of CMB lensing with CMB temperature and E-mode polarization respectively, are non-zero only on very large scales due to correlations induced by the late-universe effects on the CMB such as the Integrated Sachs--Wolfe (ISW) effect~\cite{Cooray:2001ab} and polarization generated after reionization~\cite{Lewis:2011fk}, but can be neglected in our analysis as we restrict to multipoles $\ell>300$ for the primary CMB. 

We also include a prior on $\tau$, the optical depth to reionization, which will be an important limiting factor in the inference of $M_\nu$ from lensing surveys~\cite{Allison:2015qca}. We consider two different scenarios: the {\ti{Planck}} design sensitivity $\sigma_{\rm prior}(\tau)=0.006$  (equal to the value achieved by the analysis of {\ti{Planck}} data in Ref.~\cite{Pagano:2019tci}) and the cosmic variance limit $\sigma_{\rm prior}(\tau)=0.002$~\cite{Abazajian:2016yjj}.  We include this as a Gaussian prior with width $\sigma_{\rm prior}(\tau)$. The final Fisher matrix we use for forecasting is
\be
F = F^{\rm CMB} + C_{\rm prior}^{-1}
\ee 
where $C_{\rm prior}$ is the sum of the BAO prior and the $\tau$ prior. Note that $F^{\rm CMB}$ contains both primary CMB and lensing information, as it is calculated from the covariance matrix in Eq.~\eqref{Cl_CMB}; however, as the cross power spectra between the primary CMB and the CMB lensing potential are set to zero, this can be separated as a sum of an inverse prior from the primary CMB and a Fisher matrix due to lensing alone.

Within this setup, the lower bound on the marginalized constraint on parameter $i$ is
\be
\sigma _i=  \sqrt{(F^{-1})_{ii}}.
\ee

\subsection{Calculating the baryonic bias}

Predictions for the CMB lensing power spectrum are typically computed with the dark-matter-only (DMO)\footnote{``Dark-matter-only'' computations could perhaps be more accurately described as ``gravity-only", since these computations do not neglect the baryonic contribution to the universe's matter content, but instead treat baryonic identically to dark matter, with only gravitational forces at play. However, ``dark-matter-only" is the term most commonly seen in the literature, so we also adopt it in this work.} nonlinear matter power spectrum $P_{\rm DMO}(k,z)$. However, since the true power spectrum includes (unknown) baryonic effects, the deviation from the DMO prediction that these effects induce might mimic the neutrino mass signal and result in an incorrect (``biased'') inference of the mass. If we can compute the power spectrum incorporating a given model of baryonic effects ($P_{\rm bary}(k,z)$), we can calculate the bias that would be induced in the inference of the parameter $\theta^i$ by (e.g.~\cite{Natarajan:2014xba})
\begin{widetext}
\be
B_i = F^{-1}\sum_\ell \frac{2\ell+1}{2}f_{\rm sky}\Tr\lsb\frac{\partial C^{\rm CMB}_\ell}{\partial \theta^i}\lb C_\ell^{\rm CMB}\rb^{-1}\Delta C_\ell\lb C_\ell^{\rm CMB}\rb^{-1}\rsb,
\label{eq:Bi-general}
\ee
\end{widetext}
where $\Delta C_\ell$ is the change in the covariance matrix due to baryonic effects
\be
\Delta C_\ell \equiv C_\ell\big{|}_{\rm bary}- C_\ell\big{|}_{\rm DMO}.
\ee
Note that of all the power spectra in the covariance matrix~\eqref{Cl_CMB}, only $C_\ell^{\kappa\kappa}$ is affected by the baryons, and $\Delta C_\ell^{XY}=0$ for $XY\ne \kappa\kappa$ (the $\kappa T$ and $\kappa E$ correlations are too small to be relevant). Thus, Eq.~\eqref{eq:Bi-general} simplifies to
\be
B_i = F^{-1}\sum_\ell  \frac{2\ell+1}{2} f_{\rm sky} \frac{\partial C_\ell^{\kappa\kappa}}{d\theta^i}
	\frac{1}{(C_\ell^{\kappa\kappa})^2} \Delta C_\ell^{\kappa\kappa}\ .
\ee
We use the forecasting code from Ref.~\cite{Li:2018zdm}\footnote{\url{https://github.com/msyriac/pyfisher}} to compute the Fisher matrices and biases in our forecasts.

To obtain a range of possible forms for $P_{\rm bary}$, and therefore $\Delta C_\ell^{\kappa\kappa}$, we turn to hydrodynamical simulations, as described in the next subsection.

\subsection{Simulations}\label{sec:sims}

We use $C_L^{\kappa\kappa}$ computations from Ref.~\cite{Chung:2019bsk}, which considers 7 baryonic scenarios from 4 different families of large hydrodynamical simulations (see Ref.~\cite{Chung:2019bsk} for further descriptions):
\begin{itemize}
\item the ``AGN" member of the OWLS simulation suite~\cite{vanDaalen2011,vanDaalen:2011xb,vanDaalen:2019pst}; 
\item the base BAHAMAS simulation~\cite{McCarthy:2016mry,McCarthy:2017csu,vanDaalen:2019pst}, along with the ``Low-AGN" and ``High-AGN" versions that respectively contain weaker and stronger AGN feedback than the base simulation;
\item the ``AGN" member of the Horizon simulation suite~\cite{Dubois:2014lxa,Dubois:2016,Chisari:2018prw}; and
\item the TNG100 and TNG300 runs of the IllustrisTNG simulations~\cite{Pillepich:2017fcc,Springel:2017tpz,Nelson:2017cxy,Naiman:2018,Marinacci:2017wew,Nelson:2018uso}.
\end{itemize}

The matter power spectrum $\hat{P}(k,z)$ is measured from the simulation outputs at several different redshifts, both from DMO runs (which treat baryons and dark matter identically) and from runs that include baryonic processes along with gravity. The measured power spectra have considerable uncertainty due to sample variance arising from the finite number of modes within each simulated volume, but the majority of this sample variance arises from randomness in the initial conditions that manifests primarily at large scales. Each pair of DMO and full-hydro runs begins with the same initial conditions (i.e.\ amplitudes and phases of modes at the initial time), and therefore the sample variance errors mostly cancel\footnote{The ratio $\hat{R}(k,z)$ will itself have some sample variance, because it is dominated by baryonic effects on the highest-mass halos within a given simulation volume, and the set of such halos will depend on the initial conditions. Ref.~\cite{Foreman:2019ahr} quantified the sample variance in $\hat{R}(k,z)$ for a subset of the simulations considered in this work, finding it to be at the few-percent level for $k\lesssim 20\invMpc$ (Refs.~\cite{Chisari:2018prw,vanDaalen:2019pst} reached similar conclusions.). This is acceptable for our work, which is focused on the range of $\hat{R}(k,z)$ between different simulations rather than the absolute precision of any one simulation.} in the ratio 
\be
\hat R(k,z) \equiv \frac{\hat P_{\rm bary}(k,z)}{\hat P_{\rm DMO}(k,z)}\ .
\label{baryonic_ratio}
\ee
The corresponding CMB lensing power spectrum $C_\ell^{\kappa\kappa}\big{|}_{\rm bary}$ can then be computed by using 
\be
P_{\rm bary}(k,z) = P_{\rm fid}(k,z)\hat R(k,z)
\label{eq:Pbary-with-R}
\ee
in Eq. \eqref{clkappakappa}, where $P_{\rm fid}(k,z)$ is the fiducial DMO prediction for the matter power spectrum. Note that the different simulations have been run with different cosmological models, while we compute $P_{\rm fid}$ using a single cosmology in our forecasts. Refs.~\cite{Mead:2015yca,Mummery:2017lcn,vanDaalen:2019pst} have found that $\hat{R}(k,z)$ has only a weak dependence on background cosmology, so Eq.~\eqref{eq:Pbary-with-R} is sufficient for our forecasts, while for work requiring percent-level accuracy, the cosmology-dependence of $\hat{R}(k,z)$ should be carefully accounted for.\footnote{An example is precise comparisons of $P_{\rm bary}$ to predictions from perturbation theory~\cite{Braganca:2020nhv}. We thank Matthew Lewandowski for discussions on this point.}

\subsection{Experimental configurations}\label{sec:cmbexp}

Several CMB experiments are planned or being built that will begin observations this decade and that are aimed at measuring CMB fluctuations on small scales, for gravitational lensing reconstruction and other secondary anisotropies.  Here, we consider an experiment similar to the Simons Observatory (SO)~\cite{Ade:2018sbj}, due to begin taking data in the first half of the 2020s.  The large aperture telescope for this experiment will have a 6\,m diameter and will observe large fractions of the sky at high angular resolution in six frequency channels.  We also consider an experiment like CMB-S4~\cite{Abazajian:2016yjj,Abazajian:2019eic}, which will have comparable angular resolution and frequency coverage, but higher sensitivity; it is expected to begin taking data on a later timeline than SO. 
We include Gaussian instrumental white noise on the CMB power spectra:
\be
N_\ell=N_T e^{\frac{\ell(\ell+1)\Theta_{\rm FWHM}^2}{8\ln2}},
\ee
where $N_T$ is the noise variance and $\Theta_{\rm FWHM}$ is the beam size of the experiment. For both experiments, we use $\Theta_{\rm FWHM} =1.4'$; for SO, we use $N_T$ corresponding to a map noise level of $6\,\mu\text{K-arcmin}$ and for S4 we use $N_T$ corresponding to $1\,\mu\text{K-arcmin}$.

For $C_L^{\kappa \kappa}$, we include reconstruction noise $N_L^{\kappa\kappa}$ corresponding to the minimum-variance reconstruction of Ref.~\cite{Okamoto:2003zw} relevant to the experiment we are considering (SO or S4). We include multipoles $90\le L \le 3100$ in the lensing power spectra, with the upper limit chosen based on where the statistical sensitivity drops off, and the lower limit having negligible impact on the results. For the primary CMB,  we include multipoles $300\le\ell\le3000$ for $C_\ell^{TT}$ and $300\le\ell\le5000$ for $C_\ell^{EE}$, with the upper limits based on where uncleaned foregrounds are expected to become significant in the lensing reconstruction~\cite{vanEngelen:2013rla,Osborne:2013nna}. We assume a sky fraction of $f_\mathrm{sky} = 0.4$ and full overlap between all the fields we consider.  For the $EB$-based reconstruction, which dominates the information at S4 noise, we include iterated delensing~\cite{Smith:2010gu}. 


\section{Strategy 1: Small angular-scale cut-off}\label{sec:Lmax_cutoff}

    The forecasts in Ref.~\cite{Chung:2019bsk} considered all scales over which future CMB  surveys will have appreciable sensitivity to the lensing power spectrum---ie, a summation over multipoles $90<L<3100$.  However,  baryonic effects are concentrated at a different (though not disjoint) range of scales than the neutrino mass constraint (see Fig.~\ref{fig:baryons_clkk}).

\begin{figure*}
\includegraphics[width=0.32\textwidth]{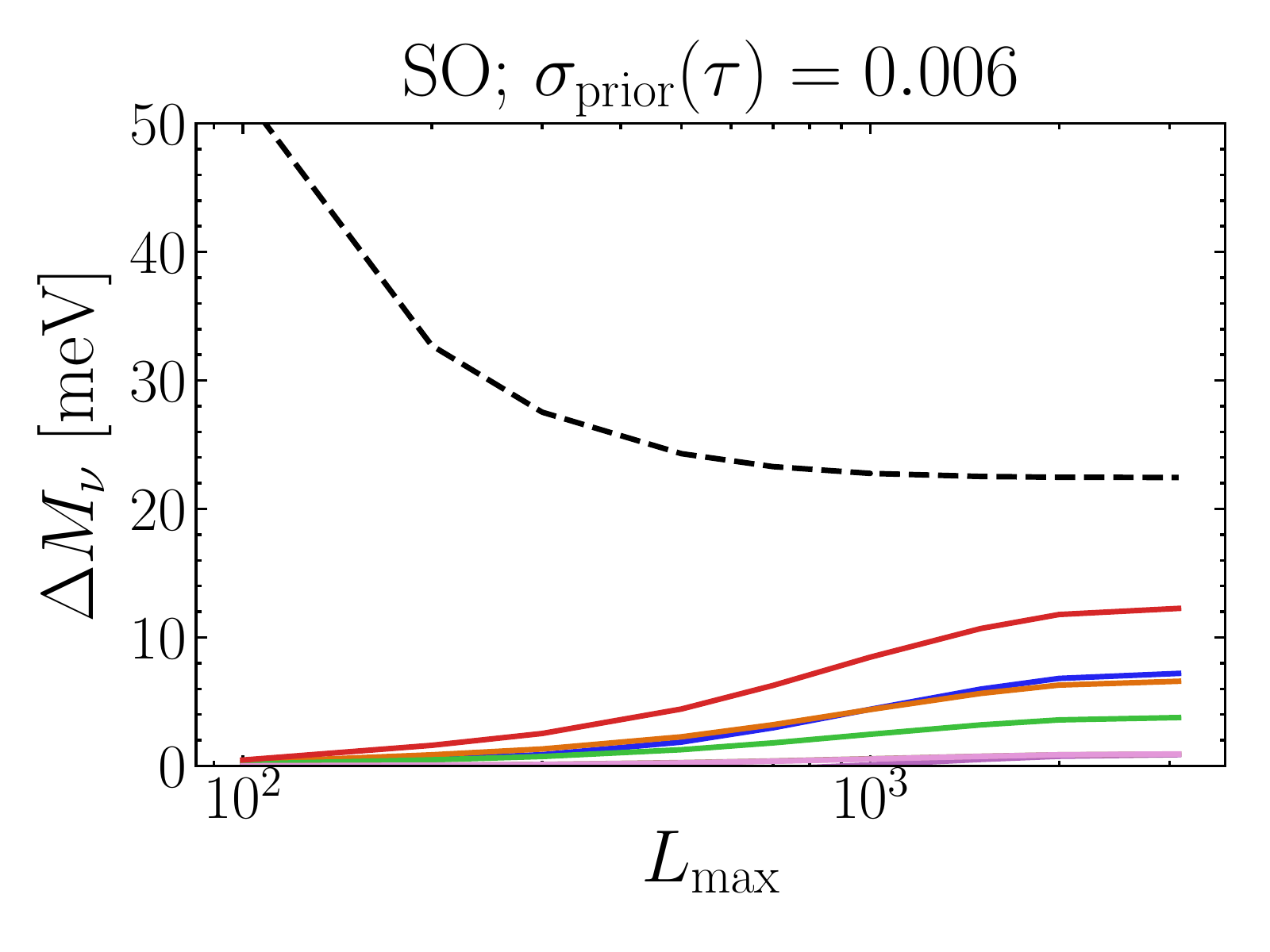}
\includegraphics[width=0.32\textwidth]{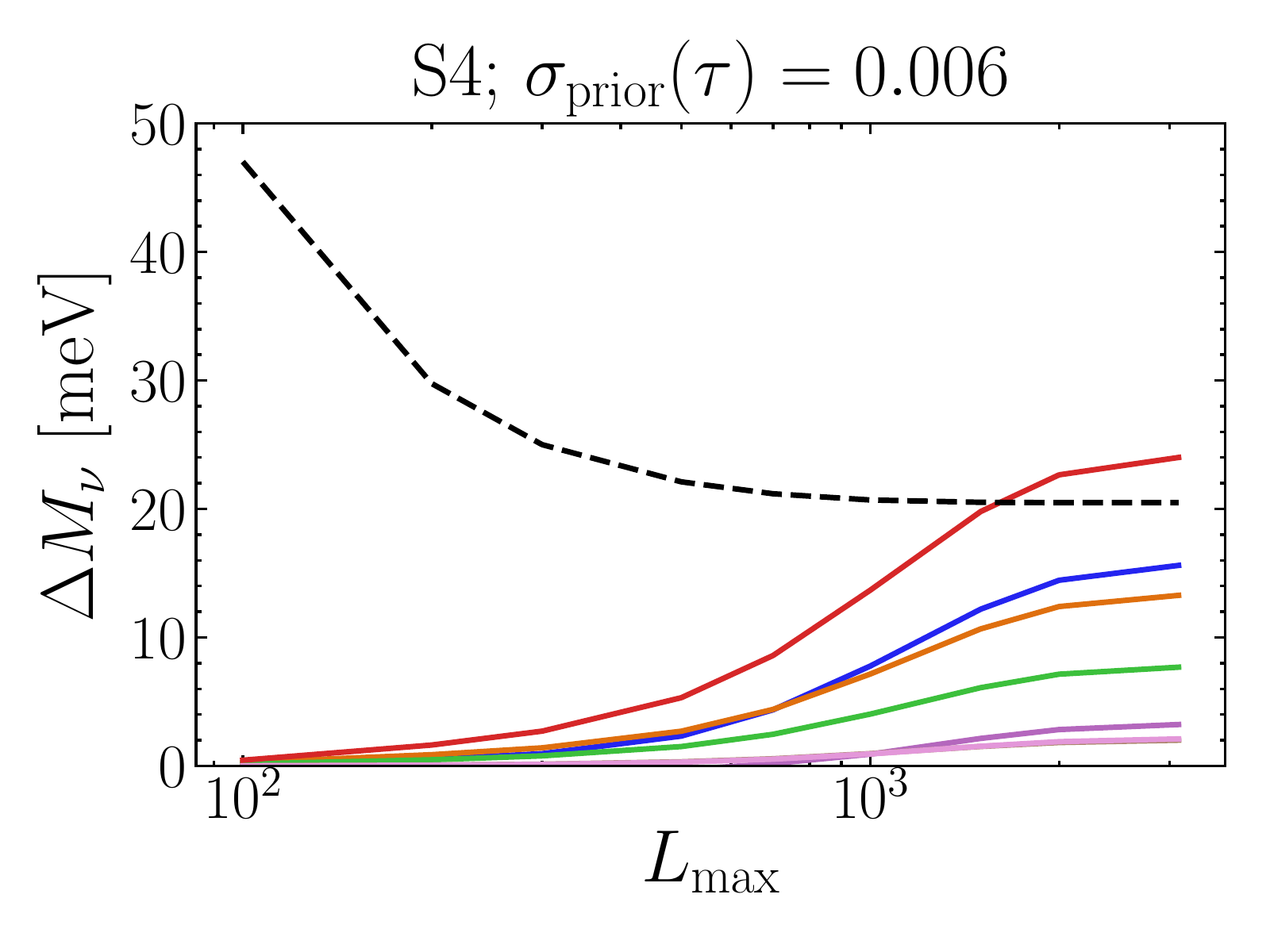}
\includegraphics[width=0.32\textwidth]{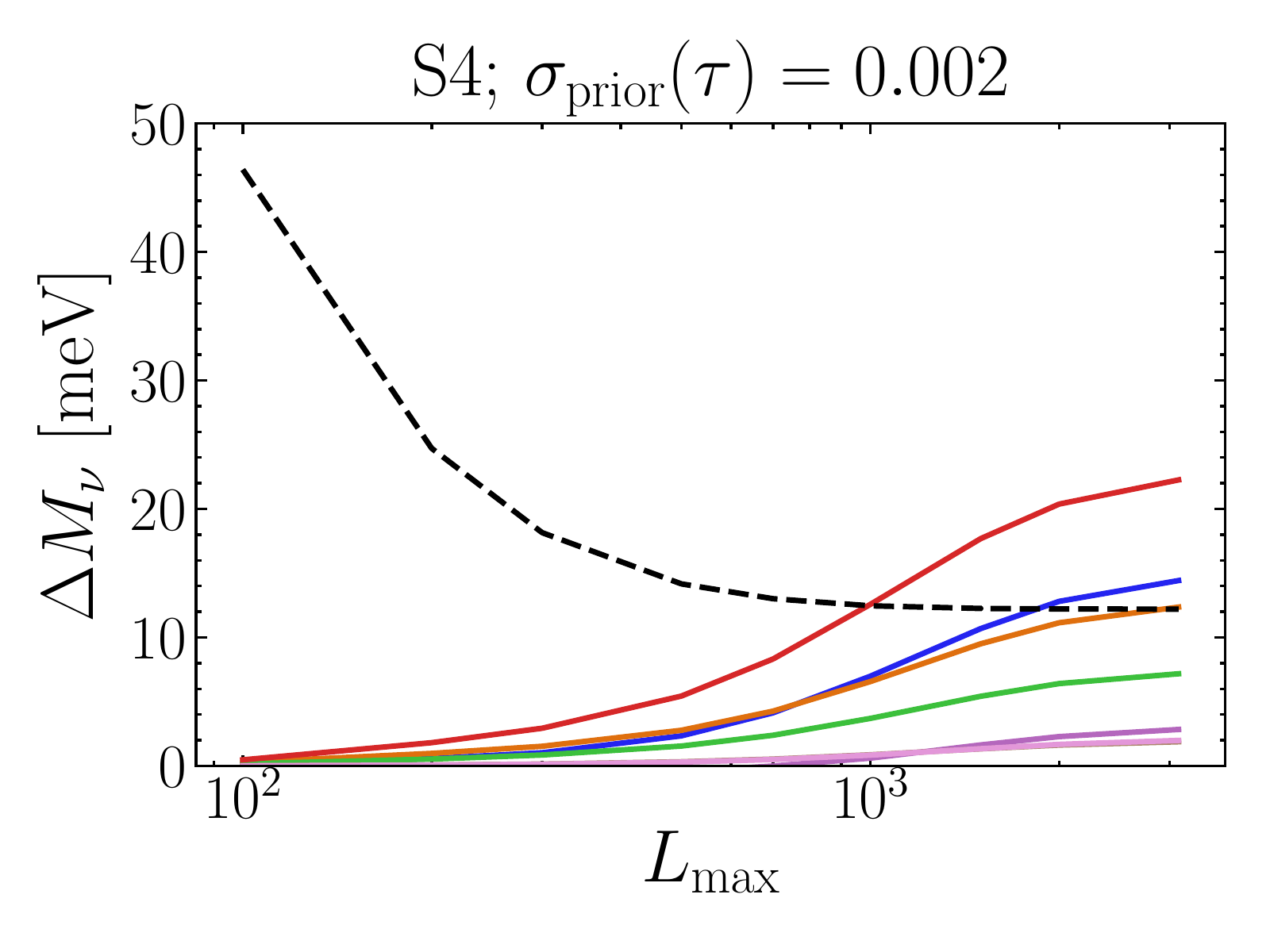}\\
\includegraphics[width=0.65\textwidth]{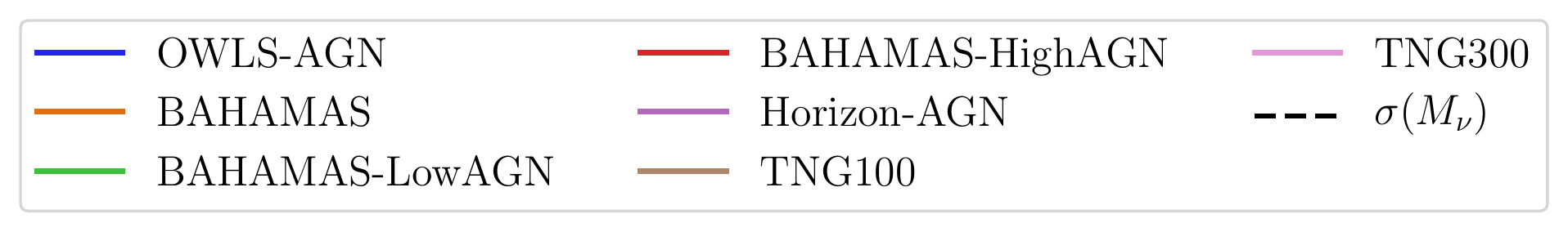}
\caption{Demonstrating mitigation strategy 1: the calculated biases  on $M_\nu$ plotted against the maximum CMB lensing  multipole $L_{\rm max}$ included in the forecast for the different simulations. In each case the $1\sigma$ constraint is shown as a dotted line. Note that the TNG100 and TNG300 lines are almost identical. \label{fig:biases_lmax} }
\end{figure*}
    
    With this in mind, the first mitigation strategy we implement is a simple $L_{\rm max}$ cut-off, where $L_{\rm max}$ is the maximum lensing multipole included in the analysis.  In Fig.~\ref{fig:biases_lmax}, we show the behaviour of the $1\sigma$ constraint on $M_\nu$ as well as the biases from different models of baryonic effects as we introduce this cut-off. It is clear that for all of the experimental setups, the constraints saturate at around $L_{\rm max}\sim 1000$ and there is no benefit to including multipoles $L\greaterthanapprox1000$; this happens because the suppression of $C_L^{\kappa\kappa}$ is roughly constant for $L\gtrsim 1000$, while the experimental errorbars on~$C_L^{\kappa\kappa}$ increase with $L$ over the same range (compare Fig.~\ref{fig:baryons_clkk} of this work with Fig.~1 of Ref.~\cite{Chung:2019bsk}). Meanwhile, we see that including higher multipoles does indeed increase the bias, and imposing $L_{\rm max} \sim 1000$ can reduce the bias by a factor of $\sim 2$ in some cases. However, in particular for the most advanced experimental configuration, the biases can still be of the same order of magnitude as the expected constraint, and so further mitigation methods will be needed to reduce the bias to an acceptable level.


\section{Strategy 2: Subtraction of external tracers}\label{sec:subtraction_cosmicshear}

\subsection{Isolating the low-$z$ contribution to the CMB lensing potential}
Baryonic phenomena begin to imprint themselves on structure formation  at a much later time in cosmological history than neutrino mass effects. As the CMB lensing kernel is an integral over all redshifts, we receive (weighted) information from all of cosmological history since recombination. However, if we could ``subtract'' the low-$z$ contribution to the lensing map to isolate the high-$z$ effects, we could potentially remove most of the bias while still being sensitive to $M_\nu$. 

To illustrate the ideal outcome of such a procedure, we define a high-$z$ CMB lensing field by
\be
C_L^{\kappa_{h}\kappa_{h}} = \int_{\chi_{min}}^{\chi_\mathrm{CMB}} d\chi \frac{W^\kappa(\chi)^2}{\chi^2}P_m\lb k=\frac{L+1/2}{\chi},z\rb,
\ee
where $\chi_{min}$  is some lower bound of the integration; $\chi_{min}=0$ corresponds to the standard CMB lensing scenario. We forecast the $1\sigma$ errors and baryonic biases on $M_\nu$ as in Section \ref{sec:bias_constraints} but replacing  $C_L^{\kappa\kappa}$ with $C_L^{\kappa_{h}\kappa_{h}}$, and we consider their dependences on the lower limit of integration $\chi_{min}$. 

\begin{figure*}[t]
\begin{center}
\includegraphics[width=0.32\textwidth]{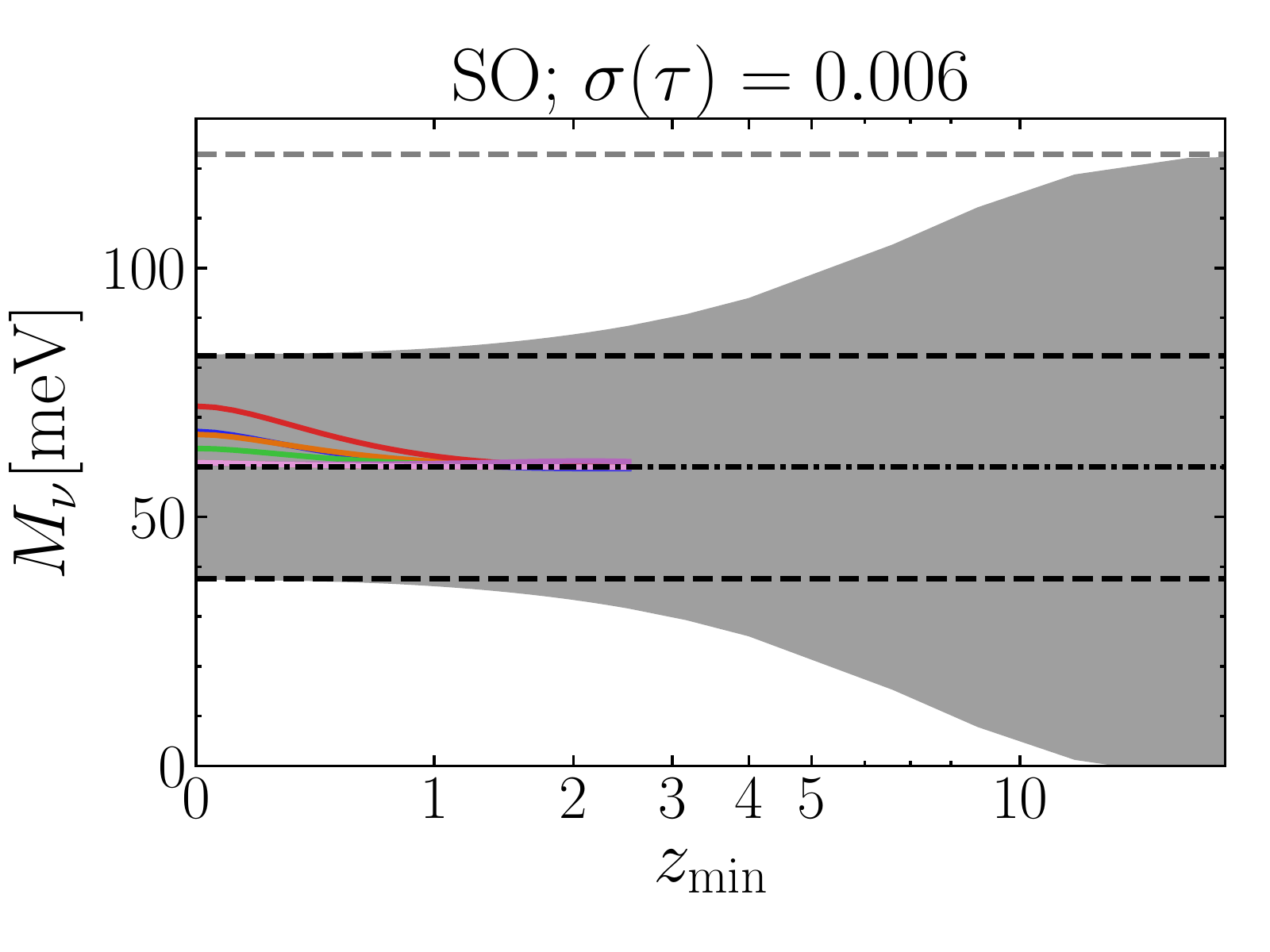}
\includegraphics[width=0.32\textwidth]{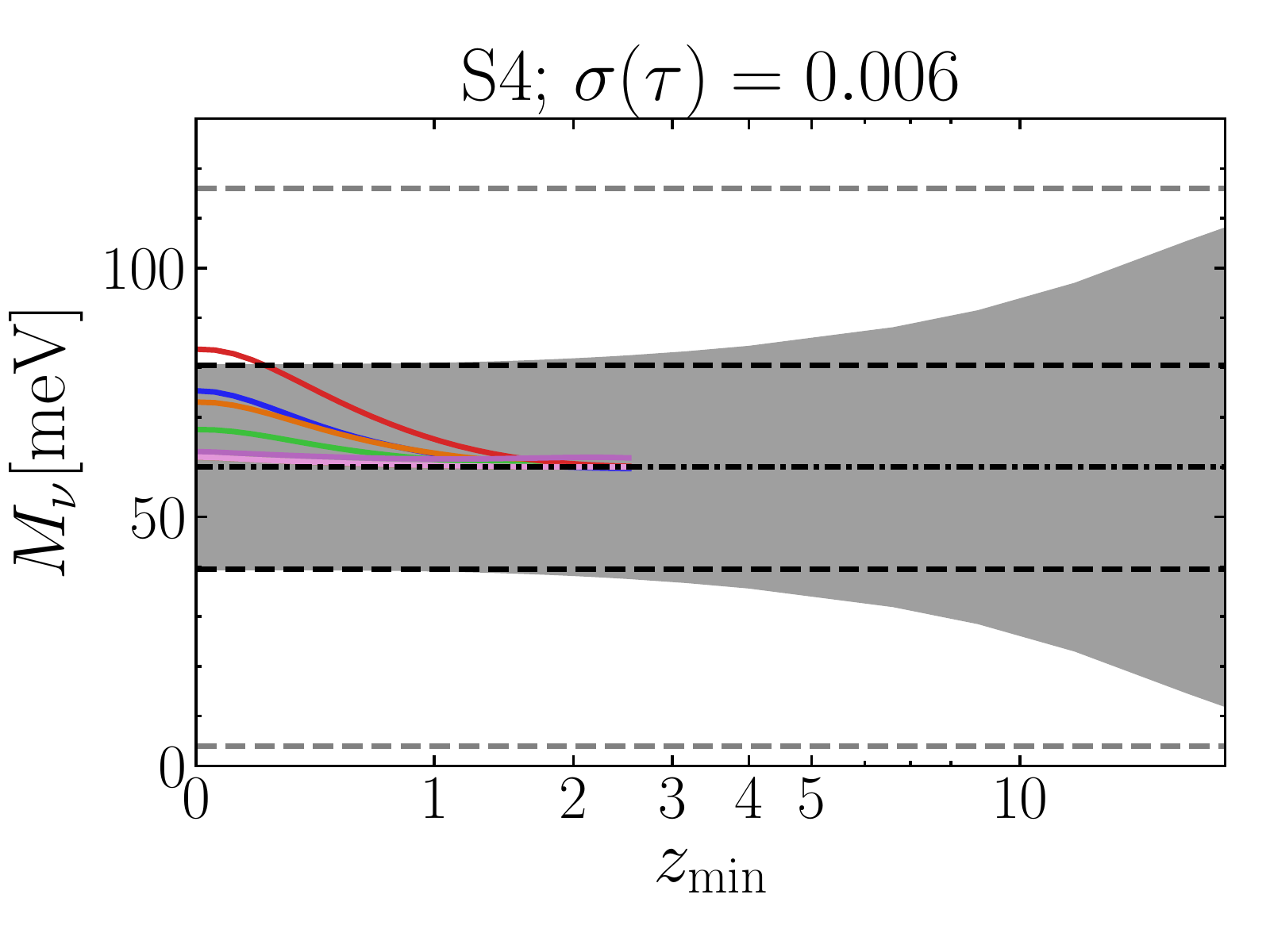}
\includegraphics[width=0.32\textwidth]{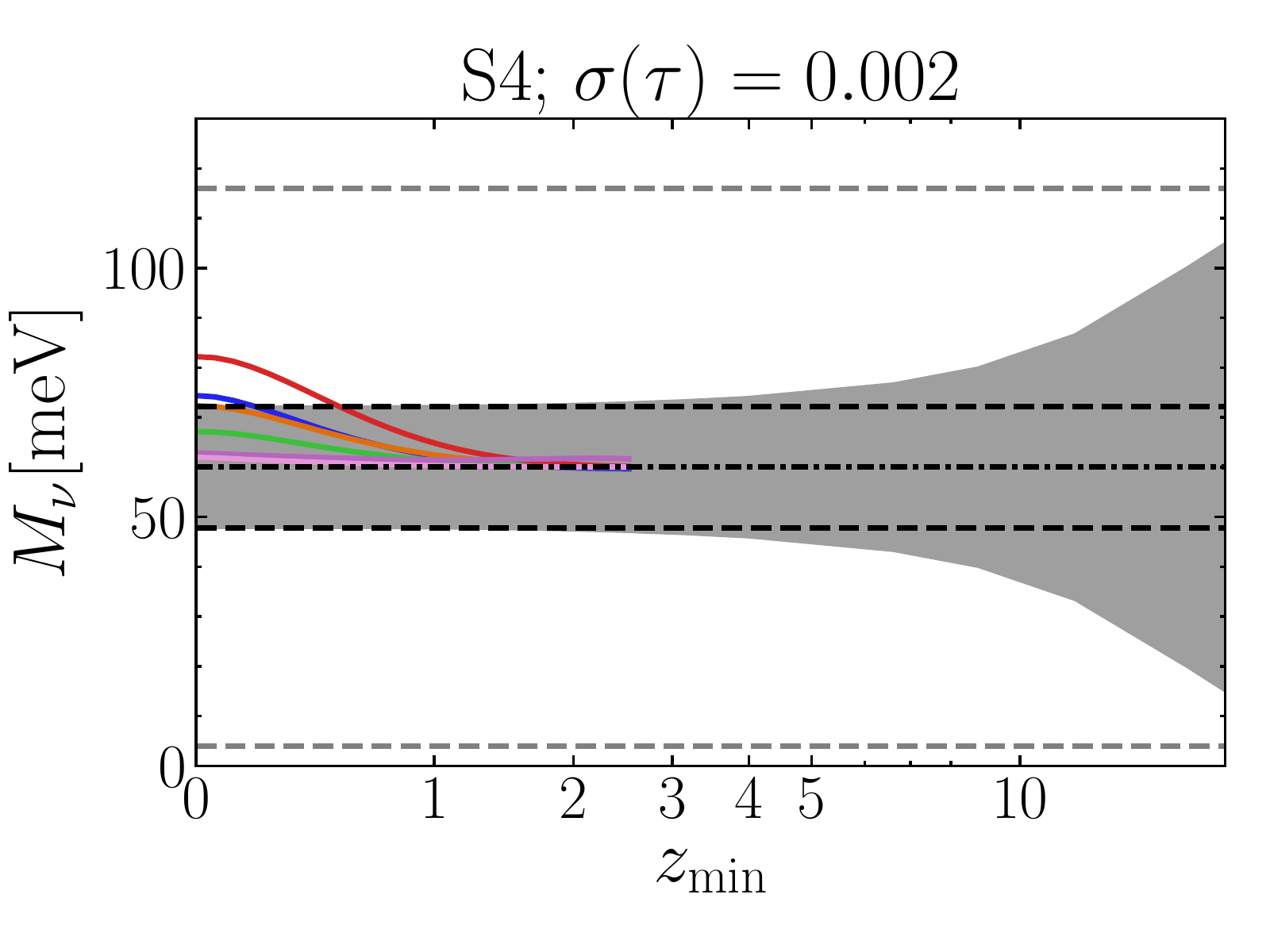}\\
\includegraphics[width=0.9\textwidth]{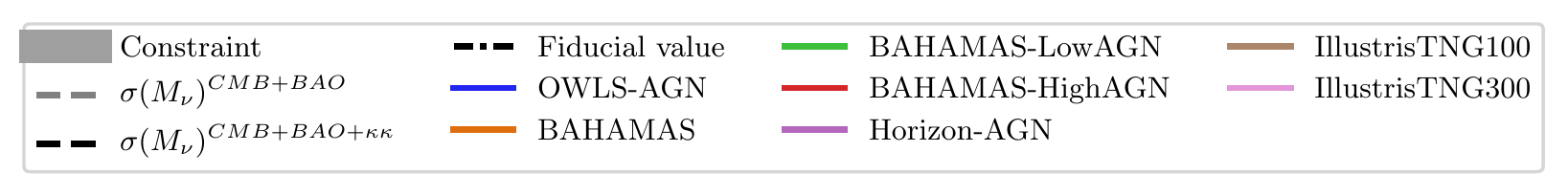}
\end{center}
\caption{The biases (solid lines) and the forecast $1\sigma$ constraints on $M_\nu$, plotted against the minimum $z$ $z_{\mathrm {min}}$ used to integrate the CMB lensing kernel, assuming that perfect removal of the $z<z_{\rm min}$ contribution is possible. We see that the baryonic bias starts to become relevant at $z_{\mathrm {min}}\sim2$. However, constraining information for $M_\nu$ comes from higher $z$.  Note that the x-axis is scaled logarithmically in $(1+z)$.  }\label{fig:z_constraints}
\end{figure*}

We show in Figure \ref{fig:z_constraints} the behaviour of the constraints and the biases plotted against $z_{min}=z(\chi_{min})$. Here we see explicitly that the biases are introduced in the late universe at around $z\lessthanapprox2$, while the constraints on $M_\nu$ come from a much larger redshift range. It is clear that if we could isolate the portion of the lensing map that is sourced at $z\greaterthanapprox2$ we could remove a significant portion of the baryonic bias on $M_\nu$ without sacrificing much constraining power on $M_\nu$.

Of course, we do not have direct access to measurements of the field $\kappa_h$  and so the configuration in Fig.~\ref{fig:z_constraints} is simply a toy model for illustrative purposes. However, through cross-correlation with a low-$z$ external tracer $\hat X$ (such as a cosmic shear map), we could remove the low-$z$ portion of the CMB lensing field by defining a new field
\be
\hat\kappa^{\rm sub}(\vl) = \hat \kappa(\vl) - \frac{C_L^{X\kappa}+N_L^{X\kappa}}{C_L^{XX}+N_L^{XX}}\hat X(\vl),
\label{eq:kappasub}
\ee
where $\hat \kappa(\vl)$ is the original CMB lensing convergence, and in this subsection we use hats to denote quantities that include noise (i.e. $C_L^{\hat{A}\hat{B}} = C_L^{AB} + N_L^{AB}$). In Eq.~\eqref{eq:kappasub}, we weight $\hat{X}$ with a matched filter designed to extract the portion of $X$ correlated with $\kappa$, assuming that this correlation is dominated by low redshifts. Note that in our implementation, we will assume that the filter is computed using theoretical expressions with the fixed fiducial cosmological parameters, such that it is not varied in our Fisher calculations.

To see how Eq.~\eqref{eq:kappasub} accomplishes our goal, observe that if the weights perfectly match the true statistics of~$\hat{X}$ and~$\hat{\kappa}$,  the power spectrum of $\hat{\kappa}_{\rm sub}$ reduces to
\be
C_L^{\hat{\kappa}_{\rm sub}\hat{\kappa}_{\rm sub}}
	= C_L^{ \kappa \kappa}+N_L^{ \kappa \kappa}
	-  \frac{\lb C_L^{X\kappa}+N_L^{X\kappa}\rb^2}{C_L^{XX}+N_L^{XX}}.  
	\label{eq:Cellkappasub}
\ee
Furthermore, if we decompose $\kappa$ into uncorrelated pieces sourced by low and high redshifts, $\kappa=\kappa_{\rm low}+\kappa_{\rm high}$, and assume that $X(\vl)=T(L)\kappa_{\rm low}(\vl)$, so that $X$ is perfectly correlated with $\kappa_{\rm low}$ with a transfer function $T(L)$, Eq.~\eqref{eq:Cellkappasub} becomes
\begin{align}
C_L^{\hat{\kappa}_{\rm sub}\hat{\kappa}_{\rm sub}}
	 &= C_L^{\kappa_{\rm low}\kappa_{\rm low}} + C_L^{\kappa_{\rm high}\kappa_{\rm high}} + N_L^{ \kappa \kappa} \nonumber \\ 
	 &\quad-  \frac{\lb T(L) C_L^{\kappa_{\rm low}\kappa_{\rm low}}+N_L^{X\kappa}\rb^2}
	{T(L)^2 C_L^{\kappa_{\rm low}\kappa_{\rm low}}+N_L^{XX}}.
\end{align}
With high noise on $X$, we recover $C_L^{\hat{\kappa}\hat{\kappa}}$, but in the low-noise limit ($N_L^{X\kappa}, N_L^{XX} \to 0$), we obtain
\be
C_L^{\hat{\kappa}_{\rm sub}\hat{\kappa}_{\rm sub}}
	=  C_L^{\kappa_{\rm high}\kappa_{\rm high}} + N_L^{ \kappa \kappa},
\ee
and therefore the low-$z$ contribution to the lensing power spectrum is perfectly subtracted.\footnote{
An alternative strategy to the map-level subtraction we have considered here would be to perform the forecasts when including all auto- and cross-power spectra between the lensing map and the other tracer, as performed in, e.g., Refs.\cite{Das:2013aia,Schaan:2016ois,Schaan:2020qox} in the context of using CMB lensing to mitigate systematic effects seen in cosmic shear surveys.  However, this method is \ti{more} sensitive to the baryonic biases sourced at low-$z$, and should be implemented with  a mechanism for marginalizing over baryonic models, which we do not consider in this section for simplicity (although we will consider such a mechanism for the CMB-lensing alone case in Section \ref{sec:marginalisation_mead}). }

In reality, for a tracer we can directly measure, such as cosmic shear, the assumption of perfect and exclusive correlation with the low-$z$ contribution to CMB lensing does not exactly hold; however, if this correlation is sufficiently high, we expect that we can still subtract a significant portion of the unwanted low-$z$ contribution to a lensing map. We consider an explicit example in the following sections.

\subsection{Cosmic shear from the Rubin Observatory}\label{sec:lsst_shear} 

Cosmic shear is an ideal candidate for an external tracer $\hat{X}$ that we can use to isolate and subtract the low-redshift contribution to CMB lensing maps. As light rays from distant galaxies travel through the universe, their paths are deflected by the intervening matter (just as  the CMB is lensed), and this introduces correlated ellipticities in the observed images of these galaxies. These correlations, either amongst these galaxies or between the galaxies and another tracer of large-scale structure, are most commonly measured directly from catalogs of observed galaxy shapes. In our forecasts, we will assume that these catalogs can be converted into lensing convergence maps\footnote{For recent work on such ``mass-mapping" techniques for cosmic shear, see Refs.~\cite{Jeffrey:2019fag,Mawdsley:2019had,Pires:2019zcc,Price:2020mry}, several of which are extensions to the method first presented in Ref.~\cite{Kaiser:1992ps}. Alternatively, it may be possible to implement our proposal in Sec.~\ref{sec:subtraction_cosmicshear} starting directly from shear catalogs, but we leave this to future work.}, and take $\hat{X}$ to refer to such a map constructed from galaxies in a given redshift bin.

The lensing efficiency for a source galaxy at comoving distance $\chi_S$ is given by Eq.~\eqref{efficiency_CMB} with $\chi_{\rm CMB}$ replaced by $\chi_S$:
\be
W^\kappa(\chi,\chi_S) = \frac{3}{2}\Omega_m\lb \frac{H_0}{c}\rb^2\frac{\chi}{a(\chi)}\frac{\chi_{S}-\chi}{\chi_{S}}.
\ee
In practice, galaxies are binned into photometric redshift bins with a finite extent in redshift space, and so we measure the cosmic shear from galaxies at a range of sources. To calculate the shear efficiency for such a bin of galaxies, we integrate over the redshift extent of the bin and weight by the galaxy distribution $\frac{dn}{d\chi}$:
\be
W^{i}(\chi) =\frac{1}{n_i}\int_{\chi^i_i}^{\chi^f_i} d\chi_S\frac{dn}{d\chi_S}W^\kappa(\chi,\chi_S) 
\ee
for a bin $i$ between $\chi^i_i$ and $\chi^f_i$, where $n_i$ is the total number density of the bin $n_i=\int_{\chi^i_i}^{\chi^f_i} d\chi \frac{dn}{d\chi}$.
We consider explicitly the distribution predicted for the Rubin Observatory's LSST (Legacy Survey of Space and Time) Gold sample of galaxies~\cite{2009arXiv0912.0201L}, a sample that will be used to measure cosmic shear. We take the distribution from Ref.~\cite{Chang:2013xja} 
\be
\frac{dn}{dz}= n_s z^{1.24} \exp{\left[-\left(\frac{z}{0.5}\right)^{1.01}\right]},
\label{galaxy_distribution}
\ee
with a total number density $n_s=26\text{ arcmin}^{-2}$. $\frac{dn}{d\chi}$ can be found from Eq.~\eqref{galaxy_distribution} by computing $\frac{dn}{d\chi}=\frac{dn}{dz}\frac{dz}{d\chi}$.

The shear power spectrum of bin $i$ is given by
\be
C_L^{\kappa_i\kappa_i}=\int_0^{\chi_i^f} d\chi  \frac{W^{i}(\chi) ^2}{\chi^2}P_m\lb k=\frac{L+1/2}{\chi},z\rb .
\ee
We consider a survey with $N$ source bins containing equal numbers of galaxies for the cosmic shear fields, and as we wish to use them to subtract as much of the low-$z$ contribution to the CMB lensing kernel as possible, we combine them in such a way to maximise their correlation with the CMB lensing convergence.  As such, we consider a linear combination of shear fields
\be
\hat X = \sum_i c_i \hat X_i
\ee 
where $\hat X_i$ is the convergence map of bin $i$. The coefficients $c_i$ are chosen to maximise the correlation coefficient between $\hat X$ and the CMB lensing potential; we compute them following the linear-algebraic methods of Ref.~\cite{Sherwin:2015baa} (see their Appendix A).\footnote{Such a map, combining samples in the mid-infrared and far-infrared from \textit{WISE} and \textit{Planck} to be maximally correlated with CMB lensing, was generated in Ref.~\cite{Yu:2017djs}.}  The $c_i$ that maximise the correlation coefficient of $\hat X$ with the CMB lensing potential $r_L\equiv\frac{C_L^{X\kappa}}{\sqrt{C_L^{XX}C_L^{\kappa\kappa}}}$ are 
\be
c_i =\sum_j\lb C_L^{\kappa_A\kappa_B}\rb^{-1}_{ij}C_L^{\kappa_j \kappa_{\rm CMB}},
\ee 
where $C_L^{\kappa_A\kappa_B}$ is the covariance matrix of the cosmic shear fields (including noise), and $C_L^{\kappa_j \kappa_{\rm CMB}}$ is the cross-power spectrum between the cosmic shear field $j$ and the CMB lensing convergence. The elements of $C_L^{\kappa_A\kappa_B}$ are 
\begin{align}
\nonumber
C_L^{\kappa_{A}\kappa_B}{}_{ij}
&=\int d\chi  \frac{W^{i}(\chi) W^{j}(\chi) }{\chi^2}P_m\lb k=\frac{L+1/2}{\chi},z\rb    \\
&\quad+N_L^{\kappa_{i}\kappa_{j}},\label{clkappaab}
\end{align}
with shear noise power spectrum $N_L^{\kappa_{i}\kappa_{j}}$ given by 
\be
N_L^{\kappa_{i}\kappa_{j}}=\delta_{ij}\frac{\sigma_\epsilon^2}{n_i}
\ee 
where $\sigma_\epsilon$ is the intrinsic shape noise (we take $\sigma_\epsilon=0.26$~\cite{Chang:2013xja}) and $n_i$ is the total angular number density of bin $i$.  The cross power spectrum between CMB lensing convergence and cosmic shear in bin $i$ is given by
\be
C_L^{\kappa_{j}\kappa_{\rm CMB}}=\int d\chi  \frac{W^{j}(\chi) W_{\rm CMB}^{\kappa}(\chi) }{\chi^2}P_m\lb k=\frac{L+1/2}{\chi},z\rb \label{cellkappacmbcross}
\ee
where $ W_{\rm CMB}^{\kappa}(\chi)$ is the CMB lensing efficiency kernel given in Eq.~\eqref{efficiency_CMB}.

\subsection{Intrinsic alignments}

It is a non-trivial exercise to separate the apparent ellipticities induced by cosmic shear from the inherent ellipticities of galaxies. Under the assumption that the galaxies have a random distribution of ellipticities, this is not a problem, as taking a high number of galaxies in the sample ensures that the average intrinsic alignment averages to zero and there is no bias to the signal (although Poissonian noise remains). However, if there is an intrinsic alignment to the galaxies' true ellipticities (e.g.\ as caused by alignment with the large-scale tidal field), this will bias one's inference of a lensing signal. With this in mind, we introduce an intrinsic alignment contribution to our forecasts our forecasts involving cosmic shear to ensure that the effect of intrinsic alignments on the ability of cosmic shear to subtract the low-$z$ information of the CMB lensing kernel is accounted for.

 To quantify the intrinsic alignment contribution, the observed ellipticity $\gamma$ can be separated into a part induced by gravity $\gamma^G$ and a part that is intrinsic $\gamma^I$,
\be
\gamma = \gamma^G+\gamma^I,
\ee
such that the two-point correlation of $\gamma$ with itself is
\be
\left<\gamma\gamma\right>=\left<\gamma^G\gamma^G\right>+\left<\gamma^G\gamma^I\right>+\left<\gamma^I\gamma^G\right>+\left<\gamma^I\gamma^I\right>.
\ee
In terms of the angular power spectra between the ellipticities of galaxies in two redshift bins labelled by $i$ and $j$, we can write
\be
C_L^{\gamma^i\gamma^j}=C_L^{{G_i} {G_j}} + C_L^{G_i I_j} + C_L^{G_j I_i} + C_L^{I_i I_j}.
\ee
Correlations between the tidal field responsible for the intrinsic alignments of foreground galaxies and the gravitational field lensing the images of distant galaxies can cause $C_L^{G_i I_j}$ to be non-zero when the redshift bin $j$ is in front of bin $i$ (negligibly small contributions, which are exactly zero in the Limber approximation, come in the case when $i$ is in front of $j$). This extra contribution to the correlation is also present in the CMB lensing-cosmic shear cross power spectra:
\be
C_L^{\kappa_{j}\kappa_{\rm CMB}}=C_L^{G_{j}\kappa_{\rm CMB}}+C_L^{I_{j}\kappa_{\rm CMB}}.
\ee

The gravitational contributions $C_\ell^{G_i G_j}$ and $C_\ell^{G_{j}\kappa_{\rm CMB}}$ are the cosmic shear expressions given in Eqs.~\eqref{clkappaab}  and~\eqref{cellkappacmbcross}. For the other terms, the power spectrum of intrinsic shear $P_{II}(k,z)$ must be introduced. Then, the $I-I$ correlations in redshift bin $i$ are given (within the Limber approximation) by an integral over the redshift bin, weighted by the galaxy density:
\be
C_L^{I_i I_i}=\frac{1}{n_i^2}\int_{\chi^i_j}^{\chi^f_j}{d\chi} \lb\frac{dn}{d\chi_S}\rb^2 P_{II}\lb k=\frac{L+1/2}{\chi},z\rb.
\ee
The cross power spectra between bins $C_L^{I_i I_j}$ is zero (within the Limber approximation) for $i\ne j$, as the redshift window functions do not overlap. For the cross-term $C_L^{GI}$ we introduce the cross power spectrum between matter and intrinsic ellipticities $P_{I,m}(k,z)$ such that
\be
C_L^{G_i I_j}=\frac{1}{n_j}\int_{\chi^i_j}^{\chi^f_j}d\chi \lb\frac{dn}{d\chi_S}\rb  \frac{W^{i}(\chi) }{\chi^2}P_{I,m}\lb k=\frac{L+1/2}{\chi},z\rb,\label{cross_gi}
\ee
where $W^{i}(\chi)$ can be replaced by $W_{\rm CMB}^{\kappa}(\chi)$ to get $C_L^{I_{j}\kappa_{\rm CMB}}$. As mentioned earlier, this is only non-zero for $j$ in front of $i$ as ${W^{i}(\chi) }$ is zero for $\chi>\chi^f_i$, ie over the integration range of Eq.~\eqref{cross_gi} if bin $j$ is behind bin $i$.

A simple model for the power spectrum $P_{II}(k,z)$ assumes that (on large scales) galaxies are aligned with their host dark matter halos, which are given ellipticities by their local tidal field, implying that, in the linear regime~\cite{PhysRevD.70.063526,2007NJPh....9..444B},
\be
P^{\rm lin}_{II}(k,z)=\lb\frac{A_{IA}C_1 \bar \rho a^2}{\bar D}\rb^2 P^{\rm lin}_m(k,z)
\ee
where $\bar \rho$ is the mean matter density of the universe at redshift $z$, $\bar D=\frac{D(z)}{a}$ where $D(z)$ is the growth factor normalized to 1 today; and $P^{\rm lin}_m(k,z)$ is the \ti{linear} matter power spectrum. $C_1$ is a constant amplitude  $C_1=5\times 10^{-14}h^{-2}M_\odot^{-1}\mathrm{Mpc}^{3}$,, chosen to match the IA amplitude of superCOSMOS in~\cite{Brown:2000gt} and $A_{IA}$ is an overall normalisation parameter with a fiducial value of unity which should be marginalized over due to uncertainty in the overall amplitude of the intrinsic alignment power.  The cross spectrum between the intrinsic ellipticities and the matter power spectrum $P_{I,m}(k,z)$ is given (again in the linear regime) by
\be
P^{\rm lin}_{I,m}(k,z)=-\lb\frac{A_{IA}C_1 \bar \rho a^2}{\bar D}\rb P^{\rm lin}_{m}(k,z).
\ee
In \cite{2007NJPh....9..444B} this linear alignment model was extended to a ``non-linear linear alignment model'' by replacing $P^{\rm lin}_{m}(k,z)$ with the non-linear $P_{m}(k,z)$. We use this latter model in our forecasts.

In cosmic shear analyses, it is customary to use a very wide prior on the normalization $A_{IA}$. However, recent shear surveys have been successful in constraining this parameter in combination with photometric or spectroscopic galaxy clustering: for example, when ignoring galaxy colors, the Dark Energy Survey obtained $\sigma(A_{IA})\approx 0.33$ with its year-1 data~\cite{Samuroff:2018xuo}, and the KiDS survey achieved $\sigma(A_{IA})\approx 0.5$ in Ref.~\cite{Johnston:2018nfi}, with both surveys also able to distinguish the amount of intrinsic alignments associated with red and blue galaxies (see also Ref.~\cite{Yao:2020jpj}). Future shear surveys are expected to improve upon this, especially if accompanied by coordinated wide-field spectroscopy (e.g.~\cite{Mandelbaum:2019zej}). Thus, in our baseline forecasts we include a prior of 25\% of the fiducial value of $A_{IA}$, but we also explore the dependence of our results on this choice.

\subsection{Impact of shear subtraction on neutrino mass inference}
\label{sec:shearsub-results}

\begin{figure*}[t]
\begin{center}

\includegraphics[width=0.32\textwidth]{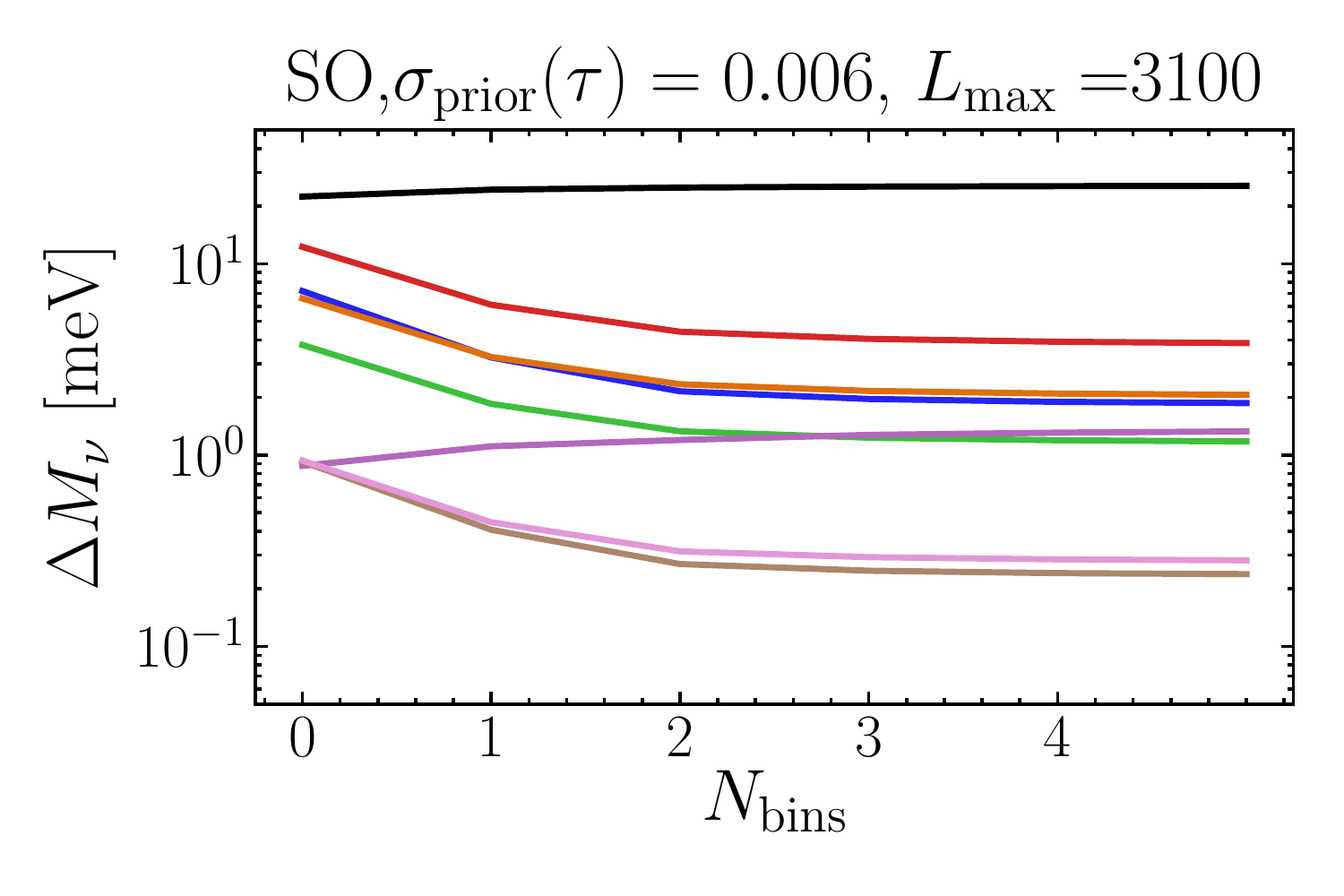}
\includegraphics[width=0.32\textwidth]{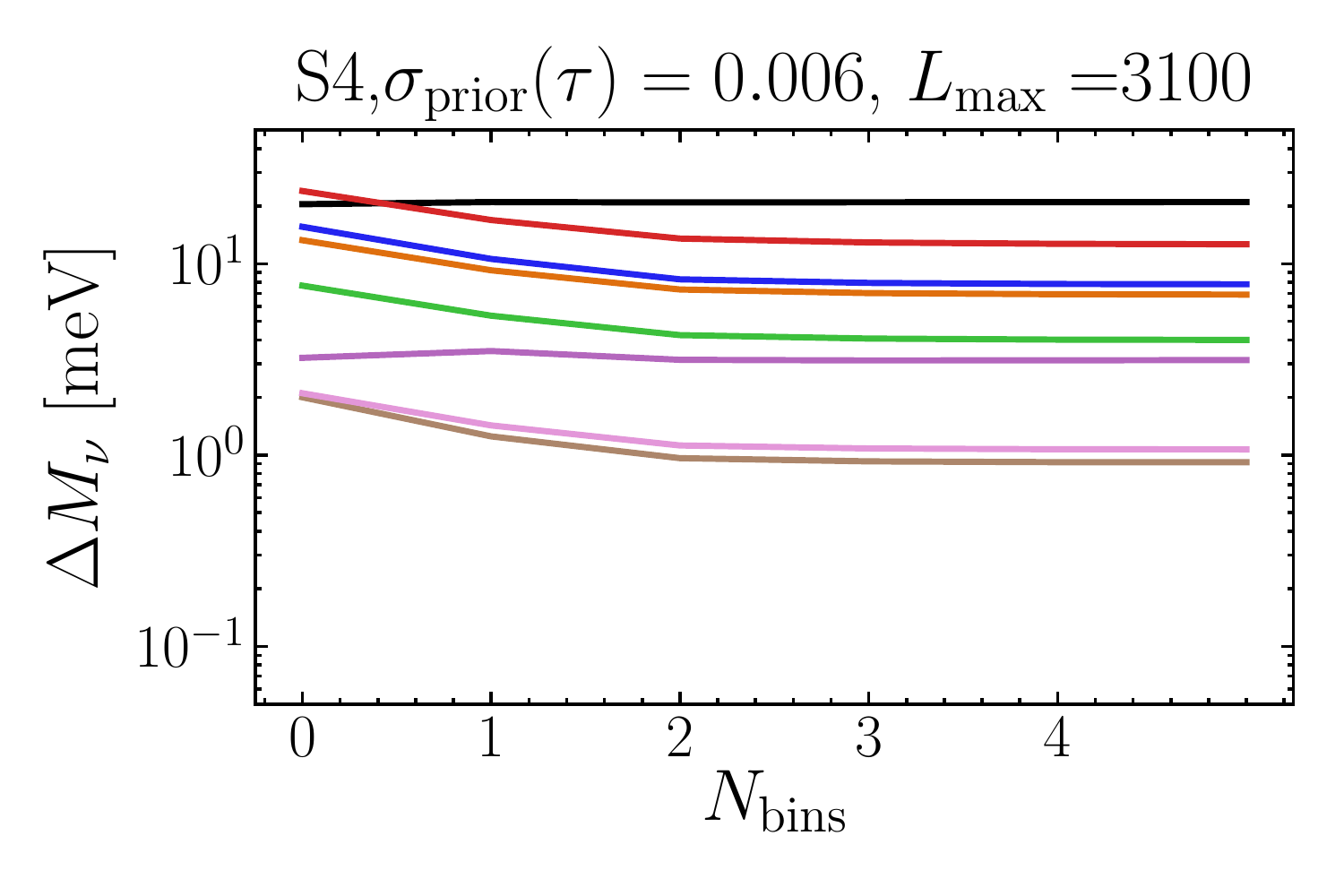}
\includegraphics[width=0.32\textwidth]{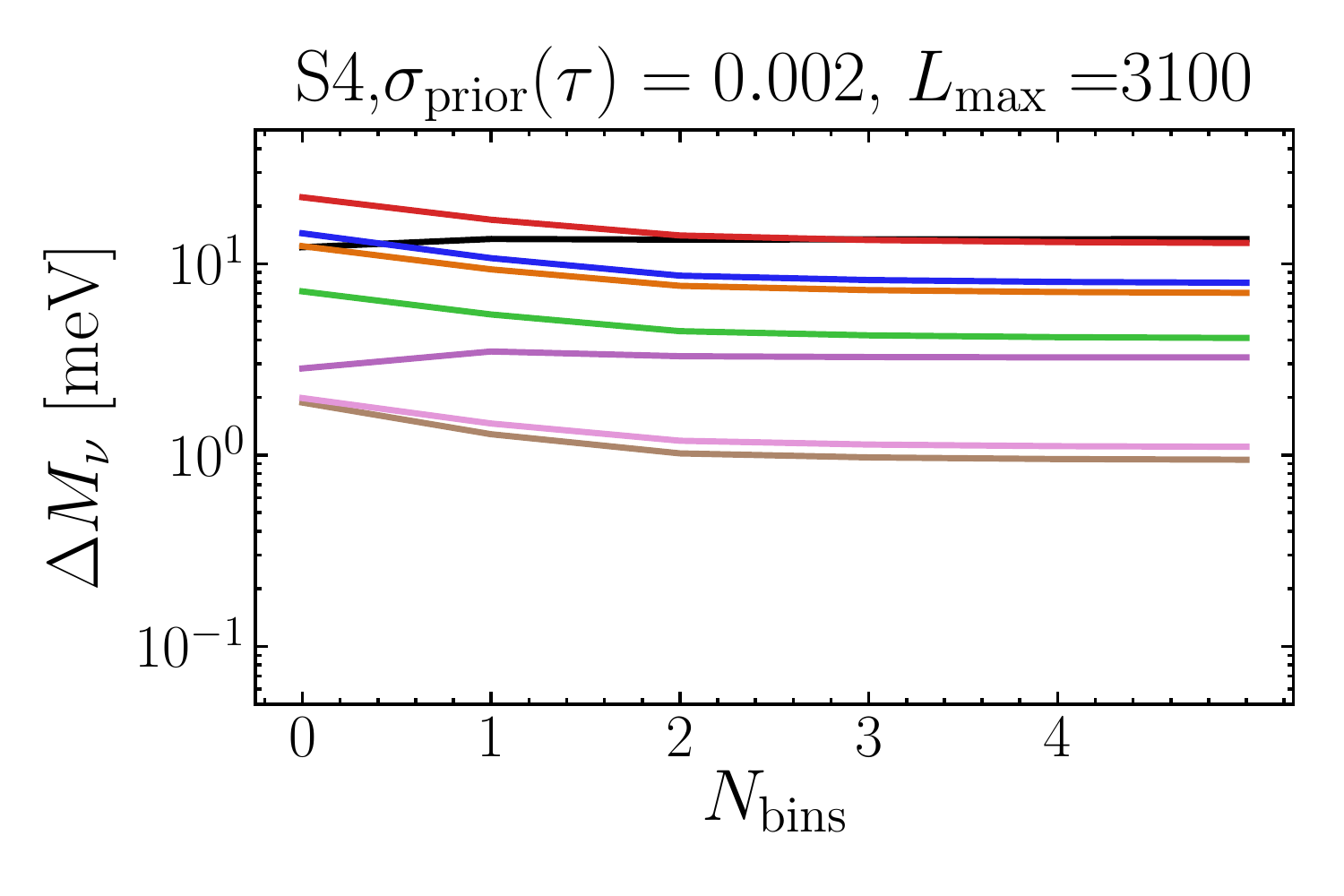}\\
\includegraphics[width=0.32\textwidth]{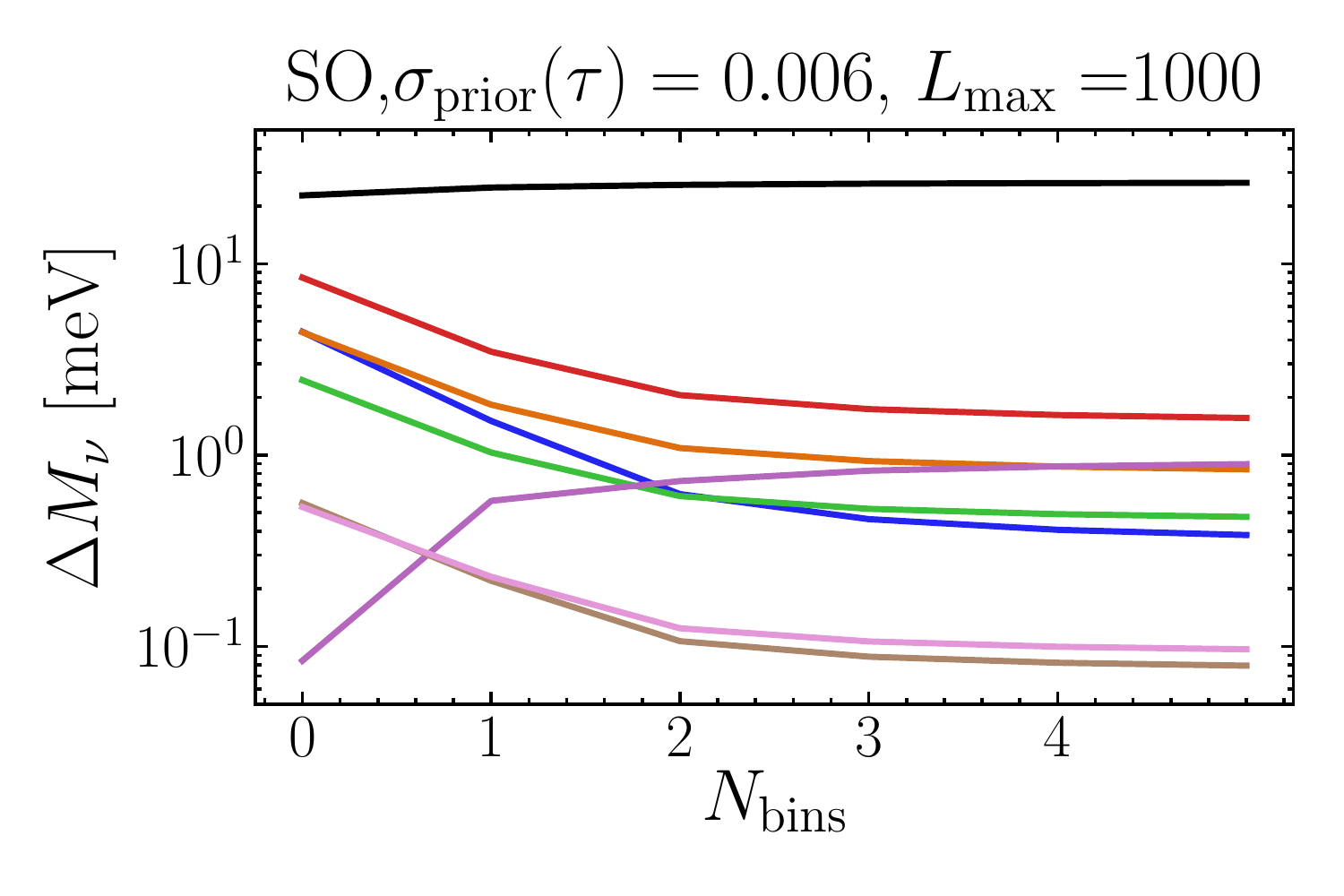}
\includegraphics[width=0.32\textwidth]{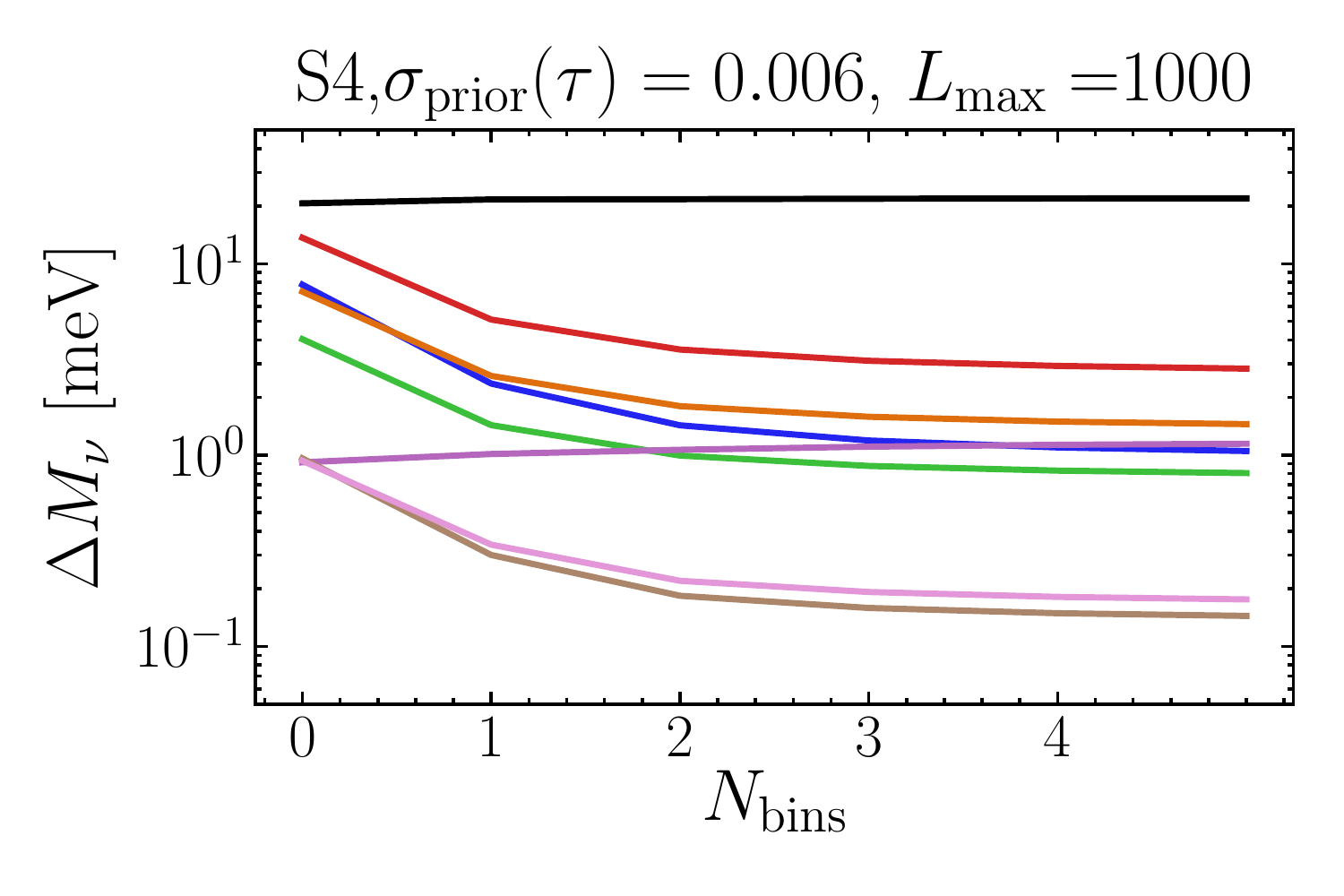}
\includegraphics[width=0.32\textwidth]{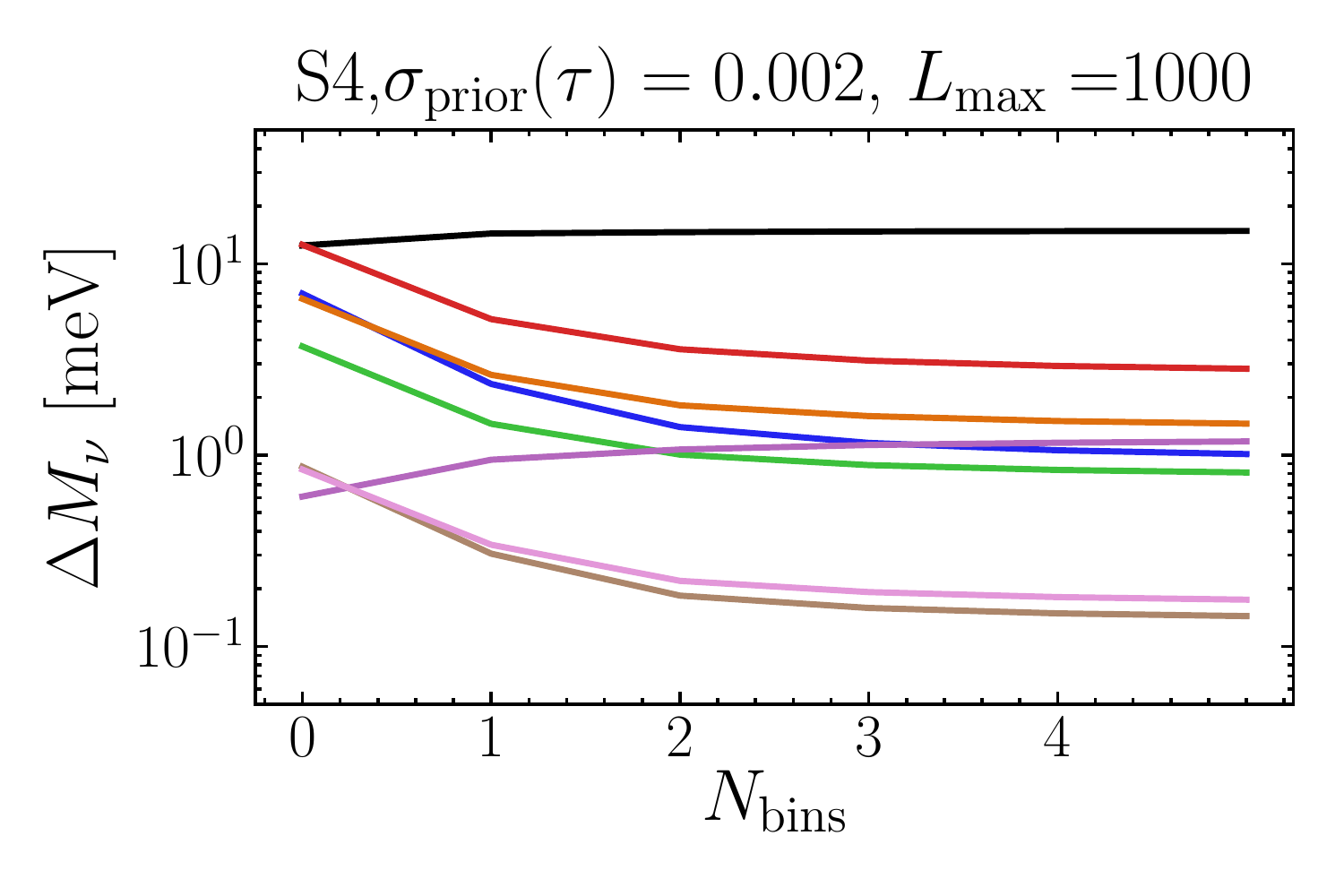}\\
\includegraphics[width=0.55\textwidth]{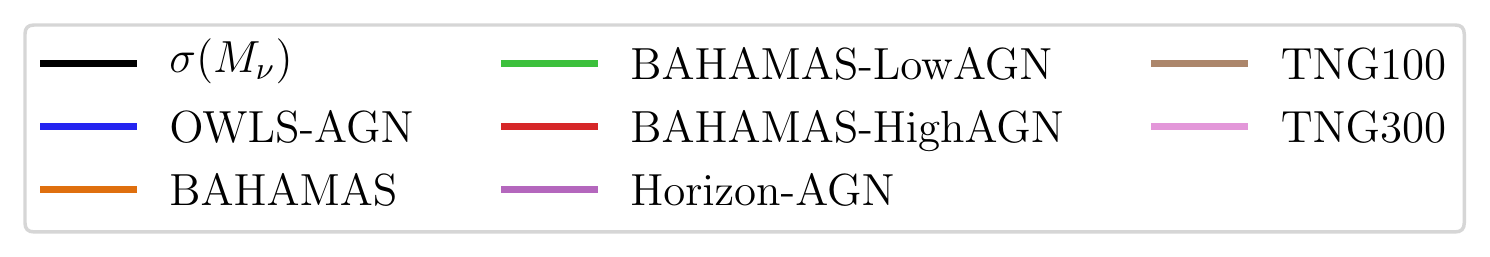}
\caption{Demonstrating mitigation strategy 2: the decrease in the bias when an optimal combination of $N_\mathrm{bins}$ cosmic shear bins is subtracted from the CMB lensing map.  With the exception of $N_\mathrm{bins} = 0$ (which corresponds to the no-subtraction case), the total redshift extent and number of galaxies are held constant, with  $N_\mathrm{bins}$ controlling the slice thickness.  On top is for an analysis with $L_{\rm max}$ of 3100; on bottom we have applied a scale cut of $L_{\rm max}=1000$ (as well as subtracting cosmic shear).  All curves shown include marginalization over intrinsic alignments for cosmic shear, with a 25\% prior on the intrinsic alignment amplitude. It is clear that the bias on $M_\nu$ can be reduced significantly in the latter case, without appreciably loosening the constraint on $M_\nu$.  }\label{fig:shear_subtraction}
\end{center}
\end{figure*}

Subtracting the optimally-combined shear map of Sec.~\ref{sec:lsst_shear} from the CMB lensing map results in a significant reduction of the baryonic bias on $M_\nu$ without an appreciable increase in the statistical uncertainty. We show some results in Figure~\ref{fig:shear_subtraction}, where we plot the constraints and bias on $M_\nu$ against $N_{\rm bins}$, the number of galaxy redshift bins  we use to construct $\hat X$. 
Increasing $N_{\rm bins}$ does not increase the number of galaxies in the analysis, but it does have the effect of increasing our redshift resolution and allowing us to weight the galaxy kernels to match the CMB lensing kernel better. Note, however, that at some value of $N_{\rm bins}$, photometric redshift errors will not allow for increasing $N_{\rm bins}$ to give us better redshift resolution, and we expect the curves to saturate around this point in a treatment where photometric redshifts are included. However, for the redshift errors expected of the Rubin Observatory's LSST, we expect to be able to significantly subtract the bias on $M_\nu$ with this method even with a small number of bins, in which case photometric redshift errors will likely be insignificant due to the wide extents of the bins. 

Figure~\ref{fig:shear_subtraction} shows that, with $L_{\rm max}=3100$ for CMB lensing, implementing the shear subtraction generally decreases the bias on $M_\nu$ associated with all but one simulation by factors of 3 to 7 for SO, and a factor of $\sim$2 for S4 (we denote the no-subtraction case by $N_{\rm bins}=0$ in the figure). With $L_{\rm max}=1000$, we similar improvements for SO, but closer to a factor of 5 improvement for S4, in addition to the factor of 2 improvement arising from the $L_{\rm max}$ cut. In all cases, the bulk of the improvement can be obtained with only 2 shear redshift bins (with boundaries $z=[0,0.93,4]$) for the LSST specifications we use. Concurrent with the decrease in bias is an increase in the statistical uncertainty on $M_\nu$, by at most $\sim$20\%. 

Most simulations in Fig.~\ref{fig:shear_subtraction} display similar behavior, with the bias on $M_\nu$ decreasing with $N_{\rm bins}$, but Horizon-AGN exhibits the opposite trend, with the bias {\em increasing} as more redshift bins are used for the subtraction. This is due to the stronger effect of baryons on high-redshift ($z\gtrsim 3$) clustering observed in Horizon-AGN as compared to the other simulations (contrast Fig.~2 of Ref.~\cite{Chisari:2018prw} with results from other simulations summarized in Refs.~\cite{vanDaalen:2019pst, Foreman:2019ahr}). If the high-redshift clustering in Horizon-AGN (primarily driven by gas pressure delaying the collapse of dark matter into halos, rather than AGN feedback~\cite{Chisari:2018prw}), as opposed to that in the other simulations (in which the effect of baryons at high $z$ is much more mild) is reproduced in the actual universe, isolating the high-$z$ part of a CMB lensing map will not be sufficient to mitigate the impact of baryons on a neutrino mass constraint.

\begin{figure*}
\includegraphics[width=0.49\textwidth]{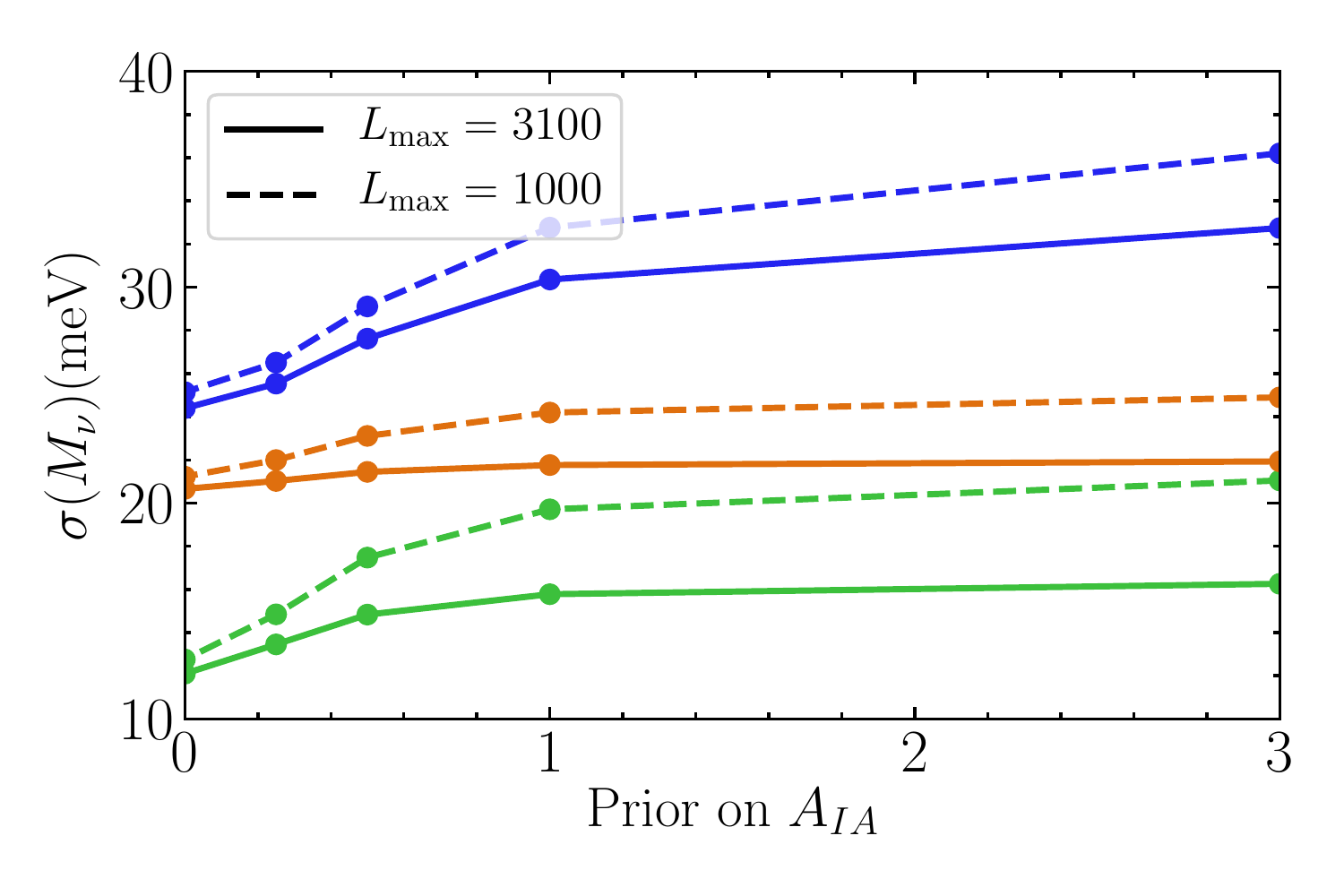}
\includegraphics[width=0.49\textwidth]{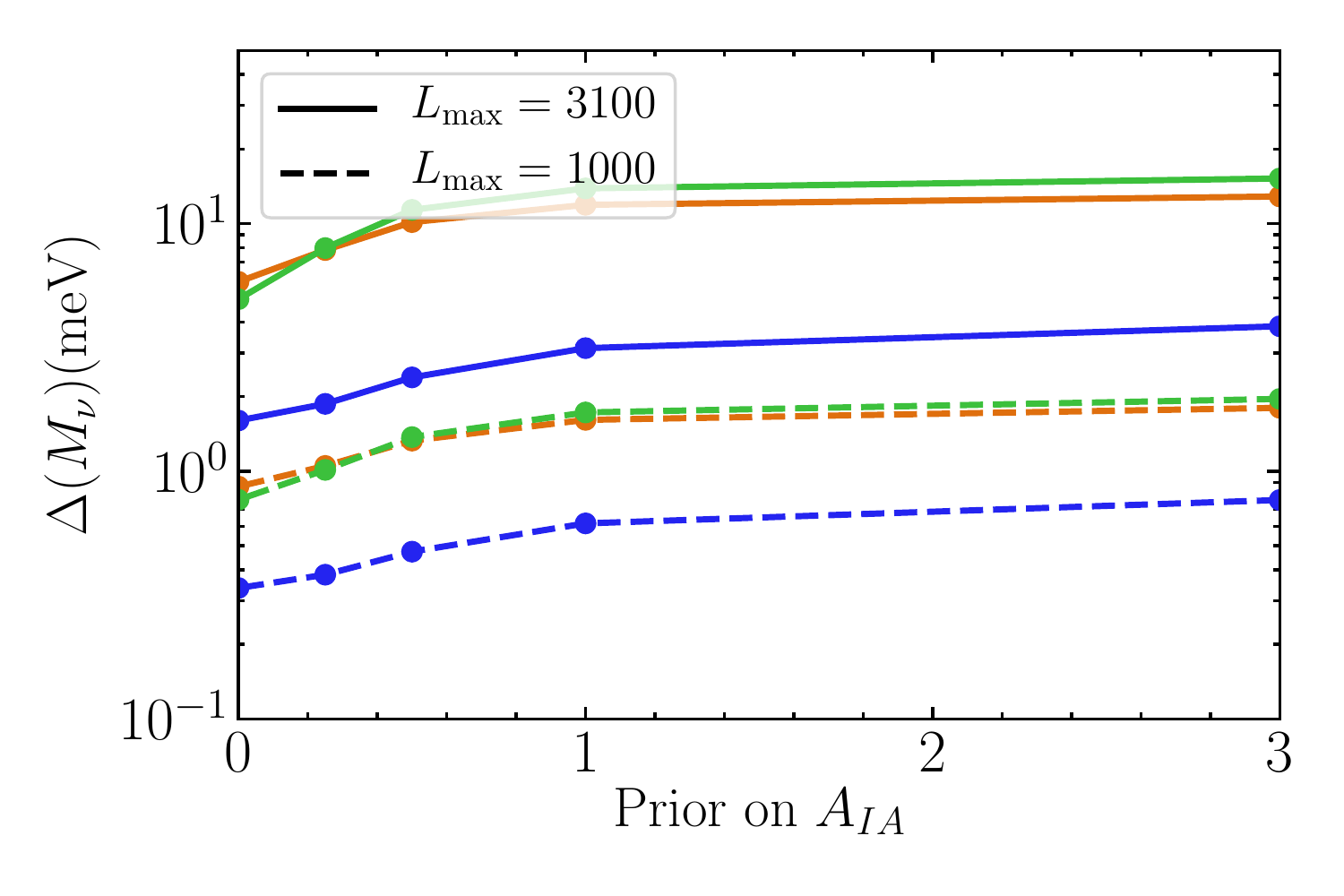}
\includegraphics[width=0.55\textwidth]{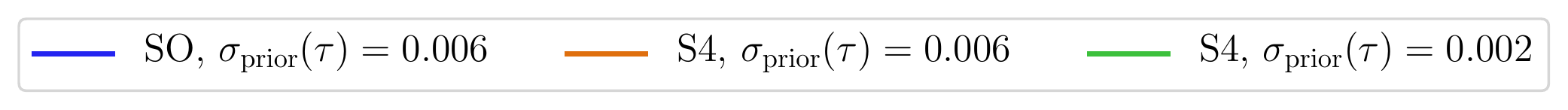}
\caption{The effect of marginalizing over the intrinsic alignment amplitude. If a prior is not included the constraint on $\sigma(M_\nu)$ can increase significantly, as is shown on the left; the bias also increases, as is shown on the right for OWLS-AGN. }\label{fig:aia_plot}
\end{figure*}

In Fig.~\ref{fig:aia_plot}, we show the effect of varying the prior on the intrinsic alignment amplitude $A_{IA}$ on the uncertainty and bias on $M_\nu$, for $N_{\rm bins}=5$ and the bias associated with the OWLS-AGN simulation. For CMB-S4 with the tighter $\tau$ prior, loosening $\sigma(A_{IA})$ from $0.25$ to $3$ degrades the expected errorbar on $M_\nu$ by 15\% for $L_{\rm max}=3100$ or 40\% for $L_{\rm max}=1000$, and increases the bias on $M_\nu$ by a factor of $\sim$2 compared to the $\sigma(A_{IA})=0.25$ case. Thus, with the wider $A_{IA}$ prior, the shear subtraction combined with the $L_{\rm max}$ cut is generally still able to reduce the bias on $M_\nu$ by a factor of $\sim$5-6 compared to the case with no mitigation strategy. On the other hand, exact knowledge of $A_{IA}$ leads to improvements in the uncertainty and bias on $M_\nu$ of a few tens of percents. Note that the nonlinear alignment model we have used for intrinsic alignments will likely be superseded by more detailed models for future shear surveys (e.g.~\cite{Vlah:2019byq,Fortuna:2020vsz}), but our exploration of priors on $A_{IA}$ can be taken as indicative of how one's knowledge of intrinsic alignments affects the shear subtraction procedure we have described.


\section{Strategy 3: marginalization over additional parameters}\label{sec:marginalisation_mead}

The mitigation strategies we considered above involved removing portions of the data most sensitive to baryons, via scale cuts and/or subtracting proxies for low-redshift information in CMB lensing, while retaining as much constraining power on $M_\nu$ as possible. However, as biased constraints on $M_\nu$ arise from neglecting baryonic effects in the theoretical modelling of the matter power spectrum, one can instead incorporate a model for these effects; by marginalizing over the associated parameters, we can hope to reduce the bias on $M_\nu$ without requiring precise knowledge of the impact of baryons. Examples of such models include perturbation theory~\cite{Lewandowski:2014rca,Chen:2019cfu,Braganca:2020nhv}, extended halo models~\cite{Semboloni:2012yh,Mead:2015yca,Mead:2016zqy,Mead:2020qgo,Mead:2020vgs,Debackere:2019cec}, empirical fitting functions~\cite{Harnois-Deraps:2014sva,vanDaalen:2019pst} or principal-component decompositions~\cite{Eifler:2014iva,Mohammed:2017nei,Huang:2018wpy} for the matter power spectrum from simulations, ``baryonification" algorithms that modify the outputs of N-body simulations~\cite{Schneider:2015wta,Schneider:2018pfw,Arico:2019ykw,Arico:2020yyf}, emulators~\cite{Schneider:2019snl}, or approaches based on machine learning~\cite{Troster:2019mys,Villaescusa-Navarro:2020rxg}.   

We choose to test this marginalization approach using the model from Ref.~\cite{Mead:2016zqy}. This model is based on the halo model (e.g.~\cite{Cooray:2002dia}), with a modified 1-halo term for the matter power spectrum. Baryonic effects are parameterized by two parameters. The first, $A$, is the amplitude of the halo concentration-mass relationship $c(M,z)$:
\be
c(M,z)=A\frac{1+z_{\rm f}(M)}{1+z},
\ee
where $z_{\rm f}(M)$ is the formation redshift of halos of mass $M$; this is designed to capture the effects of processes such as gas cooling, which can cause increased halo concentration. The second parameter, $\eta$, alters the (Fourier-transformed) halo density profile $u(k,M)$ via
\be
u(k,M) \rightarrow u(\nu^\eta k,M),
\label{ukmeta}
\ee
where $\nu=\frac{\delta_c}{\sigma(M)}$ with $\delta_c$ the critical density required for spherical collapse and $\sigma(M)$ variance in the initial density fluctuation field when smoothed with a tophat filter with the size of the virial radius of the halo. The Fourier-transformed halo profile is given by
\be
u(k,M)\equiv \frac{1}{M}\int_0^{r_v} dr 4\pi r^2 \rho(r,M)\frac{\sin(kr)}{kr} ,
\ee
where $r_v$ is the virial radius of the halo and $\rho(r,M)$ is the halo density profile in real space. For positive $\eta$, the modification \eqref{ukmeta} ``puffs out'' higher-mass halos ($\nu>1$) and contracts lower mass halos, and as such $\eta$ is referred to as the ``halo bloating parameter''. This is intended to capture some of the effects of AGN feedback on halo profiles. By fitting~$A$ and $\eta$ to the OWLS simulations, Ref.~\cite{Mead:2015yca} found that it was adequate to use a single redshift-independent value for~$A$, while the redshift dependence of $\eta$ was well-captured by
\be
 \eta(z) = \eta_0-0.3\sigma_8(z)
 \label{eq:eta-zdep}
 \ee 
 with a single $\eta_0$, where $\sigma_8$ is the variance of density fluctuations over a sphere with radius $8\invMpc$, and we use those choices in our calculations.

To ensure that the parameter space of $A$ and $\eta_0$ sufficiently describes the baryonic effects in the simulations we are considering, in Appendix~\ref{App:model_fits} we fit these parameters to the $P_{\rm bary}/P_{\rm DMO}$ ratios from each simulation, and compare the ``best-fit'' predictions with the simulations' measurements. We indeed find that the model is able to reproduce all simulation results with a precision of $\sim$5\% over the scales we are concerned with, which is an acceptable level since we are more concerned with the range of simulation results rather than exactly reproducing any one simulation. This agreement also justifies our use of the model from Ref.~\cite{Mead:2016zqy} as opposed to turning to more recent updates (e.g.~\cite{Mead:2020qgo,Mead:2020vgs}).

\begin{table*}
\begin{center}
\begin{tabular}{|c|c|c|c||c|c|c||c|c|}
\hline
Expt	& $\sigma_{\rm prior}(\tau)$ & \multicolumn{2}{c||}{$\sigma(M_\nu)$  [meV]}
	& Simulation &\multicolumn{2}{c||}{Bias $\Delta M_\nu$ [meV] } 
	&\multicolumn{2}{c|}{$\Delta M_\nu/\sigma(M_\nu)$ }\\
	
	&	& no marg. & after marg. 
	&	& no marg. & after marg. & no marg. & after marg. \\
\hline
\hline

\multirow{7}{*}{SO} &\multirow{7}{*}{0.006}&\multirow{7}{*}{ $ 22$}&\multirow{7}{*}{ $  24$}

& OWLS-AGN & 7.2 &  0.14 & 0.32 & 0.0060\\\cline{5-9}
&&&& BAHAMAS & 6.6 &  0.74 & 0.29 & 0.031\\\cline{5-9}
&&&& BAHAMAS-LowAGN & 3.8 &  0.43 & 0.17 & 0.018\\\cline{5-9}
&&&& BAHAMAS-HighAGN & 12 &  1.4 & 0.55 & 0.059\\\cline{5-9}
&&&& Horizon-AGN & 0.88 &  -0.81 & 0.039 & -0.033\\\cline{5-9}
&&&& TNG100 & 0.92 &  0.090 & 0.041 & 0.0038\\\cline{5-9}
&&&& TNG300 & 0.92 &  0.088 & 0.042 & 0.0037  \\\hline\hline
\multirow{7}{*}{S4} &\multirow{7}{*}{0.006}&\multirow{7}{*}{ $ 20 $}&\multirow{7}{*}{ $ 22$}

& OWLS-AGN & 16 &  0.39 & 0.76 & 0.018\\\cline{5-9}
&&&& BAHAMAS & 13 &  1.1 & 0.65 & 0.052\\\cline{5-9}
&&&& BAHAMAS-LowAGN & 7.7 &  0.67 & 0.38 & 0.031\\\cline{5-9}
&&&& BAHAMAS-HighAGN & 24 &  2.4 & 1.2 & 0.11\\\cline{5-9}
&&&& Horizon-AGN & 3.2 &  -0.78 & 0.16 & -0.036\\\cline{5-9}
&&&& TNG100 & 2.0 &  0.22 & 0.098 & 0.010\\\cline{5-9}
&&&& TNG300 & 2.1 &  0.20 & 0.10 & 0.0091  \\
\hline
\hline
\multirow{7}{*}{S4} &\multirow{7}{*}{0.002}&\multirow{7}{*}{ $ 12 $}&\multirow{7}{*}{ $ 14$}

& OWLS-AGN & 14 &  0.37 & 1.2 & 0.027\\\cline{5-9}
&&&& BAHAMAS & 12 &  1.2 & 1.0 & 0.083\\\cline{5-9}
&&&& BAHAMAS-LowAGN & 7.2 &  0.68 & 0.59 & 0.049\\\cline{5-9}
&&&& BAHAMAS-HighAGN & 22 &  2.4 & 1.8 & 0.18\\\cline{5-9}
&&&& Horizon-AGN & 2.8 &  -0.89 & 0.23 & -0.064\\\cline{5-9}
&&&& TNG100 & 1.9 &  0.21 & 0.15 & 0.015\\\cline{5-9}
&&&& TNG300 & 2.0 &  0.19 & 0.16 & 0.014  \\\hline
\end{tabular}
\caption{Expected uncertainty and bias on $M_\nu$ with and without marginalization over the $A$ and $\eta_0$ parameters of modified halo model from Ref.~\cite{Mead:2016zqy}, using CMB lensing multipoles up to $L_{\rm max}=3100$. With marginalization, the biases are reduced by a factor of $\sim$10 in most cases, while the uncertainty is only degraded by $\sim$15\% at most.
}\label{tab:bias_mead}

\end{center}

\end{table*}

To include this model in our forecasts, we use the same formalism as in Sec.~\ref{sec:bias_constraints}, but expanding the vector of parameters in Eq.~\eqref{parameter_vector} to include $A$ and $\eta_0$:
\be
\vec \theta = \left(h, \Omega_bh^2,\Omega_ch^2,\tau, n_s,A_s,M_\nu;A,\eta_0\right).
\ee
The fiducial values we use for $A$ and $\eta_0$ are $A=3.13$ and $\eta_0=0.60$, corresponding to those fit to the DMO run of OWLS in Ref.~\cite{Mead:2015yca}. We do not assume any prior knowledge of the true values of these parameters, which is a conservative choice given the multitude of other datasets which could likely constrain them at some level. Upon marginalizing over $A$ and $\eta_0$, along with the other cosmological parameters, we find the results in Table~\ref{tab:bias_mead}. In particular, we find that the uncertainty on $M_\nu$ is degraded by only $\sim$10\% compared to the case where baryonic effects are ignored in the modelling (equivalent to fixing $A$ and $\eta_0$ to their fiducial values), while the biases on $M_\nu$ are drastically reduced, by factors of $\sim$10 or more. In particular, for CMB-S4 with the tightest $\tau$ prior, the bias corresponding to BAHAMAS-HighAGN is roughly 0.2$\sigma$, while for all other simulations it is less than 0.1$\sigma$; without marginalization, there are 3 simulations that induce a bias exceeding $1\sigma$.

\begin{figure}[t]
\includegraphics[width=0.5\textwidth]{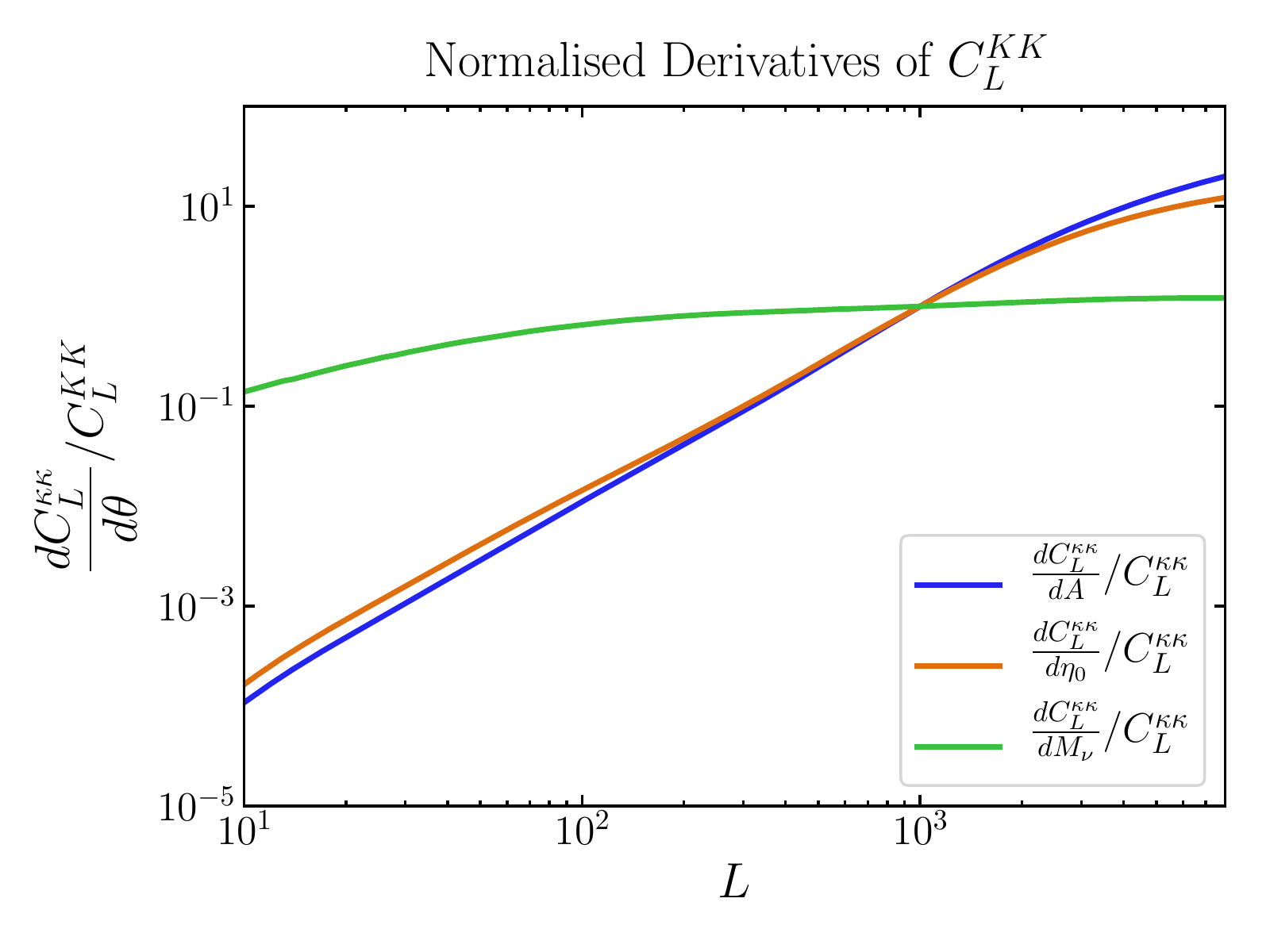}
\caption{The derivatives of the lensing power spectrum with respect to the parameters of the model for baryonic effects from Ref.~\cite{Mead:2016zqy}, and also $M_\nu$, all normalised by their own values at $L=1000$. $M_\nu$ is tending towards a  scale-independent effect on $C_L^{\kappa\kappa}$ across a wide range of scales, while $A$ and $\eta_0$ are significantly scale dependent, becoming increasingly important on small scales.}\label{fig:cl_derivs_mead}
\end{figure}

To understand why this prescription works so well at removing the bias while preserving the constraining power, it is helpful to plot the derivatives of $C_L^{\kappa\kappa}$ with respect to $A$, $\eta_0$, and $M_\nu$; see Fig.~\ref{fig:cl_derivs_mead}. We see that $A$ and $\eta_0$ have significantly different effects on the shape of the lensing power spectrum than $M_\nu$: since they only modify the 1-halo term in the matter power spectrum, they have the strongest impact at small scales, while neutrino mass suppresses structure growth over a wider range of scales (recall Fig.~\ref{fig:baryons_clkk}). This lack of degeneracy implies that the $M_\nu$ constraint is not degraded when $A$ and $\eta_0$ are marginalized over; furthermore, since the model covers the space of baryonic effects well, the marginalization is effective at removing the associated bias from a determination of $M_\nu$.
These conclusions are consistent with other studies of cosmic shear~\cite{Copeland:2019bho,Schneider:2019xpf,Parimbelli:2018yzv}, which have found that when priors from the primary CMB or other observations are included, marginalizing over a baryonic model with only a few parameters enables unbiased constraints on neutrino mass without large increases in uncertainty.

\section{Discussion \& Conclusion}\label{sec:discussion}

Upcoming measurements of CMB lensing have great promise to measure the sum of neutrino masses ($M_\nu$), but this measurement will only be possible if each of several systematic effects are tightly controlled. In this work, we considered one such systematic, related to the impact of ``baryonic effects" (the name given to astrophysical processes like gas cooling and AGN feedback) on the lensing power spectrum. Recent simulations indicate that uncertainty in these effects can bias a neutrino mass measurement from CMB lensing by a sizeable fraction of the statistical errorbar if they are not incorporated in the modelling or mitigated in some other way~\cite{Chung:2019bsk}. 

\begin{figure*}[t]
\begin{center}
\includegraphics[width=0.32\textwidth]{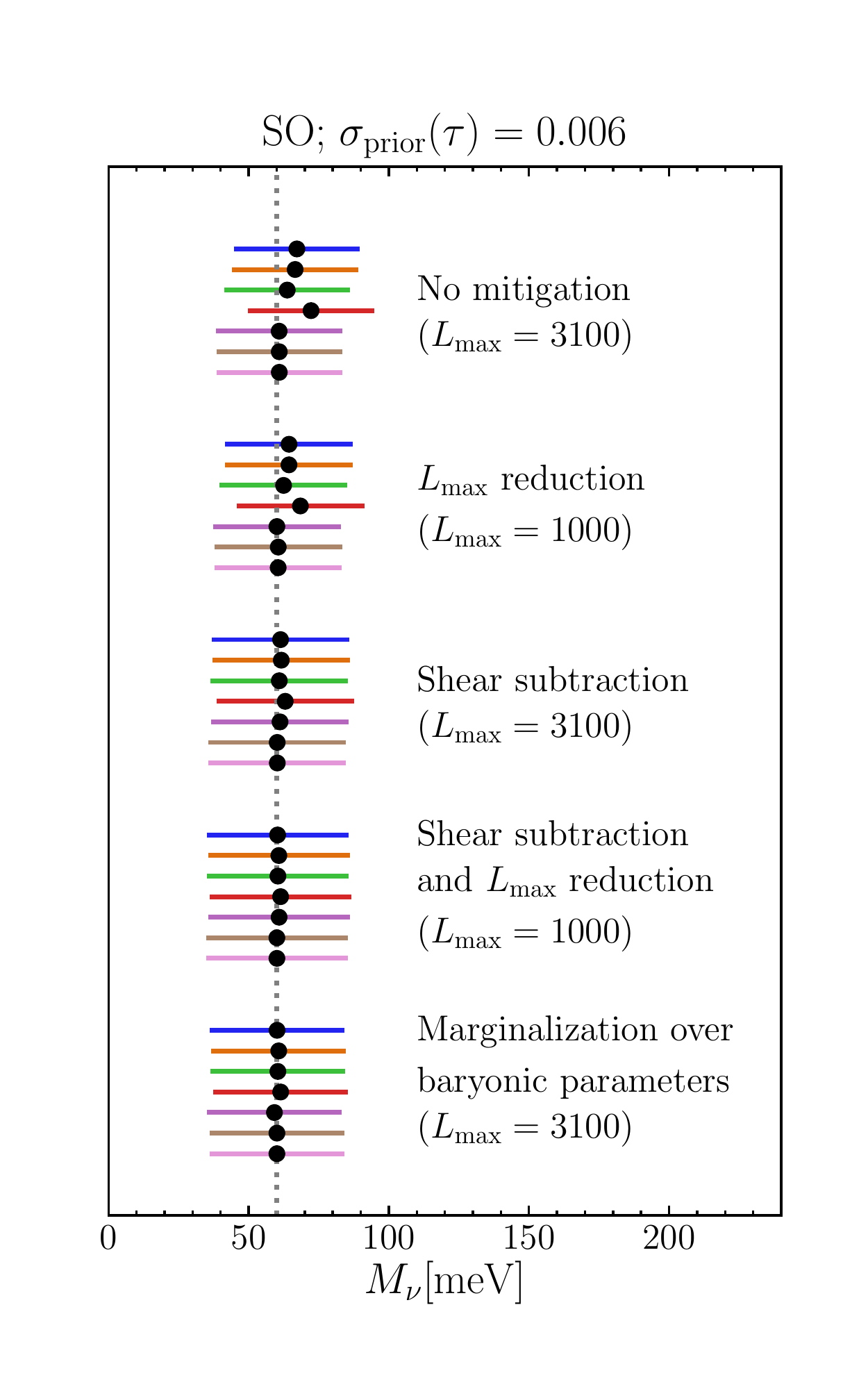}
\includegraphics[width=0.32\textwidth]{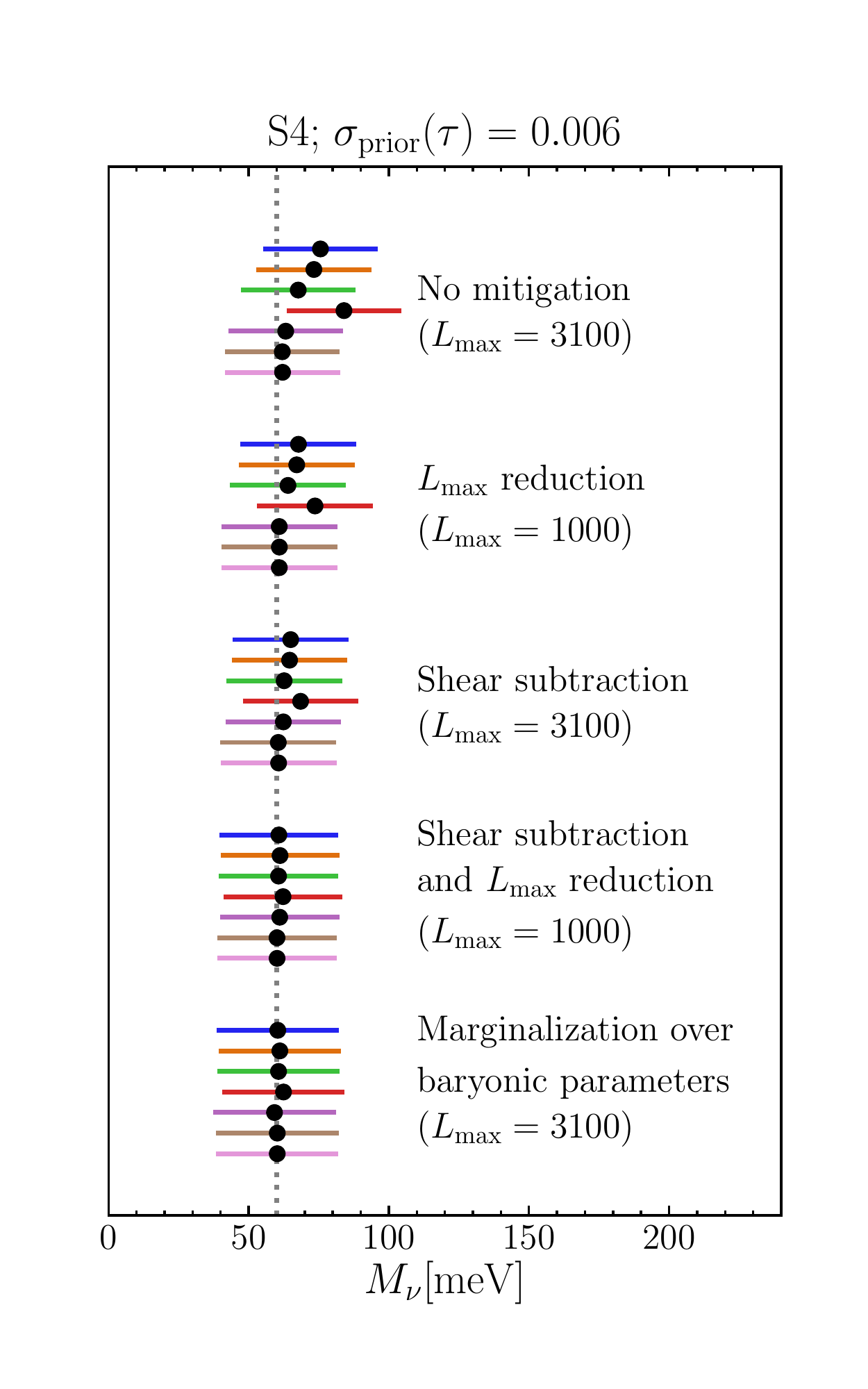}
\includegraphics[width=0.32\textwidth]{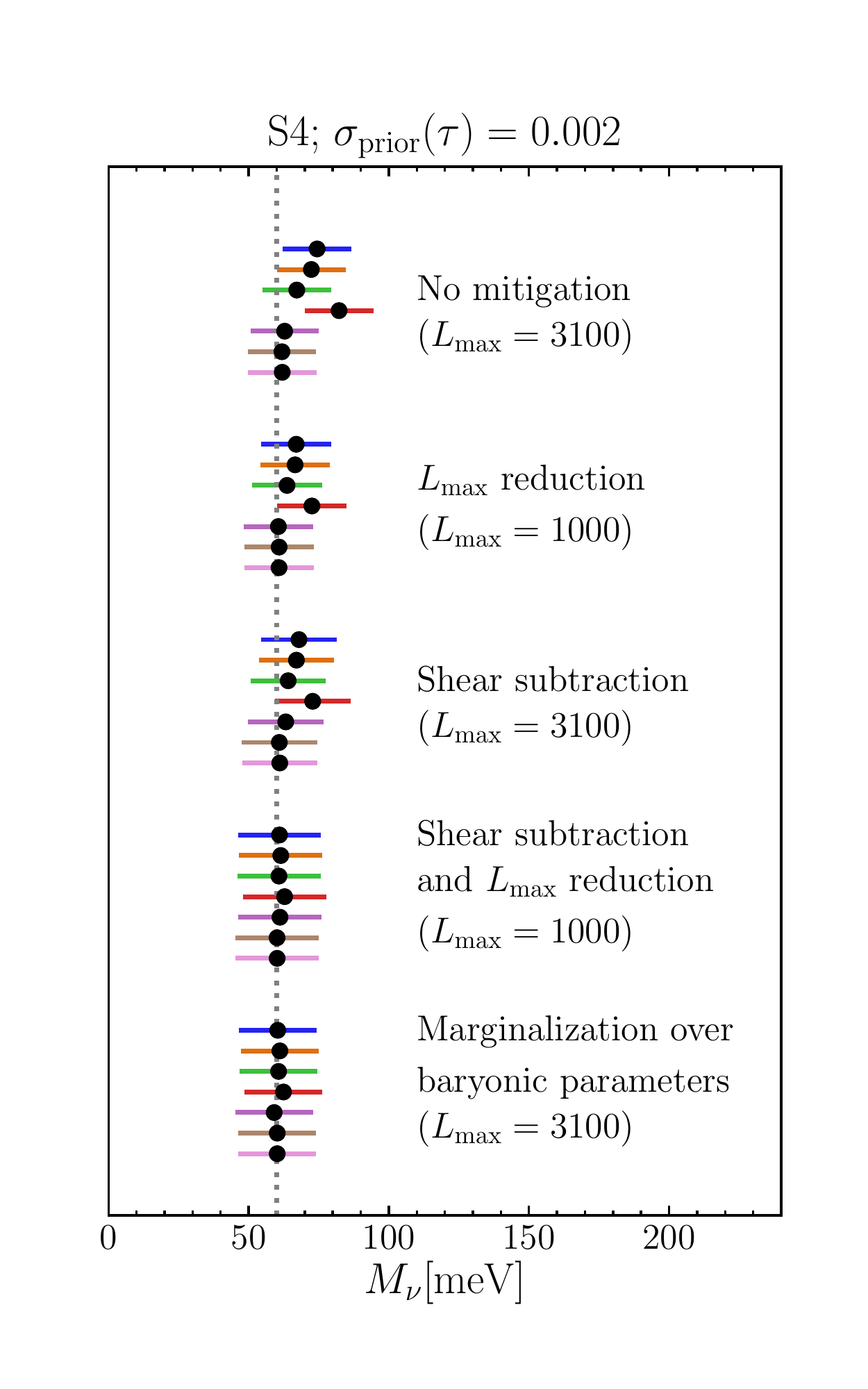}
\includegraphics[scale=0.55]{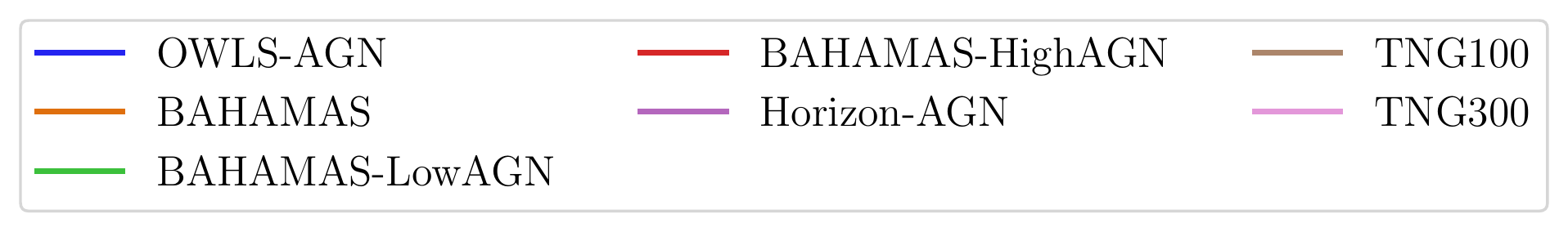}
\caption{A summary of the efficacy of our mitigation methods. For each experimental configuration and each mitigation method, we show the forecast error bars for $M_\nu$, centred on the fiducial value of 60 meV but offset by the bias forecast for each simulation.  Without any mitigation, the biases are substantial and have differing values for different simulations \cite{Chung:2019bsk}.  However, it can be seen that $L_\mathrm{max}$ reduction and shear subtraction each reduce the biases to some extent, and that when both methods are performed the biases are almost completely removed.  Marginalization over baryonic parameters also does an excellent job of almost completely removing the biases.  }\label{fig:bias_summary_plot}
\end{center}
\end{figure*}

We investigated three strategies for mitigating this bias, using Fisher forecasts that combine expected CMB lensing and primary CMB measurements from upcoming experiments, BAO constraints expected from DESI, and a prior on the mean optical depth $\tau$, either from current {\ti {Planck}} constraints or assuming a cosmic-variance-limited measurement. Our chosen strategies and results are as follows:
\begin{enumerate}
\item {\bf Decreasing the largest multipole $L_{\rm max}$ (smallest physical distance) used for neutrino mass constraints (Sec.~\ref{sec:Lmax_cutoff})}:
Baryonic effects are stronger at smaller scales, so one can attempt to reduce sensitivity to these effects by excluding smaller scales from an analysis. We found that keeping $L_{\rm max}\gtrsim 1000$ is necessary to avoid sacrificing significant constraining power on $M_\nu$, and that setting $L_{\rm max}=1000$ can reduce the bias on $M_\nu$ by as much as a factor of 2 (compared to our fiducial forecast with $L_{\rm max}=3100$). However, this reduction still allows the bias be of the same order as the statistical errorbar on $M_\nu$, so further mitigation is needed.
\item {\bf Removing the low-redshift contribution to the lensing map using external tracers (Sec.~\ref{sec:subtraction_cosmicshear})}: CMB lensing probes structures over a wide redshift range, while simulations indicate that baryonic effects should only have a sizeable effect on clustering at the lower end of this range. Thus, one can attempt to remove the low-$z$ contribution from a CMB lensing map, and use the resulting map for a neutrino mass analysis. We have implemented this proposal using a weighted sum of cosmic shear measurements in different source redshift bins. We found that, with $L_{\rm max}=1000$,  subtracting a combination of 2 shear redshift bins (modeled on the Rubin Observatory's LSST) from a CMB lensing map is sufficient to reduce the bias on $M_\nu$ by factors of 5 or more, down to 3 meV for the most extreme simulation we consider. We also considered the effect of intrinsic alignments in the LSST-like data on this procedure, and found a minor impact, when assuming the priors that are expected.
\item {\bf Marginalizing over a parameterization of baryonic effects (Sec.~\ref{sec:marginalisation_mead})}:
If the effects of neutrino mass and baryonic processes are sufficiently nondegenerate in the lensing power spectrum, one can consider including a simple parameterization of the latter in the matter power spectrum model, and marginalizing over the corresponding parameters. After checking that the modified halo model from Ref.~\cite{Mead:2016zqy}  can describe our set of simulations with appropriate accuracy, we found that marginalizing over the model's two parameters only degrades the constraining power on $M_\nu$ by about 15\%, while reducing the bias on $M_\nu$ by factors of 10 or more, down to $\sim$2$\,$meV for the most extreme simulation. This is true even without a reduction in $L_{\rm max}$.
\end{enumerate}
We conclude that either of strategies 2 or 3 should be sufficient to reduce the bias in $M_\nu$ from baryons to an acceptable level ($\Delta M_\nu \lesssim 0.2\sigma(M_\nu)$) for CMB-S4. These results are summarized in Fig.~\ref{fig:bias_summary_plot}.

There is a fourth mitigation strategy that we have not explored in this work: using external measurements to constrain or fix the form of baryonic effects on CMB lensing. It has been shown that the matter power spectrum suppression seen in a range of hydrodynamical simulations is strongly correlated with the mean baryon fraction of group- and cluster-scale halos ($M_{\rm halo} \sim 10^{14} h^{-1}M_\odot$)~\cite{vanDaalen:2019pst,Debackere:2019cec}, so an external constraint on this quantity would likely give a much sharper picture of how much power suppression to include in the modelling. This constraint may be achievable with future X-ray observations of groups and clusters (e.g.~\cite{Gonzalez:2013awy}). For example, when using the model for baryonic effects from Ref.~\cite{Schneider:2018pfw}, Ref.~\cite{Schneider:2019xpf} finds that gas fraction measurements by the upcoming {\ti{eROSITA}} telescope~\cite{Grandis:2018mle}, combined with cluster mass estimates from Euclid weak lensing, will significantly reduce the uncertainty on $M_\nu$ from a combination of cosmic shear and {\ti {Planck}} CMB measurements. However, inference of gas fractions from X-ray measurements requires a measure of the total halo mass, and the various methods to obtain this (e.g.\ assuming hydrostatic equilibrium or using weak lensing of background galaxies) each come with their own caveats. Furthermore, X-ray measurements are most sensitive to hot gas in a halo's interior, while Refs.~\cite{Schneider:2015wta,Debackere:2019cec} have shown that gas at the outskirts of groups and clusters has an important effect on the matter power spectrum.

This fainter, more diffuse gas can be probed using the thermal and kinetic Sunyaev-Zel'dovich (SZ) effects. Mean gas fractions can be extracted from cross-correlations between CMB maps and group/cluster catalogs~\cite{Soergel:2016mce,Vikram:2016dpo,Tanimura:2020une,Lim:2019jfn}, or more generally, stacked gas and pressure profiles can be measured directly from these cross-correlations~\cite{Battaglia:2017neq,Tanimura:2019whh,Amodeo:2020mmu,Ma:2020cir}, with upcoming surveys promising to provide much more powerful measurements~\cite{Pandey:2019uxy,Battaglia:2019dew}. Further constraints are possible by correlating thermal SZ maps and cosmic shear~\cite{Hojjati:2014usa,Hojjati:2016nbx} or CMB lensing itself~\cite{Hill:2013dxa}, helping to break degeneracies between baryonic effects and neutrino mass. Clearly, there are many avenues for independent constraints of baryonic effects, which can be incorporated into an analysis of CMB lensing. However, the strategies presented in this paper do not depend on any such constraints, and therefore represent a promising approach to pursue in parallel.

Finally, it is worth noting that we have made specific choices when implementing these strategies in our forecasts, but other choices are possible. For instance, one could choose a low-redshift tracer other than cosmic shear to implement the map-level subtraction from Sec.~\ref{sec:subtraction_cosmicshear}; another option would be to use spectroscopic or photometric galaxy catalogs, although galaxy bias and selection effects would need to be carefully accounted for in the subtraction procedure. One could also consider marginalizing over other models for baryonic effects, such as effective-field-theory--based perturbation theory, which Ref.~\cite{Braganca:2020nhv} found to be capable of describing baryonic effects on CMB lensing at $L\lesssim 2000$ with suitable accuracy for CMB-S4. Regardless of these specific choices, we expect our general conclusions about these strategies to hold. Therefore, it appears that baryonic effects on a neutrino mass constraint from CMB lensing can be straightforwardly reduced to a negligible level.


\begin{acknowledgments}
We thank Mat Madhavacheril and Blake Sherwin for interesting discussions that provided motivation for this work, and we also thank Shahab Joudaki, Matthew Lewandowski, and Emmanuel Schaan for useful discussions. 
We thank Francisco Villaescusa-Navarro for measuring and providing IllustrisTNG power spectra used in this work, the OWLS and Horizon-AGN teams for making their power spectra publicly available, Mat Madhavacheril for making his forecasting code publicly available, and Eegene Chung for sharing her lensing and forecasting codes.  Research at Perimeter Institute is supported in part by the Government of Canada through the Department of Innovation, Science and Industry Canada and by the Province of Ontario through the Ministry of Colleges and Universities. FMcC acknowledges support from the Vanier Canada Graduate Scholarships program.
\end{acknowledgments}

\bibliography{references}

\begin{thebibliography}{115}%
\makeatletter
\providecommand \@ifxundefined [1]{%
 \@ifx{#1\undefined}
}%
\providecommand \@ifnum [1]{%
 \ifnum #1\expandafter \@firstoftwo
 \else \expandafter \@secondoftwo
 \fi
}%
\providecommand \@ifx [1]{%
 \ifx #1\expandafter \@firstoftwo
 \else \expandafter \@secondoftwo
 \fi
}%
\providecommand \natexlab [1]{#1}%
\providecommand \enquote  [1]{``#1''}%
\providecommand \bibnamefont  [1]{#1}%
\providecommand \bibfnamefont [1]{#1}%
\providecommand \citenamefont [1]{#1}%
\providecommand \href@noop [0]{\@secondoftwo}%
\providecommand \href [0]{\begingroup \@sanitize@url \@href}%
\providecommand \@href[1]{\@@startlink{#1}\@@href}%
\providecommand \@@href[1]{\endgroup#1\@@endlink}%
\providecommand \@sanitize@url [0]{\catcode `\\12\catcode `\$12\catcode
  `\&12\catcode `\#12\catcode `\^12\catcode `\_12\catcode `\%12\relax}%
\providecommand \@@startlink[1]{}%
\providecommand \@@endlink[0]{}%
\providecommand \url  [0]{\begingroup\@sanitize@url \@url }%
\providecommand \@url [1]{\endgroup\@href {#1}{\urlprefix }}%
\providecommand \urlprefix  [0]{URL }%
\providecommand \Eprint [0]{\href }%
\providecommand \doibase [0]{https://doi.org/}%
\providecommand \selectlanguage [0]{\@gobble}%
\providecommand \bibinfo  [0]{\@secondoftwo}%
\providecommand \bibfield  [0]{\@secondoftwo}%
\providecommand \translation [1]{[#1]}%
\providecommand \BibitemOpen [0]{}%
\providecommand \bibitemStop [0]{}%
\providecommand \bibitemNoStop [0]{.\EOS\space}%
\providecommand \EOS [0]{\spacefactor3000\relax}%
\providecommand \BibitemShut  [1]{\csname bibitem#1\endcsname}%
\let\auto@bib@innerbib\@empty
\bibitem [{\citenamefont {Fukuda}\ \emph {et~al.}(1998)\citenamefont {Fukuda}
  \emph {et~al.}}]{PhysRevLett.81.1562}%
  \BibitemOpen
  \bibfield  {author} {\bibinfo {author} {\bibfnamefont {Y.}~\bibnamefont
  {Fukuda}} \emph {et~al.} (\bibinfo {collaboration} {Super-Kamiokande
  Collaboration}),\ }\href {https://doi.org/10.1103/PhysRevLett.81.1562}
  {\bibfield  {journal} {\bibinfo  {journal} {Phys. Rev. Lett.}\ }\textbf
  {\bibinfo {volume} {81}},\ \bibinfo {pages} {1562} (\bibinfo {year}
  {1998})}\BibitemShut {NoStop}%
\bibitem [{\citenamefont {Ahmad}\ \emph {et~al.}(2002)\citenamefont {Ahmad}
  \emph {et~al.}}]{PhysRevLett.89.011301}%
  \BibitemOpen
  \bibfield  {author} {\bibinfo {author} {\bibfnamefont {Q.~R.}\ \bibnamefont
  {Ahmad}} \emph {et~al.} (\bibinfo {collaboration} {SNO Collaboration}),\
  }\href {https://doi.org/10.1103/PhysRevLett.89.011301} {\bibfield  {journal}
  {\bibinfo  {journal} {Phys. Rev. Lett.}\ }\textbf {\bibinfo {volume} {89}},\
  \bibinfo {pages} {011301} (\bibinfo {year} {2002})}\BibitemShut {NoStop}%
\bibitem [{\citenamefont {Patrignani}\ \emph {et~al.}(2016)\citenamefont
  {Patrignani} \emph {et~al.}}]{Patrignani:2016xqp}%
  \BibitemOpen
  \bibfield  {author} {\bibinfo {author} {\bibfnamefont {C.}~\bibnamefont
  {Patrignani}} \emph {et~al.} (\bibinfo {collaboration} {Particle Data
  Group}),\ }\href {https://doi.org/10.1088/1674-1137/40/10/100001} {\bibfield
  {journal} {\bibinfo  {journal} {Chin. Phys. C}\ }\textbf {\bibinfo {volume}
  {40}},\ \bibinfo {pages} {100001} (\bibinfo {year} {2016})}\BibitemShut
  {NoStop}%
\bibitem [{\citenamefont {Aghanim}\ \emph {et~al.}(2020)\citenamefont {Aghanim}
  \emph {et~al.}}]{Aghanim:2018eyx}%
  \BibitemOpen
  \bibfield  {author} {\bibinfo {author} {\bibfnamefont {N.}~\bibnamefont
  {Aghanim}} \emph {et~al.} (\bibinfo {collaboration} {Planck}),\ }\href
  {https://doi.org/10.1051/0004-6361/201833910} {\bibfield  {journal} {\bibinfo
   {journal} {Astron. Astrophys.}\ }\textbf {\bibinfo {volume} {641}},\
  \bibinfo {pages} {A6} (\bibinfo {year} {2020})},\ \Eprint
  {https://arxiv.org/abs/1807.06209} {arXiv:1807.06209 [astro-ph.CO]}
  \BibitemShut {NoStop}%
\bibitem [{\citenamefont {Lesgourgues}\ and\ \citenamefont
  {Pastor}(2006)}]{Lesgourgues:2006nd}%
  \BibitemOpen
  \bibfield  {author} {\bibinfo {author} {\bibfnamefont {J.}~\bibnamefont
  {Lesgourgues}}\ and\ \bibinfo {author} {\bibfnamefont {S.}~\bibnamefont
  {Pastor}},\ }\href {https://doi.org/10.1016/j.physrep.2006.04.001} {\bibfield
   {journal} {\bibinfo  {journal} {Phys. Rept.}\ }\textbf {\bibinfo {volume}
  {429}},\ \bibinfo {pages} {307} (\bibinfo {year} {2006})},\ \Eprint
  {https://arxiv.org/abs/astro-ph/0603494} {arXiv:astro-ph/0603494}
  \BibitemShut {NoStop}%
\bibitem [{\citenamefont {Lesgourgues}\ and\ \citenamefont
  {Pastor}(2012)}]{Lesgourgues:2012uu}%
  \BibitemOpen
  \bibfield  {author} {\bibinfo {author} {\bibfnamefont {J.}~\bibnamefont
  {Lesgourgues}}\ and\ \bibinfo {author} {\bibfnamefont {S.}~\bibnamefont
  {Pastor}},\ }\href {https://doi.org/10.1155/2012/608515} {\bibfield
  {journal} {\bibinfo  {journal} {Adv. High Energy Phys.}\ }\textbf {\bibinfo
  {volume} {2012}},\ \bibinfo {pages} {608515} (\bibinfo {year} {2012})},\
  \Eprint {https://arxiv.org/abs/1212.6154} {arXiv:1212.6154 [hep-ph]}
  \BibitemShut {NoStop}%
\bibitem [{\citenamefont {Ade}\ \emph {et~al.}(2019)\citenamefont {Ade} \emph
  {et~al.}}]{Ade:2018sbj}%
  \BibitemOpen
  \bibfield  {author} {\bibinfo {author} {\bibfnamefont {P.}~\bibnamefont
  {Ade}} \emph {et~al.} (\bibinfo {collaboration} {Simons Observatory}),\
  }\href {https://doi.org/10.1088/1475-7516/2019/02/056} {\bibfield  {journal}
  {\bibinfo  {journal} {JCAP}\ }\textbf {\bibinfo {volume} {02}},\ \bibinfo
  {pages} {056}},\ \Eprint {https://arxiv.org/abs/1808.07445} {arXiv:1808.07445
  [astro-ph.CO]} \BibitemShut {NoStop}%
\bibitem [{\citenamefont {Benson}\ \emph {et~al.}(2014)\citenamefont {Benson}
  \emph {et~al.}}]{Benson:2014qhw}%
  \BibitemOpen
  \bibfield  {author} {\bibinfo {author} {\bibfnamefont {B.}~\bibnamefont
  {Benson}} \emph {et~al.} (\bibinfo {collaboration} {SPT-3G}),\ }\href
  {https://doi.org/10.1117/12.2057305} {\bibfield  {journal} {\bibinfo
  {journal} {Proc. SPIE Int. Soc. Opt. Eng.}\ }\textbf {\bibinfo {volume}
  {9153}},\ \bibinfo {pages} {91531P} (\bibinfo {year} {2014})},\ \Eprint
  {https://arxiv.org/abs/1407.2973} {arXiv:1407.2973 [astro-ph.IM]}
  \BibitemShut {NoStop}%
\bibitem [{\citenamefont {Abazajian}\ \emph {et~al.}(2016)\citenamefont
  {Abazajian} \emph {et~al.}}]{Abazajian:2016yjj}%
  \BibitemOpen
  \bibfield  {author} {\bibinfo {author} {\bibfnamefont {K.~N.}\ \bibnamefont
  {Abazajian}} \emph {et~al.} (\bibinfo {collaboration} {CMB-S4}),\ }\href@noop
  {} {\  (\bibinfo {year} {2016})},\ \Eprint {https://arxiv.org/abs/1610.02743}
  {arXiv:1610.02743 [astro-ph.CO]} \BibitemShut {NoStop}%
\bibitem [{\citenamefont {Allison}\ \emph {et~al.}(2015)\citenamefont
  {Allison}, \citenamefont {Caucal}, \citenamefont {Calabrese}, \citenamefont
  {Dunkley},\ and\ \citenamefont {Louis}}]{Allison:2015qca}%
  \BibitemOpen
  \bibfield  {author} {\bibinfo {author} {\bibfnamefont {R.}~\bibnamefont
  {Allison}}, \bibinfo {author} {\bibfnamefont {P.}~\bibnamefont {Caucal}},
  \bibinfo {author} {\bibfnamefont {E.}~\bibnamefont {Calabrese}}, \bibinfo
  {author} {\bibfnamefont {J.}~\bibnamefont {Dunkley}},\ and\ \bibinfo {author}
  {\bibfnamefont {T.}~\bibnamefont {Louis}},\ }\href
  {https://doi.org/10.1103/PhysRevD.92.123535} {\bibfield  {journal} {\bibinfo
  {journal} {Phys. Rev. D}\ }\textbf {\bibinfo {volume} {92}},\ \bibinfo
  {pages} {123535} (\bibinfo {year} {2015})},\ \Eprint
  {https://arxiv.org/abs/1509.07471} {arXiv:1509.07471 [astro-ph.CO]}
  \BibitemShut {NoStop}%
\bibitem [{\citenamefont {White}(2004)}]{White:2004kv}%
  \BibitemOpen
  \bibfield  {author} {\bibinfo {author} {\bibfnamefont {M.~J.}\ \bibnamefont
  {White}},\ }\href {https://doi.org/10.1016/j.astropartphys.2004.06.001}
  {\bibfield  {journal} {\bibinfo  {journal} {Astropart. Phys.}\ }\textbf
  {\bibinfo {volume} {22}},\ \bibinfo {pages} {211} (\bibinfo {year} {2004})},\
  \Eprint {https://arxiv.org/abs/astro-ph/0405593} {arXiv:astro-ph/0405593}
  \BibitemShut {NoStop}%
\bibitem [{\citenamefont {Zhan}\ and\ \citenamefont
  {Knox}(2004)}]{Zhan:2004wq}%
  \BibitemOpen
  \bibfield  {author} {\bibinfo {author} {\bibfnamefont {H.}~\bibnamefont
  {Zhan}}\ and\ \bibinfo {author} {\bibfnamefont {L.}~\bibnamefont {Knox}},\
  }\href {https://doi.org/10.1086/426712} {\bibfield  {journal} {\bibinfo
  {journal} {Astrophys. J.}\ }\textbf {\bibinfo {volume} {616}},\ \bibinfo
  {pages} {L75} (\bibinfo {year} {2004})},\ \Eprint
  {https://arxiv.org/abs/astro-ph/0409198} {arXiv:astro-ph/0409198}
  \BibitemShut {NoStop}%
\bibitem [{\citenamefont {Jing}\ \emph {et~al.}(2006)\citenamefont {Jing},
  \citenamefont {Zhang}, \citenamefont {Lin}, \citenamefont {Gao},\ and\
  \citenamefont {Springel}}]{Jing:2005gm}%
  \BibitemOpen
  \bibfield  {author} {\bibinfo {author} {\bibfnamefont {Y.~P.}\ \bibnamefont
  {Jing}}, \bibinfo {author} {\bibfnamefont {P.}~\bibnamefont {Zhang}},
  \bibinfo {author} {\bibfnamefont {W.~P.}\ \bibnamefont {Lin}}, \bibinfo
  {author} {\bibfnamefont {L.}~\bibnamefont {Gao}},\ and\ \bibinfo {author}
  {\bibfnamefont {V.}~\bibnamefont {Springel}},\ }\href
  {https://doi.org/10.1086/503547} {\bibfield  {journal} {\bibinfo  {journal}
  {Astrophys. J.}\ }\textbf {\bibinfo {volume} {640}},\ \bibinfo {pages} {L119}
  (\bibinfo {year} {2006})},\ \Eprint {https://arxiv.org/abs/astro-ph/0512426}
  {arXiv:astro-ph/0512426} \BibitemShut {NoStop}%
\bibitem [{\citenamefont {Rudd}\ \emph {et~al.}(2008)\citenamefont {Rudd},
  \citenamefont {Zentner},\ and\ \citenamefont {Kravtsov}}]{Rudd:2007zx}%
  \BibitemOpen
  \bibfield  {author} {\bibinfo {author} {\bibfnamefont {D.~H.}\ \bibnamefont
  {Rudd}}, \bibinfo {author} {\bibfnamefont {A.~R.}\ \bibnamefont {Zentner}},\
  and\ \bibinfo {author} {\bibfnamefont {A.~V.}\ \bibnamefont {Kravtsov}},\
  }\href {https://doi.org/10.1086/523836} {\bibfield  {journal} {\bibinfo
  {journal} {Astrophys. J.}\ }\textbf {\bibinfo {volume} {672}},\ \bibinfo
  {pages} {19} (\bibinfo {year} {2008})},\ \Eprint
  {https://arxiv.org/abs/astro-ph/0703741} {arXiv:astro-ph/0703741}
  \BibitemShut {NoStop}%
\bibitem [{\citenamefont {Semboloni}\ \emph {et~al.}(2011)\citenamefont
  {Semboloni}, \citenamefont {Hoekstra}, \citenamefont {Schaye}, \citenamefont
  {van Daalen},\ and\ \citenamefont {McCarthy}}]{Semboloni:2011fe}%
  \BibitemOpen
  \bibfield  {author} {\bibinfo {author} {\bibfnamefont {E.}~\bibnamefont
  {Semboloni}}, \bibinfo {author} {\bibfnamefont {H.}~\bibnamefont {Hoekstra}},
  \bibinfo {author} {\bibfnamefont {J.}~\bibnamefont {Schaye}}, \bibinfo
  {author} {\bibfnamefont {M.~P.}\ \bibnamefont {van Daalen}},\ and\ \bibinfo
  {author} {\bibfnamefont {I.~J.}\ \bibnamefont {McCarthy}},\ }\href
  {https://doi.org/10.1111/j.1365-2966.2011.19385.x} {\bibfield  {journal}
  {\bibinfo  {journal} {Mon. Not. Roy. Astron. Soc.}\ }\textbf {\bibinfo
  {volume} {417}},\ \bibinfo {pages} {2020} (\bibinfo {year} {2011})},\ \Eprint
  {https://arxiv.org/abs/1105.1075} {arXiv:1105.1075 [astro-ph.CO]}
  \BibitemShut {NoStop}%
\bibitem [{\citenamefont {Natarajan}\ \emph {et~al.}(2014)\citenamefont
  {Natarajan}, \citenamefont {Zentner}, \citenamefont {Battaglia},\ and\
  \citenamefont {Trac}}]{Natarajan:2014xba}%
  \BibitemOpen
  \bibfield  {author} {\bibinfo {author} {\bibfnamefont {A.}~\bibnamefont
  {Natarajan}}, \bibinfo {author} {\bibfnamefont {A.~R.}\ \bibnamefont
  {Zentner}}, \bibinfo {author} {\bibfnamefont {N.}~\bibnamefont {Battaglia}},\
  and\ \bibinfo {author} {\bibfnamefont {H.}~\bibnamefont {Trac}},\ }\href
  {https://doi.org/10.1103/PhysRevD.90.063516} {\bibfield  {journal} {\bibinfo
  {journal} {Phys. Rev. D}\ }\textbf {\bibinfo {volume} {90}},\ \bibinfo
  {pages} {063516} (\bibinfo {year} {2014})},\ \Eprint
  {https://arxiv.org/abs/1405.6205} {arXiv:1405.6205 [astro-ph.CO]}
  \BibitemShut {NoStop}%
\bibitem [{\citenamefont {Copeland}\ \emph {et~al.}(2020)\citenamefont
  {Copeland}, \citenamefont {Taylor},\ and\ \citenamefont
  {Hall}}]{Copeland:2019bho}%
  \BibitemOpen
  \bibfield  {author} {\bibinfo {author} {\bibfnamefont {D.}~\bibnamefont
  {Copeland}}, \bibinfo {author} {\bibfnamefont {A.}~\bibnamefont {Taylor}},\
  and\ \bibinfo {author} {\bibfnamefont {A.}~\bibnamefont {Hall}},\ }\href
  {https://doi.org/10.1093/mnras/staa314} {\bibfield  {journal} {\bibinfo
  {journal} {Mon. Not. Roy. Astron. Soc.}\ }\textbf {\bibinfo {volume} {493}},\
  \bibinfo {pages} {1640} (\bibinfo {year} {2020})},\ \Eprint
  {https://arxiv.org/abs/1905.08754} {arXiv:1905.08754 [astro-ph.CO]}
  \BibitemShut {NoStop}%
\bibitem [{\citenamefont {Schneider}\ \emph
  {et~al.}(2020{\natexlab{a}})\citenamefont {Schneider}, \citenamefont
  {Refregier}, \citenamefont {Grandis}, \citenamefont {Eckert}, \citenamefont
  {Stoira}, \citenamefont {Kacprzak}, \citenamefont {Knabenhans}, \citenamefont
  {Stadel},\ and\ \citenamefont {Teyssier}}]{Schneider:2019xpf}%
  \BibitemOpen
  \bibfield  {author} {\bibinfo {author} {\bibfnamefont {A.}~\bibnamefont
  {Schneider}}, \bibinfo {author} {\bibfnamefont {A.}~\bibnamefont
  {Refregier}}, \bibinfo {author} {\bibfnamefont {S.}~\bibnamefont {Grandis}},
  \bibinfo {author} {\bibfnamefont {D.}~\bibnamefont {Eckert}}, \bibinfo
  {author} {\bibfnamefont {N.}~\bibnamefont {Stoira}}, \bibinfo {author}
  {\bibfnamefont {T.}~\bibnamefont {Kacprzak}}, \bibinfo {author}
  {\bibfnamefont {M.}~\bibnamefont {Knabenhans}}, \bibinfo {author}
  {\bibfnamefont {J.}~\bibnamefont {Stadel}},\ and\ \bibinfo {author}
  {\bibfnamefont {R.}~\bibnamefont {Teyssier}},\ }\href
  {https://doi.org/10.1088/1475-7516/2020/04/020} {\bibfield  {journal}
  {\bibinfo  {journal} {JCAP}\ }\textbf {\bibinfo {volume} {04}},\ \bibinfo
  {pages} {020}},\ \Eprint {https://arxiv.org/abs/1911.08494} {arXiv:1911.08494
  [astro-ph.CO]} \BibitemShut {NoStop}%
\bibitem [{\citenamefont {Chung}\ \emph {et~al.}(2020)\citenamefont {Chung},
  \citenamefont {Foreman},\ and\ \citenamefont {van Engelen}}]{Chung:2019bsk}%
  \BibitemOpen
  \bibfield  {author} {\bibinfo {author} {\bibfnamefont {E.}~\bibnamefont
  {Chung}}, \bibinfo {author} {\bibfnamefont {S.}~\bibnamefont {Foreman}},\
  and\ \bibinfo {author} {\bibfnamefont {A.}~\bibnamefont {van Engelen}},\
  }\href {https://doi.org/10.1103/PhysRevD.101.063534} {\bibfield  {journal}
  {\bibinfo  {journal} {Phys. Rev. D}\ }\textbf {\bibinfo {volume} {101}},\
  \bibinfo {pages} {063534} (\bibinfo {year} {2020})},\ \Eprint
  {https://arxiv.org/abs/1910.09565} {arXiv:1910.09565 [astro-ph.CO]}
  \BibitemShut {NoStop}%
\bibitem [{\citenamefont {Harnois-D\'eraps}\ \emph {et~al.}(2015)\citenamefont
  {Harnois-D\'eraps}, \citenamefont {van Waerbeke}, \citenamefont {Viola},\
  and\ \citenamefont {Heymans}}]{Harnois-Deraps:2014sva}%
  \BibitemOpen
  \bibfield  {author} {\bibinfo {author} {\bibfnamefont {J.}~\bibnamefont
  {Harnois-D\'eraps}}, \bibinfo {author} {\bibfnamefont {L.}~\bibnamefont {van
  Waerbeke}}, \bibinfo {author} {\bibfnamefont {M.}~\bibnamefont {Viola}},\
  and\ \bibinfo {author} {\bibfnamefont {C.}~\bibnamefont {Heymans}},\ }\href
  {https://doi.org/10.1093/mnras/stv646} {\bibfield  {journal} {\bibinfo
  {journal} {Mon. Not. Roy. Astron. Soc.}\ }\textbf {\bibinfo {volume} {450}},\
  \bibinfo {pages} {1212} (\bibinfo {year} {2015})},\ \Eprint
  {https://arxiv.org/abs/1407.4301} {arXiv:1407.4301 [astro-ph.CO]}
  \BibitemShut {NoStop}%
\bibitem [{\citenamefont {Foreman}\ \emph {et~al.}(2016)\citenamefont
  {Foreman}, \citenamefont {Becker},\ and\ \citenamefont
  {Wechsler}}]{Foreman:2016jzy}%
  \BibitemOpen
  \bibfield  {author} {\bibinfo {author} {\bibfnamefont {S.}~\bibnamefont
  {Foreman}}, \bibinfo {author} {\bibfnamefont {M.~R.}\ \bibnamefont
  {Becker}},\ and\ \bibinfo {author} {\bibfnamefont {R.~H.}\ \bibnamefont
  {Wechsler}},\ }\href {https://doi.org/10.1093/mnras/stw2189} {\bibfield
  {journal} {\bibinfo  {journal} {Mon. Not. Roy. Astron. Soc.}\ }\textbf
  {\bibinfo {volume} {463}},\ \bibinfo {pages} {3326} (\bibinfo {year}
  {2016})},\ \Eprint {https://arxiv.org/abs/1605.09056} {arXiv:1605.09056
  [astro-ph.CO]} \BibitemShut {NoStop}%
\bibitem [{\citenamefont {Huang}\ \emph {et~al.}(2020)\citenamefont {Huang}
  \emph {et~al.}}]{Huang:2020tpm}%
  \BibitemOpen
  \bibfield  {author} {\bibinfo {author} {\bibfnamefont {H.-J.}\ \bibnamefont
  {Huang}} \emph {et~al.} (\bibinfo {collaboration} {DES}),\ }\href@noop {} {\
  (\bibinfo {year} {2020})},\ \Eprint {https://arxiv.org/abs/2007.15026}
  {arXiv:2007.15026 [astro-ph.CO]} \BibitemShut {NoStop}%
\bibitem [{\citenamefont {Yoon}\ and\ \citenamefont
  {Jee}(2020)}]{Yoon:2020bop}%
  \BibitemOpen
  \bibfield  {author} {\bibinfo {author} {\bibfnamefont {M.}~\bibnamefont
  {Yoon}}\ and\ \bibinfo {author} {\bibfnamefont {M.~J.}\ \bibnamefont {Jee}},\
  }\href@noop {} {\  (\bibinfo {year} {2020})},\ \Eprint
  {https://arxiv.org/abs/2007.16166} {arXiv:2007.16166 [astro-ph.CO]}
  \BibitemShut {NoStop}%
\bibitem [{\citenamefont {Lewis}\ and\ \citenamefont
  {Challinor}(2006)}]{Lewis:2006fu}%
  \BibitemOpen
  \bibfield  {author} {\bibinfo {author} {\bibfnamefont {A.}~\bibnamefont
  {Lewis}}\ and\ \bibinfo {author} {\bibfnamefont {A.}~\bibnamefont
  {Challinor}},\ }\href {https://doi.org/10.1016/j.physrep.2006.03.002}
  {\bibfield  {journal} {\bibinfo  {journal} {Phys. Rept.}\ }\textbf {\bibinfo
  {volume} {429}},\ \bibinfo {pages} {1} (\bibinfo {year} {2006})},\ \Eprint
  {https://arxiv.org/abs/astro-ph/0601594} {arXiv:astro-ph/0601594 [astro-ph]}
  \BibitemShut {NoStop}%
\bibitem [{\citenamefont {{Limber}}(1953)}]{1953ApJ...117..134L}%
  \BibitemOpen
  \bibfield  {author} {\bibinfo {author} {\bibfnamefont {D.~N.}\ \bibnamefont
  {{Limber}}},\ }\href {https://doi.org/10.1086/145672} {\bibfield  {journal}
  {\bibinfo  {journal} {\apj}\ }\textbf {\bibinfo {volume} {117}},\ \bibinfo
  {pages} {134} (\bibinfo {year} {1953})}\BibitemShut {NoStop}%
\bibitem [{\citenamefont {LoVerde}\ and\ \citenamefont
  {Afshordi}(2008)}]{LoVerde:2008re}%
  \BibitemOpen
  \bibfield  {author} {\bibinfo {author} {\bibfnamefont {M.}~\bibnamefont
  {LoVerde}}\ and\ \bibinfo {author} {\bibfnamefont {N.}~\bibnamefont
  {Afshordi}},\ }\href {https://doi.org/10.1103/PhysRevD.78.123506} {\bibfield
  {journal} {\bibinfo  {journal} {Phys. Rev.}\ }\textbf {\bibinfo {volume}
  {D78}},\ \bibinfo {pages} {123506} (\bibinfo {year} {2008})},\ \Eprint
  {https://arxiv.org/abs/0809.5112} {arXiv:0809.5112 [astro-ph]} \BibitemShut
  {NoStop}%
\bibitem [{\citenamefont {Pratten}\ and\ \citenamefont
  {Lewis}(2016)}]{Pratten:2016dsm}%
  \BibitemOpen
  \bibfield  {author} {\bibinfo {author} {\bibfnamefont {G.}~\bibnamefont
  {Pratten}}\ and\ \bibinfo {author} {\bibfnamefont {A.}~\bibnamefont
  {Lewis}},\ }\href {https://doi.org/10.1088/1475-7516/2016/08/047} {\bibfield
  {journal} {\bibinfo  {journal} {JCAP}\ }\textbf {\bibinfo {volume}
  {1608}}\bibfield  {number} {\bibinfo  {number} { (08)},\ \bibinfo {pages}
  {047}},\ }\Eprint {https://arxiv.org/abs/1605.05662} {arXiv:1605.05662
  [astro-ph.CO]} \BibitemShut {NoStop}%
\bibitem [{\citenamefont {Fabbian}\ \emph {et~al.}(2018)\citenamefont
  {Fabbian}, \citenamefont {Calabrese},\ and\ \citenamefont
  {Carbone}}]{Fabbian:2017wfp}%
  \BibitemOpen
  \bibfield  {author} {\bibinfo {author} {\bibfnamefont {G.}~\bibnamefont
  {Fabbian}}, \bibinfo {author} {\bibfnamefont {M.}~\bibnamefont {Calabrese}},\
  and\ \bibinfo {author} {\bibfnamefont {C.}~\bibnamefont {Carbone}},\ }\href
  {https://doi.org/10.1088/1475-7516/2018/02/050} {\bibfield  {journal}
  {\bibinfo  {journal} {JCAP}\ }\textbf {\bibinfo {volume} {1802}}\bibfield
  {number} {\bibinfo  {number} { (02)},\ \bibinfo {pages} {050}},\ }\Eprint
  {https://arxiv.org/abs/1702.03317} {arXiv:1702.03317 [astro-ph.CO]}
  \BibitemShut {NoStop}%
\bibitem [{\citenamefont {Duffy}\ \emph {et~al.}(2010)\citenamefont {Duffy},
  \citenamefont {Schaye}, \citenamefont {Kay}, \citenamefont {Dalla~Vecchia},
  \citenamefont {Battye},\ and\ \citenamefont {Booth}}]{Duffy:2010hf}%
  \BibitemOpen
  \bibfield  {author} {\bibinfo {author} {\bibfnamefont {A.~R.}\ \bibnamefont
  {Duffy}}, \bibinfo {author} {\bibfnamefont {J.}~\bibnamefont {Schaye}},
  \bibinfo {author} {\bibfnamefont {S.~T.}\ \bibnamefont {Kay}}, \bibinfo
  {author} {\bibfnamefont {C.}~\bibnamefont {Dalla~Vecchia}}, \bibinfo {author}
  {\bibfnamefont {R.~A.}\ \bibnamefont {Battye}},\ and\ \bibinfo {author}
  {\bibfnamefont {C.}~\bibnamefont {Booth}},\ }\href
  {https://doi.org/10.1111/j.1365-2966.2010.16613.x} {\bibfield  {journal}
  {\bibinfo  {journal} {Mon. Not. Roy. Astron. Soc.}\ }\textbf {\bibinfo
  {volume} {405}},\ \bibinfo {pages} {2161} (\bibinfo {year} {2010})},\ \Eprint
  {https://arxiv.org/abs/1001.3447} {arXiv:1001.3447 [astro-ph.CO]}
  \BibitemShut {NoStop}%
\bibitem [{\citenamefont {Lewis}\ \emph {et~al.}(2000)\citenamefont {Lewis},
  \citenamefont {Challinor},\ and\ \citenamefont {Lasenby}}]{Lewis:1999bs}%
  \BibitemOpen
  \bibfield  {author} {\bibinfo {author} {\bibfnamefont {A.}~\bibnamefont
  {Lewis}}, \bibinfo {author} {\bibfnamefont {A.}~\bibnamefont {Challinor}},\
  and\ \bibinfo {author} {\bibfnamefont {A.}~\bibnamefont {Lasenby}},\ }\href
  {https://doi.org/10.1086/309179} {\bibfield  {journal} {\bibinfo  {journal}
  {apj}\ }\textbf {\bibinfo {volume} {538}},\ \bibinfo {pages} {473} (\bibinfo
  {year} {2000})},\ \Eprint {https://arxiv.org/abs/astro-ph/9911177}
  {arXiv:astro-ph/9911177 [astro-ph]} \BibitemShut {NoStop}%
\bibitem [{\citenamefont {Mead}\ \emph {et~al.}(2016)\citenamefont {Mead},
  \citenamefont {Heymans}, \citenamefont {Lombriser}, \citenamefont {Peacock},
  \citenamefont {Steele},\ and\ \citenamefont {Winther}}]{Mead:2016zqy}%
  \BibitemOpen
  \bibfield  {author} {\bibinfo {author} {\bibfnamefont {A.}~\bibnamefont
  {Mead}}, \bibinfo {author} {\bibfnamefont {C.}~\bibnamefont {Heymans}},
  \bibinfo {author} {\bibfnamefont {L.}~\bibnamefont {Lombriser}}, \bibinfo
  {author} {\bibfnamefont {J.}~\bibnamefont {Peacock}}, \bibinfo {author}
  {\bibfnamefont {O.}~\bibnamefont {Steele}},\ and\ \bibinfo {author}
  {\bibfnamefont {H.}~\bibnamefont {Winther}},\ }\href
  {https://doi.org/10.1093/mnras/stw681} {\bibfield  {journal} {\bibinfo
  {journal} {Mon. Not. Roy. Astron. Soc.}\ }\textbf {\bibinfo {volume} {459}},\
  \bibinfo {pages} {1468} (\bibinfo {year} {2016})},\ \Eprint
  {https://arxiv.org/abs/1602.02154} {arXiv:1602.02154 [astro-ph.CO]}
  \BibitemShut {NoStop}%
\bibitem [{\citenamefont {Aghamousa}\ \emph {et~al.}(2016)\citenamefont
  {Aghamousa} \emph {et~al.}}]{Aghamousa:2016zmz}%
  \BibitemOpen
  \bibfield  {author} {\bibinfo {author} {\bibfnamefont {A.}~\bibnamefont
  {Aghamousa}} \emph {et~al.} (\bibinfo {collaboration} {DESI}),\ }\href@noop
  {} {\  (\bibinfo {year} {2016})},\ \Eprint {https://arxiv.org/abs/1611.00036}
  {arXiv:1611.00036 [astro-ph.IM]} \BibitemShut {NoStop}%
\bibitem [{\citenamefont {Peloton}\ \emph {et~al.}(2017)\citenamefont
  {Peloton}, \citenamefont {Schmittfull}, \citenamefont {Lewis}, \citenamefont
  {Carron},\ and\ \citenamefont {Zahn}}]{Peloton:2016kbw}%
  \BibitemOpen
  \bibfield  {author} {\bibinfo {author} {\bibfnamefont {J.}~\bibnamefont
  {Peloton}}, \bibinfo {author} {\bibfnamefont {M.}~\bibnamefont
  {Schmittfull}}, \bibinfo {author} {\bibfnamefont {A.}~\bibnamefont {Lewis}},
  \bibinfo {author} {\bibfnamefont {J.}~\bibnamefont {Carron}},\ and\ \bibinfo
  {author} {\bibfnamefont {O.}~\bibnamefont {Zahn}},\ }\href
  {https://doi.org/10.1103/PhysRevD.95.043508} {\bibfield  {journal} {\bibinfo
  {journal} {Phys. Rev. D}\ }\textbf {\bibinfo {volume} {95}},\ \bibinfo
  {pages} {043508} (\bibinfo {year} {2017})},\ \Eprint
  {https://arxiv.org/abs/1611.01446} {arXiv:1611.01446 [astro-ph.CO]}
  \BibitemShut {NoStop}%
\bibitem [{\citenamefont {Green}\ \emph {et~al.}(2017)\citenamefont {Green},
  \citenamefont {Meyers},\ and\ \citenamefont {van Engelen}}]{Green:2016cjr}%
  \BibitemOpen
  \bibfield  {author} {\bibinfo {author} {\bibfnamefont {D.}~\bibnamefont
  {Green}}, \bibinfo {author} {\bibfnamefont {J.}~\bibnamefont {Meyers}},\ and\
  \bibinfo {author} {\bibfnamefont {A.}~\bibnamefont {van Engelen}},\ }\href
  {https://doi.org/10.1088/1475-7516/2017/12/005} {\bibfield  {journal}
  {\bibinfo  {journal} {JCAP}\ }\textbf {\bibinfo {volume} {12}},\ \bibinfo
  {pages} {005}},\ \Eprint {https://arxiv.org/abs/1609.08143} {arXiv:1609.08143
  [astro-ph.CO]} \BibitemShut {NoStop}%
\bibitem [{\citenamefont {Cooray}(2002)}]{Cooray:2001ab}%
  \BibitemOpen
  \bibfield  {author} {\bibinfo {author} {\bibfnamefont {A.}~\bibnamefont
  {Cooray}},\ }\href {https://doi.org/10.1103/PhysRevD.65.103510} {\bibfield
  {journal} {\bibinfo  {journal} {Phys. Rev. D}\ }\textbf {\bibinfo {volume}
  {65}},\ \bibinfo {pages} {103510} (\bibinfo {year} {2002})},\ \Eprint
  {https://arxiv.org/abs/astro-ph/0112408} {arXiv:astro-ph/0112408}
  \BibitemShut {NoStop}%
\bibitem [{\citenamefont {Lewis}\ \emph {et~al.}(2011)\citenamefont {Lewis},
  \citenamefont {Challinor},\ and\ \citenamefont {Hanson}}]{Lewis:2011fk}%
  \BibitemOpen
  \bibfield  {author} {\bibinfo {author} {\bibfnamefont {A.}~\bibnamefont
  {Lewis}}, \bibinfo {author} {\bibfnamefont {A.}~\bibnamefont {Challinor}},\
  and\ \bibinfo {author} {\bibfnamefont {D.}~\bibnamefont {Hanson}},\ }\href
  {https://doi.org/10.1088/1475-7516/2011/03/018} {\bibfield  {journal}
  {\bibinfo  {journal} {JCAP}\ }\textbf {\bibinfo {volume} {03}},\ \bibinfo
  {pages} {018}},\ \Eprint {https://arxiv.org/abs/1101.2234} {arXiv:1101.2234
  [astro-ph.CO]} \BibitemShut {NoStop}%
\bibitem [{\citenamefont {Pagano}\ \emph {et~al.}(2019)\citenamefont {Pagano},
  \citenamefont {Delouis}, \citenamefont {Mottet}, \citenamefont {Puget},\ and\
  \citenamefont {Vibert}}]{Pagano:2019tci}%
  \BibitemOpen
  \bibfield  {author} {\bibinfo {author} {\bibfnamefont {L.}~\bibnamefont
  {Pagano}}, \bibinfo {author} {\bibfnamefont {J.~M.}\ \bibnamefont {Delouis}},
  \bibinfo {author} {\bibfnamefont {S.}~\bibnamefont {Mottet}}, \bibinfo
  {author} {\bibfnamefont {J.~L.}\ \bibnamefont {Puget}},\ and\ \bibinfo
  {author} {\bibfnamefont {L.}~\bibnamefont {Vibert}},\ }\href@noop {} {\
  (\bibinfo {year} {2019})},\ \Eprint {https://arxiv.org/abs/1908.09856}
  {arXiv:1908.09856 [astro-ph.CO]} \BibitemShut {NoStop}%
\bibitem [{\citenamefont {Li}\ \emph {et~al.}(2018)\citenamefont {Li},
  \citenamefont {Gluscevic}, \citenamefont {Boddy},\ and\ \citenamefont
  {Madhavacheril}}]{Li:2018zdm}%
  \BibitemOpen
  \bibfield  {author} {\bibinfo {author} {\bibfnamefont {Z.}~\bibnamefont
  {Li}}, \bibinfo {author} {\bibfnamefont {V.}~\bibnamefont {Gluscevic}},
  \bibinfo {author} {\bibfnamefont {K.~K.}\ \bibnamefont {Boddy}},\ and\
  \bibinfo {author} {\bibfnamefont {M.~S.}\ \bibnamefont {Madhavacheril}},\
  }\href {https://doi.org/10.1103/PhysRevD.98.123524} {\bibfield  {journal}
  {\bibinfo  {journal} {Phys. Rev.}\ }\textbf {\bibinfo {volume} {D98}},\
  \bibinfo {pages} {123524} (\bibinfo {year} {2018})},\ \Eprint
  {https://arxiv.org/abs/1806.10165} {arXiv:1806.10165 [astro-ph.CO]}
  \BibitemShut {NoStop}%
\bibitem [{\citenamefont {van Daalen}\ \emph
  {et~al.}(2011{\natexlab{a}})\citenamefont {van Daalen}, \citenamefont
  {Schaye}, \citenamefont {Booth},\ and\ \citenamefont
  {Vecchia}}]{vanDaalen2011}%
  \BibitemOpen
  \bibfield  {author} {\bibinfo {author} {\bibfnamefont {M.~P.}\ \bibnamefont
  {van Daalen}}, \bibinfo {author} {\bibfnamefont {J.}~\bibnamefont {Schaye}},
  \bibinfo {author} {\bibfnamefont {C.~M.}\ \bibnamefont {Booth}},\ and\
  \bibinfo {author} {\bibfnamefont {C.~D.}\ \bibnamefont {Vecchia}},\ }\href
  {https://doi.org/10.1111/j.1365-2966.2011.18981.x} {\bibfield  {journal}
  {\bibinfo  {journal} {Mon. Not. Roy. Astron. Soc.}\ }\textbf {\bibinfo
  {volume} {415}},\ \bibinfo {pages} {3649} (\bibinfo {year}
  {2011}{\natexlab{a}})},\ \Eprint {https://arxiv.org/abs/1104.1174}
  {arXiv:1104.1174 [astro-ph.CO]} \BibitemShut {NoStop}%
\bibitem [{\citenamefont {van Daalen}\ \emph
  {et~al.}(2011{\natexlab{b}})\citenamefont {van Daalen}, \citenamefont
  {Schaye}, \citenamefont {Booth},\ and\ \citenamefont
  {Vecchia}}]{vanDaalen:2011xb}%
  \BibitemOpen
  \bibfield  {author} {\bibinfo {author} {\bibfnamefont {M.~P.}\ \bibnamefont
  {van Daalen}}, \bibinfo {author} {\bibfnamefont {J.}~\bibnamefont {Schaye}},
  \bibinfo {author} {\bibfnamefont {C.}~\bibnamefont {Booth}},\ and\ \bibinfo
  {author} {\bibfnamefont {C.~D.}\ \bibnamefont {Vecchia}},\ }\href
  {https://doi.org/10.1111/j.1365-2966.2011.18981.x} {\bibfield  {journal}
  {\bibinfo  {journal} {Mon. Not. Roy. Astron. Soc.}\ }\textbf {\bibinfo
  {volume} {415}},\ \bibinfo {pages} {3649} (\bibinfo {year}
  {2011}{\natexlab{b}})},\ \Eprint {https://arxiv.org/abs/1104.1174}
  {arXiv:1104.1174 [astro-ph.CO]} \BibitemShut {NoStop}%
\bibitem [{\citenamefont {van Daalen}\ \emph {et~al.}(2020)\citenamefont {van
  Daalen}, \citenamefont {McCarthy},\ and\ \citenamefont
  {Schaye}}]{vanDaalen:2019pst}%
  \BibitemOpen
  \bibfield  {author} {\bibinfo {author} {\bibfnamefont {M.~P.}\ \bibnamefont
  {van Daalen}}, \bibinfo {author} {\bibfnamefont {I.~G.}\ \bibnamefont
  {McCarthy}},\ and\ \bibinfo {author} {\bibfnamefont {J.}~\bibnamefont
  {Schaye}},\ }\href {https://doi.org/10.1093/mnras/stz3199} {\bibfield
  {journal} {\bibinfo  {journal} {Mon. Not. Roy. Astron. Soc.}\ }\textbf
  {\bibinfo {volume} {491}},\ \bibinfo {pages} {2424} (\bibinfo {year}
  {2020})},\ \Eprint {https://arxiv.org/abs/1906.00968} {arXiv:1906.00968
  [astro-ph.CO]} \BibitemShut {NoStop}%
\bibitem [{\citenamefont {McCarthy}\ \emph {et~al.}(2017)\citenamefont
  {McCarthy}, \citenamefont {Schaye}, \citenamefont {Bird},\ and\ \citenamefont
  {Le~Brun}}]{McCarthy:2016mry}%
  \BibitemOpen
  \bibfield  {author} {\bibinfo {author} {\bibfnamefont {I.~G.}\ \bibnamefont
  {McCarthy}}, \bibinfo {author} {\bibfnamefont {J.}~\bibnamefont {Schaye}},
  \bibinfo {author} {\bibfnamefont {S.}~\bibnamefont {Bird}},\ and\ \bibinfo
  {author} {\bibfnamefont {A.~M.}\ \bibnamefont {Le~Brun}},\ }\href
  {https://doi.org/10.1093/mnras/stw2792} {\bibfield  {journal} {\bibinfo
  {journal} {Mon. Not. Roy. Astron. Soc.}\ }\textbf {\bibinfo {volume} {465}},\
  \bibinfo {pages} {2936} (\bibinfo {year} {2017})},\ \Eprint
  {https://arxiv.org/abs/1603.02702} {arXiv:1603.02702 [astro-ph.CO]}
  \BibitemShut {NoStop}%
\bibitem [{\citenamefont {Mccarthy}\ \emph {et~al.}(2018)\citenamefont
  {Mccarthy}, \citenamefont {Bird}, \citenamefont {Schaye}, \citenamefont
  {Harnois-Deraps}, \citenamefont {Font},\ and\ \citenamefont
  {Van~Waerbeke}}]{McCarthy:2017csu}%
  \BibitemOpen
  \bibfield  {author} {\bibinfo {author} {\bibfnamefont {I.~G.}\ \bibnamefont
  {Mccarthy}}, \bibinfo {author} {\bibfnamefont {S.}~\bibnamefont {Bird}},
  \bibinfo {author} {\bibfnamefont {J.}~\bibnamefont {Schaye}}, \bibinfo
  {author} {\bibfnamefont {J.}~\bibnamefont {Harnois-Deraps}}, \bibinfo
  {author} {\bibfnamefont {A.~S.}\ \bibnamefont {Font}},\ and\ \bibinfo
  {author} {\bibfnamefont {L.}~\bibnamefont {Van~Waerbeke}},\ }\href
  {https://doi.org/10.1093/mnras/sty377} {\bibfield  {journal} {\bibinfo
  {journal} {Mon. Not. Roy. Astron. Soc.}\ }\textbf {\bibinfo {volume} {476}},\
  \bibinfo {pages} {2999} (\bibinfo {year} {2018})},\ \Eprint
  {https://arxiv.org/abs/1712.02411} {arXiv:1712.02411 [astro-ph.CO]}
  \BibitemShut {NoStop}%
\bibitem [{\citenamefont {Dubois}\ \emph {et~al.}(2014)\citenamefont {Dubois}
  \emph {et~al.}}]{Dubois:2014lxa}%
  \BibitemOpen
  \bibfield  {author} {\bibinfo {author} {\bibfnamefont {Y.}~\bibnamefont
  {Dubois}} \emph {et~al.},\ }\href {https://doi.org/10.1093/mnras/stu1227}
  {\bibfield  {journal} {\bibinfo  {journal} {Mon. Not. Roy. Astron. Soc.}\
  }\textbf {\bibinfo {volume} {444}},\ \bibinfo {pages} {1453} (\bibinfo {year}
  {2014})},\ \Eprint {https://arxiv.org/abs/1402.1165} {arXiv:1402.1165
  [astro-ph.CO]} \BibitemShut {NoStop}%
\bibitem [{\citenamefont {Dubois}\ \emph {et~al.}(2016)\citenamefont {Dubois}
  \emph {et~al.}}]{Dubois:2016}%
  \BibitemOpen
  \bibfield  {author} {\bibinfo {author} {\bibfnamefont {Y.}~\bibnamefont
  {Dubois}} \emph {et~al.},\ }\href {https://doi.org/10.1093/mnras/stw2265}
  {\bibfield  {journal} {\bibinfo  {journal} {Mon. Not. Roy. Astron. Soc.}\
  }\textbf {\bibinfo {volume} {463}},\ \bibinfo {pages} {3948} (\bibinfo {year}
  {2016})},\ \Eprint {https://arxiv.org/abs/1606.03086} {arXiv:1606.03086
  [astro-ph.GA]} \BibitemShut {NoStop}%
\bibitem [{\citenamefont {Chisari}\ \emph {et~al.}(2018)\citenamefont
  {Chisari}, \citenamefont {Richardson}, \citenamefont {Devriendt},
  \citenamefont {Dubois}, \citenamefont {Schneider}, \citenamefont {Brun},
  \citenamefont {Beckmann}, \citenamefont {Peirani}, \citenamefont {Slyz},\
  and\ \citenamefont {Pichon}}]{Chisari:2018prw}%
  \BibitemOpen
  \bibfield  {author} {\bibinfo {author} {\bibfnamefont {N.~E.}\ \bibnamefont
  {Chisari}}, \bibinfo {author} {\bibfnamefont {M.~L.}\ \bibnamefont
  {Richardson}}, \bibinfo {author} {\bibfnamefont {J.}~\bibnamefont
  {Devriendt}}, \bibinfo {author} {\bibfnamefont {Y.}~\bibnamefont {Dubois}},
  \bibinfo {author} {\bibfnamefont {A.}~\bibnamefont {Schneider}}, \bibinfo
  {author} {\bibfnamefont {M.}~\bibnamefont {Brun}, \bibfnamefont
  {Amandine~Le}}, \bibinfo {author} {\bibfnamefont {R.~S.}\ \bibnamefont
  {Beckmann}}, \bibinfo {author} {\bibfnamefont {S.}~\bibnamefont {Peirani}},
  \bibinfo {author} {\bibfnamefont {A.}~\bibnamefont {Slyz}},\ and\ \bibinfo
  {author} {\bibfnamefont {C.}~\bibnamefont {Pichon}},\ }\href
  {https://doi.org/10.1093/mnras/sty2093} {\bibfield  {journal} {\bibinfo
  {journal} {Mon. Not. Roy. Astron. Soc.}\ }\textbf {\bibinfo {volume} {480}},\
  \bibinfo {pages} {3962} (\bibinfo {year} {2018})},\ \Eprint
  {https://arxiv.org/abs/1801.08559} {arXiv:1801.08559 [astro-ph.CO]}
  \BibitemShut {NoStop}%
\bibitem [{\citenamefont {Pillepich}\ \emph {et~al.}(2018)\citenamefont
  {Pillepich} \emph {et~al.}}]{Pillepich:2017fcc}%
  \BibitemOpen
  \bibfield  {author} {\bibinfo {author} {\bibfnamefont {A.}~\bibnamefont
  {Pillepich}} \emph {et~al.},\ }\href {https://doi.org/10.1093/mnras/stx3112}
  {\bibfield  {journal} {\bibinfo  {journal} {Mon. Not. Roy. Astron. Soc.}\
  }\textbf {\bibinfo {volume} {475}},\ \bibinfo {pages} {648} (\bibinfo {year}
  {2018})},\ \Eprint {https://arxiv.org/abs/1707.03406} {arXiv:1707.03406
  [astro-ph.GA]} \BibitemShut {NoStop}%
\bibitem [{\citenamefont {Springel}\ \emph {et~al.}(2018)\citenamefont
  {Springel} \emph {et~al.}}]{Springel:2017tpz}%
  \BibitemOpen
  \bibfield  {author} {\bibinfo {author} {\bibfnamefont {V.}~\bibnamefont
  {Springel}} \emph {et~al.},\ }\href {https://doi.org/10.1093/mnras/stx3304}
  {\bibfield  {journal} {\bibinfo  {journal} {Mon. Not. Roy. Astron. Soc.}\
  }\textbf {\bibinfo {volume} {475}},\ \bibinfo {pages} {676} (\bibinfo {year}
  {2018})},\ \Eprint {https://arxiv.org/abs/1707.03397} {arXiv:1707.03397
  [astro-ph.GA]} \BibitemShut {NoStop}%
\bibitem [{\citenamefont {Nelson}\ \emph
  {et~al.}(2018{\natexlab{a}})\citenamefont {Nelson} \emph
  {et~al.}}]{Nelson:2017cxy}%
  \BibitemOpen
  \bibfield  {author} {\bibinfo {author} {\bibfnamefont {D.}~\bibnamefont
  {Nelson}} \emph {et~al.},\ }\href {https://doi.org/10.1093/mnras/stx3040}
  {\bibfield  {journal} {\bibinfo  {journal} {Mon. Not. Roy. Astron. Soc.}\
  }\textbf {\bibinfo {volume} {475}},\ \bibinfo {pages} {624} (\bibinfo {year}
  {2018}{\natexlab{a}})},\ \Eprint {https://arxiv.org/abs/1707.03395}
  {arXiv:1707.03395 [astro-ph.GA]} \BibitemShut {NoStop}%
\bibitem [{\citenamefont {Naiman}\ \emph {et~al.}(2018)\citenamefont {Naiman}
  \emph {et~al.}}]{Naiman:2018}%
  \BibitemOpen
  \bibfield  {author} {\bibinfo {author} {\bibfnamefont {J.~P.}\ \bibnamefont
  {Naiman}} \emph {et~al.},\ }\href {https://doi.org/10.1093/mnras/sty618}
  {\bibfield  {journal} {\bibinfo  {journal} {Mon. Not. Roy. Astron. Soc.}\
  }\textbf {\bibinfo {volume} {480}},\ \bibinfo {pages} {1206} (\bibinfo {year}
  {2018})},\ \Eprint {https://arxiv.org/abs/1707.03401} {arXiv:1707.03401
  [astro-ph.GA]} \BibitemShut {NoStop}%
\bibitem [{\citenamefont {Marinacci}\ \emph {et~al.}(2018)\citenamefont
  {Marinacci} \emph {et~al.}}]{Marinacci:2017wew}%
  \BibitemOpen
  \bibfield  {author} {\bibinfo {author} {\bibfnamefont {F.}~\bibnamefont
  {Marinacci}} \emph {et~al.},\ }\href {https://doi.org/10.1093/mnras/sty2206}
  {\bibfield  {journal} {\bibinfo  {journal} {Mon. Not. Roy. Astron. Soc.}\
  }\textbf {\bibinfo {volume} {480}},\ \bibinfo {pages} {5113} (\bibinfo {year}
  {2018})},\ \Eprint {https://arxiv.org/abs/1707.03396} {arXiv:1707.03396
  [astro-ph.CO]} \BibitemShut {NoStop}%
\bibitem [{\citenamefont {Nelson}\ \emph
  {et~al.}(2018{\natexlab{b}})\citenamefont {Nelson} \emph
  {et~al.}}]{Nelson:2018uso}%
  \BibitemOpen
  \bibfield  {author} {\bibinfo {author} {\bibfnamefont {D.}~\bibnamefont
  {Nelson}} \emph {et~al.},\ }\href@noop {} {\  (\bibinfo {year}
  {2018}{\natexlab{b}})},\ \Eprint {https://arxiv.org/abs/1812.05609}
  {arXiv:1812.05609 [astro-ph.GA]} \BibitemShut {NoStop}%
\bibitem [{\citenamefont {Foreman}\ \emph {et~al.}(2020)\citenamefont
  {Foreman}, \citenamefont {Coulton}, \citenamefont {Villaescusa-Navarro},\
  and\ \citenamefont {Barreira}}]{Foreman:2019ahr}%
  \BibitemOpen
  \bibfield  {author} {\bibinfo {author} {\bibfnamefont {S.}~\bibnamefont
  {Foreman}}, \bibinfo {author} {\bibfnamefont {W.}~\bibnamefont {Coulton}},
  \bibinfo {author} {\bibfnamefont {F.}~\bibnamefont {Villaescusa-Navarro}},\
  and\ \bibinfo {author} {\bibfnamefont {A.}~\bibnamefont {Barreira}},\ }\href
  {https://doi.org/10.1093/mnras/staa2523} {\bibfield  {journal} {\bibinfo
  {journal} {Mon. Not. Roy. Astron. Soc.}\ }\textbf {\bibinfo {volume} {498}},\
  \bibinfo {pages} {2887} (\bibinfo {year} {2020})},\ \Eprint
  {https://arxiv.org/abs/1910.03597} {arXiv:1910.03597 [astro-ph.CO]}
  \BibitemShut {NoStop}%
\bibitem [{\citenamefont {Mead}\ \emph {et~al.}(2015)\citenamefont {Mead},
  \citenamefont {Peacock}, \citenamefont {Heymans}, \citenamefont {Joudaki},\
  and\ \citenamefont {Heavens}}]{Mead:2015yca}%
  \BibitemOpen
  \bibfield  {author} {\bibinfo {author} {\bibfnamefont {A.}~\bibnamefont
  {Mead}}, \bibinfo {author} {\bibfnamefont {J.}~\bibnamefont {Peacock}},
  \bibinfo {author} {\bibfnamefont {C.}~\bibnamefont {Heymans}}, \bibinfo
  {author} {\bibfnamefont {S.}~\bibnamefont {Joudaki}},\ and\ \bibinfo {author}
  {\bibfnamefont {A.}~\bibnamefont {Heavens}},\ }\href
  {https://doi.org/10.1093/mnras/stv2036} {\bibfield  {journal} {\bibinfo
  {journal} {Mon. Not. Roy. Astron. Soc.}\ }\textbf {\bibinfo {volume} {454}},\
  \bibinfo {pages} {1958} (\bibinfo {year} {2015})},\ \Eprint
  {https://arxiv.org/abs/1505.07833} {arXiv:1505.07833 [astro-ph.CO]}
  \BibitemShut {NoStop}%
\bibitem [{\citenamefont {Mummery}\ \emph {et~al.}(2017)\citenamefont
  {Mummery}, \citenamefont {McCarthy}, \citenamefont {Bird},\ and\
  \citenamefont {Schaye}}]{Mummery:2017lcn}%
  \BibitemOpen
  \bibfield  {author} {\bibinfo {author} {\bibfnamefont {B.~O.}\ \bibnamefont
  {Mummery}}, \bibinfo {author} {\bibfnamefont {I.~G.}\ \bibnamefont
  {McCarthy}}, \bibinfo {author} {\bibfnamefont {S.}~\bibnamefont {Bird}},\
  and\ \bibinfo {author} {\bibfnamefont {J.}~\bibnamefont {Schaye}},\ }\href
  {https://doi.org/10.1093/mnras/stx1469} {\bibfield  {journal} {\bibinfo
  {journal} {Mon. Not. Roy. Astron. Soc.}\ }\textbf {\bibinfo {volume} {471}},\
  \bibinfo {pages} {227} (\bibinfo {year} {2017})},\ \Eprint
  {https://arxiv.org/abs/1702.02064} {arXiv:1702.02064 [astro-ph.CO]}
  \BibitemShut {NoStop}%
\bibitem [{\citenamefont {Bragan\c{c}a}\ \emph {et~al.}(2020)\citenamefont
  {Bragan\c{c}a}, \citenamefont {Lewandowski}, \citenamefont {Sekera},
  \citenamefont {Senatore},\ and\ \citenamefont {Sgier}}]{Braganca:2020nhv}%
  \BibitemOpen
  \bibfield  {author} {\bibinfo {author} {\bibfnamefont {D.~P.}\ \bibnamefont
  {Bragan\c{c}a}}, \bibinfo {author} {\bibfnamefont {M.}~\bibnamefont
  {Lewandowski}}, \bibinfo {author} {\bibfnamefont {D.}~\bibnamefont {Sekera}},
  \bibinfo {author} {\bibfnamefont {L.}~\bibnamefont {Senatore}},\ and\
  \bibinfo {author} {\bibfnamefont {R.}~\bibnamefont {Sgier}},\ }\href@noop {}
  {\  (\bibinfo {year} {2020})},\ \Eprint {https://arxiv.org/abs/2010.02929}
  {arXiv:2010.02929 [astro-ph.CO]} \BibitemShut {NoStop}%
\bibitem [{\citenamefont {Abazajian}\ \emph {et~al.}(2019)\citenamefont
  {Abazajian} \emph {et~al.}}]{Abazajian:2019eic}%
  \BibitemOpen
  \bibfield  {author} {\bibinfo {author} {\bibfnamefont {K.}~\bibnamefont
  {Abazajian}} \emph {et~al.},\ }\href@noop {} {\  (\bibinfo {year} {2019})},\
  \Eprint {https://arxiv.org/abs/1907.04473} {arXiv:1907.04473 [astro-ph.IM]}
  \BibitemShut {NoStop}%
\bibitem [{\citenamefont {Okamoto}\ and\ \citenamefont
  {Hu}(2003)}]{Okamoto:2003zw}%
  \BibitemOpen
  \bibfield  {author} {\bibinfo {author} {\bibfnamefont {T.}~\bibnamefont
  {Okamoto}}\ and\ \bibinfo {author} {\bibfnamefont {W.}~\bibnamefont {Hu}},\
  }\href {https://doi.org/10.1103/PhysRevD.67.083002} {\bibfield  {journal}
  {\bibinfo  {journal} {Phys. Rev.}\ }\textbf {\bibinfo {volume} {D67}},\
  \bibinfo {pages} {083002} (\bibinfo {year} {2003})},\ \Eprint
  {https://arxiv.org/abs/astro-ph/0301031} {arXiv:astro-ph/0301031 [astro-ph]}
  \BibitemShut {NoStop}%
\bibitem [{\citenamefont {van Engelen}\ \emph {et~al.}(2014)\citenamefont {van
  Engelen}, \citenamefont {Bhattacharya}, \citenamefont {Sehgal}, \citenamefont
  {Holder}, \citenamefont {Zahn},\ and\ \citenamefont
  {Nagai}}]{vanEngelen:2013rla}%
  \BibitemOpen
  \bibfield  {author} {\bibinfo {author} {\bibfnamefont {A.}~\bibnamefont {van
  Engelen}}, \bibinfo {author} {\bibfnamefont {S.}~\bibnamefont
  {Bhattacharya}}, \bibinfo {author} {\bibfnamefont {N.}~\bibnamefont
  {Sehgal}}, \bibinfo {author} {\bibfnamefont {G.}~\bibnamefont {Holder}},
  \bibinfo {author} {\bibfnamefont {O.}~\bibnamefont {Zahn}},\ and\ \bibinfo
  {author} {\bibfnamefont {D.}~\bibnamefont {Nagai}},\ }\href
  {https://doi.org/10.1088/0004-637X/786/1/13} {\bibfield  {journal} {\bibinfo
  {journal} {Astrophys. J.}\ }\textbf {\bibinfo {volume} {786}},\ \bibinfo
  {pages} {13} (\bibinfo {year} {2014})},\ \Eprint
  {https://arxiv.org/abs/1310.7023} {arXiv:1310.7023 [astro-ph.CO]}
  \BibitemShut {NoStop}%
\bibitem [{\citenamefont {Osborne}\ \emph {et~al.}(2014)\citenamefont
  {Osborne}, \citenamefont {Hanson},\ and\ \citenamefont
  {Dor\'e}}]{Osborne:2013nna}%
  \BibitemOpen
  \bibfield  {author} {\bibinfo {author} {\bibfnamefont {S.~J.}\ \bibnamefont
  {Osborne}}, \bibinfo {author} {\bibfnamefont {D.}~\bibnamefont {Hanson}},\
  and\ \bibinfo {author} {\bibfnamefont {O.}~\bibnamefont {Dor\'e}},\ }\href
  {https://doi.org/10.1088/1475-7516/2014/03/024} {\bibfield  {journal}
  {\bibinfo  {journal} {JCAP}\ }\textbf {\bibinfo {volume} {03}},\ \bibinfo
  {pages} {024}},\ \Eprint {https://arxiv.org/abs/1310.7547} {arXiv:1310.7547
  [astro-ph.CO]} \BibitemShut {NoStop}%
\bibitem [{\citenamefont {Smith}\ \emph {et~al.}(2012)\citenamefont {Smith},
  \citenamefont {Hanson}, \citenamefont {LoVerde}, \citenamefont {Hirata},\
  and\ \citenamefont {Zahn}}]{Smith:2010gu}%
  \BibitemOpen
  \bibfield  {author} {\bibinfo {author} {\bibfnamefont {K.~M.}\ \bibnamefont
  {Smith}}, \bibinfo {author} {\bibfnamefont {D.}~\bibnamefont {Hanson}},
  \bibinfo {author} {\bibfnamefont {M.}~\bibnamefont {LoVerde}}, \bibinfo
  {author} {\bibfnamefont {C.~M.}\ \bibnamefont {Hirata}},\ and\ \bibinfo
  {author} {\bibfnamefont {O.}~\bibnamefont {Zahn}},\ }\href
  {https://doi.org/10.1088/1475-7516/2012/06/014} {\bibfield  {journal}
  {\bibinfo  {journal} {JCAP}\ }\textbf {\bibinfo {volume} {06}},\ \bibinfo
  {pages} {014}},\ \Eprint {https://arxiv.org/abs/1010.0048} {arXiv:1010.0048
  [astro-ph.CO]} \BibitemShut {NoStop}%
\bibitem [{\citenamefont {Das}\ \emph {et~al.}(2013)\citenamefont {Das},
  \citenamefont {Errard},\ and\ \citenamefont {Spergel}}]{Das:2013aia}%
  \BibitemOpen
  \bibfield  {author} {\bibinfo {author} {\bibfnamefont {S.}~\bibnamefont
  {Das}}, \bibinfo {author} {\bibfnamefont {J.}~\bibnamefont {Errard}},\ and\
  \bibinfo {author} {\bibfnamefont {D.}~\bibnamefont {Spergel}},\ }\href@noop
  {} {\  (\bibinfo {year} {2013})},\ \Eprint {https://arxiv.org/abs/1311.2338}
  {arXiv:1311.2338 [astro-ph.CO]} \BibitemShut {NoStop}%
\bibitem [{\citenamefont {Schaan}\ \emph {et~al.}(2017)\citenamefont {Schaan},
  \citenamefont {Krause}, \citenamefont {Eifler}, \citenamefont {Dor{\'e}},
  \citenamefont {Miyatake}, \citenamefont {Rhodes},\ and\ \citenamefont
  {Spergel}}]{Schaan:2016ois}%
  \BibitemOpen
  \bibfield  {author} {\bibinfo {author} {\bibfnamefont {E.}~\bibnamefont
  {Schaan}}, \bibinfo {author} {\bibfnamefont {E.}~\bibnamefont {Krause}},
  \bibinfo {author} {\bibfnamefont {T.}~\bibnamefont {Eifler}}, \bibinfo
  {author} {\bibfnamefont {O.}~\bibnamefont {Dor{\'e}}}, \bibinfo {author}
  {\bibfnamefont {H.}~\bibnamefont {Miyatake}}, \bibinfo {author}
  {\bibfnamefont {J.}~\bibnamefont {Rhodes}},\ and\ \bibinfo {author}
  {\bibfnamefont {D.~N.}\ \bibnamefont {Spergel}},\ }\href
  {https://doi.org/10.1103/PhysRevD.95.123512} {\bibfield  {journal} {\bibinfo
  {journal} {Phys. Rev.}\ }\textbf {\bibinfo {volume} {D95}},\ \bibinfo {pages}
  {123512} (\bibinfo {year} {2017})},\ \Eprint
  {https://arxiv.org/abs/1607.01761} {arXiv:1607.01761 [astro-ph.CO]}
  \BibitemShut {NoStop}%
\bibitem [{\citenamefont {Schaan}\ \emph {et~al.}(2020)\citenamefont {Schaan},
  \citenamefont {Ferraro},\ and\ \citenamefont {Seljak}}]{Schaan:2020qox}%
  \BibitemOpen
  \bibfield  {author} {\bibinfo {author} {\bibfnamefont {E.}~\bibnamefont
  {Schaan}}, \bibinfo {author} {\bibfnamefont {S.}~\bibnamefont {Ferraro}},\
  and\ \bibinfo {author} {\bibfnamefont {U.}~\bibnamefont {Seljak}},\
  }\href@noop {} {\  (\bibinfo {year} {2020})},\ \Eprint
  {https://arxiv.org/abs/2007.12795} {arXiv:2007.12795 [astro-ph.CO]}
  \BibitemShut {NoStop}%
\bibitem [{\citenamefont {Jeffrey}\ \emph {et~al.}(2020)\citenamefont
  {Jeffrey}, \citenamefont {Lanusse}, \citenamefont {Lahav},\ and\
  \citenamefont {Starck}}]{Jeffrey:2019fag}%
  \BibitemOpen
  \bibfield  {author} {\bibinfo {author} {\bibfnamefont {N.}~\bibnamefont
  {Jeffrey}}, \bibinfo {author} {\bibfnamefont {F.}~\bibnamefont {Lanusse}},
  \bibinfo {author} {\bibfnamefont {O.}~\bibnamefont {Lahav}},\ and\ \bibinfo
  {author} {\bibfnamefont {J.-L.}\ \bibnamefont {Starck}},\ }\href
  {https://doi.org/10.1093/mnras/staa127} {\bibfield  {journal} {\bibinfo
  {journal} {Mon. Not. Roy. Astron. Soc.}\ }\textbf {\bibinfo {volume} {492}},\
  \bibinfo {pages} {5023} (\bibinfo {year} {2020})},\ \Eprint
  {https://arxiv.org/abs/1908.00543} {arXiv:1908.00543 [astro-ph.CO]}
  \BibitemShut {NoStop}%
\bibitem [{\citenamefont {Mawdsley}\ \emph {et~al.}(2020)\citenamefont
  {Mawdsley} \emph {et~al.}}]{Mawdsley:2019had}%
  \BibitemOpen
  \bibfield  {author} {\bibinfo {author} {\bibfnamefont {B.}~\bibnamefont
  {Mawdsley}} \emph {et~al.} (\bibinfo {collaboration} {DES}),\ }\href
  {https://doi.org/10.1093/mnras/staa565} {\bibfield  {journal} {\bibinfo
  {journal} {Mon. Not. Roy. Astron. Soc.}\ }\textbf {\bibinfo {volume} {493}},\
  \bibinfo {pages} {5662} (\bibinfo {year} {2020})},\ \Eprint
  {https://arxiv.org/abs/1905.12682} {arXiv:1905.12682 [astro-ph.CO]}
  \BibitemShut {NoStop}%
\bibitem [{\citenamefont {Pires}\ \emph {et~al.}(2020)\citenamefont {Pires}
  \emph {et~al.}}]{Pires:2019zcc}%
  \BibitemOpen
  \bibfield  {author} {\bibinfo {author} {\bibfnamefont {S.}~\bibnamefont
  {Pires}} \emph {et~al.} (\bibinfo {collaboration} {EUCLID}),\ }\href
  {https://doi.org/10.1051/0004-6361/201936865} {\bibfield  {journal} {\bibinfo
   {journal} {Astron. Astrophys.}\ }\textbf {\bibinfo {volume} {638}},\
  \bibinfo {pages} {A141} (\bibinfo {year} {2020})},\ \Eprint
  {https://arxiv.org/abs/1910.03106} {arXiv:1910.03106 [astro-ph.CO]}
  \BibitemShut {NoStop}%
\bibitem [{\citenamefont {Price}\ \emph {et~al.}(2020)\citenamefont {Price},
  \citenamefont {McEwen}, \citenamefont {Pratley},\ and\ \citenamefont
  {Kitching}}]{Price:2020mry}%
  \BibitemOpen
  \bibfield  {author} {\bibinfo {author} {\bibfnamefont {M.~A.}\ \bibnamefont
  {Price}}, \bibinfo {author} {\bibfnamefont {J.~D.}\ \bibnamefont {McEwen}},
  \bibinfo {author} {\bibfnamefont {L.}~\bibnamefont {Pratley}},\ and\ \bibinfo
  {author} {\bibfnamefont {T.~D.}\ \bibnamefont {Kitching}},\ }\href@noop {} {\
   (\bibinfo {year} {2020})},\ \Eprint {https://arxiv.org/abs/2004.07855}
  {arXiv:2004.07855 [astro-ph.CO]} \BibitemShut {NoStop}%
\bibitem [{\citenamefont {Kaiser}\ and\ \citenamefont
  {Squires}(1993)}]{Kaiser:1992ps}%
  \BibitemOpen
  \bibfield  {author} {\bibinfo {author} {\bibfnamefont {N.}~\bibnamefont
  {Kaiser}}\ and\ \bibinfo {author} {\bibfnamefont {G.}~\bibnamefont
  {Squires}},\ }\href {https://doi.org/10.1086/172297} {\bibfield  {journal}
  {\bibinfo  {journal} {Astrophys. J.}\ }\textbf {\bibinfo {volume} {404}},\
  \bibinfo {pages} {441} (\bibinfo {year} {1993})}\BibitemShut {NoStop}%
\bibitem [{\citenamefont {{LSST Science Collaboration}}\ \emph
  {et~al.}(2009)\citenamefont {{LSST Science Collaboration}} \emph
  {et~al.}}]{2009arXiv0912.0201L}%
  \BibitemOpen
  \bibfield  {author} {\bibinfo {author} {\bibnamefont {{LSST Science
  Collaboration}}} \emph {et~al.},\ }\href@noop {} {\bibfield  {journal}
  {\bibinfo  {journal} {arXiv e-prints}\ ,\ \bibinfo {eid} {arXiv:0912.0201}}
  (\bibinfo {year} {2009})},\ \Eprint {https://arxiv.org/abs/0912.0201}
  {arXiv:0912.0201 [astro-ph.IM]} \BibitemShut {NoStop}%
\bibitem [{\citenamefont {Chang}\ \emph {et~al.}(2013)\citenamefont {Chang},
  \citenamefont {Jarvis}, \citenamefont {Jain}, \citenamefont {Kahn},
  \citenamefont {Kirkby}, \citenamefont {Connolly}, \citenamefont {Krughoff},
  \citenamefont {Peng},\ and\ \citenamefont {Peterson}}]{Chang:2013xja}%
  \BibitemOpen
  \bibfield  {author} {\bibinfo {author} {\bibfnamefont {C.}~\bibnamefont
  {Chang}}, \bibinfo {author} {\bibfnamefont {M.}~\bibnamefont {Jarvis}},
  \bibinfo {author} {\bibfnamefont {B.}~\bibnamefont {Jain}}, \bibinfo {author}
  {\bibfnamefont {S.~M.}\ \bibnamefont {Kahn}}, \bibinfo {author}
  {\bibfnamefont {D.}~\bibnamefont {Kirkby}}, \bibinfo {author} {\bibfnamefont
  {A.}~\bibnamefont {Connolly}}, \bibinfo {author} {\bibfnamefont
  {S.}~\bibnamefont {Krughoff}}, \bibinfo {author} {\bibfnamefont
  {E.}~\bibnamefont {Peng}},\ and\ \bibinfo {author} {\bibfnamefont {J.~R.}\
  \bibnamefont {Peterson}},\ }\href {https://doi.org/10.1093/mnras/stt1156}
  {\bibfield  {journal} {\bibinfo  {journal} {Mon. Not. Roy. Astron. Soc.}\
  }\textbf {\bibinfo {volume} {434}},\ \bibinfo {pages} {2121} (\bibinfo {year}
  {2013})},\ \Eprint {https://arxiv.org/abs/1305.0793} {arXiv:1305.0793
  [astro-ph.CO]} \BibitemShut {NoStop}%
\bibitem [{\citenamefont {Sherwin}\ and\ \citenamefont
  {Schmittfull}(2015)}]{Sherwin:2015baa}%
  \BibitemOpen
  \bibfield  {author} {\bibinfo {author} {\bibfnamefont {B.~D.}\ \bibnamefont
  {Sherwin}}\ and\ \bibinfo {author} {\bibfnamefont {M.}~\bibnamefont
  {Schmittfull}},\ }\href {https://doi.org/10.1103/PhysRevD.92.043005}
  {\bibfield  {journal} {\bibinfo  {journal} {Phys. Rev.}\ }\textbf {\bibinfo
  {volume} {D92}},\ \bibinfo {pages} {043005} (\bibinfo {year} {2015})},\
  \Eprint {https://arxiv.org/abs/1502.05356} {arXiv:1502.05356 [astro-ph.CO]}
  \BibitemShut {NoStop}%
\bibitem [{\citenamefont {Yu}\ \emph {et~al.}(2017)\citenamefont {Yu},
  \citenamefont {Hill},\ and\ \citenamefont {Sherwin}}]{Yu:2017djs}%
  \BibitemOpen
  \bibfield  {author} {\bibinfo {author} {\bibfnamefont {B.}~\bibnamefont
  {Yu}}, \bibinfo {author} {\bibfnamefont {J.}~\bibnamefont {Hill}},\ and\
  \bibinfo {author} {\bibfnamefont {B.~D.}\ \bibnamefont {Sherwin}},\ }\href
  {https://doi.org/10.1103/PhysRevD.96.123511} {\bibfield  {journal} {\bibinfo
  {journal} {Phys. Rev. D}\ }\textbf {\bibinfo {volume} {96}},\ \bibinfo
  {pages} {123511} (\bibinfo {year} {2017})},\ \Eprint
  {https://arxiv.org/abs/1705.02332} {arXiv:1705.02332 [astro-ph.CO]}
  \BibitemShut {NoStop}%
\bibitem [{\citenamefont {Hirata}\ and\ \citenamefont
  {Seljak}(2004)}]{PhysRevD.70.063526}%
  \BibitemOpen
  \bibfield  {author} {\bibinfo {author} {\bibfnamefont {C.~M.}\ \bibnamefont
  {Hirata}}\ and\ \bibinfo {author} {\bibfnamefont {U.~c.~v.}\ \bibnamefont
  {Seljak}},\ }\href {https://doi.org/10.1103/PhysRevD.70.063526} {\bibfield
  {journal} {\bibinfo  {journal} {Phys. Rev. D}\ }\textbf {\bibinfo {volume}
  {70}},\ \bibinfo {pages} {063526} (\bibinfo {year} {2004})}\BibitemShut
  {NoStop}%
\bibitem [{\citenamefont {{Bridle}}\ and\ \citenamefont
  {{King}}(2007)}]{2007NJPh....9..444B}%
  \BibitemOpen
  \bibfield  {author} {\bibinfo {author} {\bibfnamefont {S.}~\bibnamefont
  {{Bridle}}}\ and\ \bibinfo {author} {\bibfnamefont {L.}~\bibnamefont
  {{King}}},\ }\href {https://doi.org/10.1088/1367-2630/9/12/444} {\bibfield
  {journal} {\bibinfo  {journal} {New Journal of Physics}\ }\textbf {\bibinfo
  {volume} {9}},\ \bibinfo {pages} {444} (\bibinfo {year} {2007})},\ \Eprint
  {https://arxiv.org/abs/0705.0166} {arXiv:0705.0166 [astro-ph]} \BibitemShut
  {NoStop}%
\bibitem [{\citenamefont {Brown}\ \emph {et~al.}(2002)\citenamefont {Brown},
  \citenamefont {Taylor}, \citenamefont {Hambly},\ and\ \citenamefont
  {Dye}}]{Brown:2000gt}%
  \BibitemOpen
  \bibfield  {author} {\bibinfo {author} {\bibfnamefont {M.}~\bibnamefont
  {Brown}}, \bibinfo {author} {\bibfnamefont {A.}~\bibnamefont {Taylor}},
  \bibinfo {author} {\bibfnamefont {N.}~\bibnamefont {Hambly}},\ and\ \bibinfo
  {author} {\bibfnamefont {S.}~\bibnamefont {Dye}},\ }\href
  {https://doi.org/10.1046/j.1365-8711.2002.05354.x} {\bibfield  {journal}
  {\bibinfo  {journal} {Mon. Not. Roy. Astron. Soc.}\ }\textbf {\bibinfo
  {volume} {333}},\ \bibinfo {pages} {501} (\bibinfo {year} {2002})},\ \Eprint
  {https://arxiv.org/abs/astro-ph/0009499} {arXiv:astro-ph/0009499}
  \BibitemShut {NoStop}%
\bibitem [{\citenamefont {Samuroff}\ \emph {et~al.}(2019)\citenamefont
  {Samuroff} \emph {et~al.}}]{Samuroff:2018xuo}%
  \BibitemOpen
  \bibfield  {author} {\bibinfo {author} {\bibfnamefont {S.}~\bibnamefont
  {Samuroff}} \emph {et~al.} (\bibinfo {collaboration} {DES}),\ }\href
  {https://doi.org/10.1093/mnras/stz2197} {\bibfield  {journal} {\bibinfo
  {journal} {Mon. Not. Roy. Astron. Soc.}\ }\textbf {\bibinfo {volume} {489}},\
  \bibinfo {pages} {5453} (\bibinfo {year} {2019})},\ \Eprint
  {https://arxiv.org/abs/1811.06989} {arXiv:1811.06989 [astro-ph.CO]}
  \BibitemShut {NoStop}%
\bibitem [{\citenamefont {Johnston}\ \emph {et~al.}(2019)\citenamefont
  {Johnston} \emph {et~al.}}]{Johnston:2018nfi}%
  \BibitemOpen
  \bibfield  {author} {\bibinfo {author} {\bibfnamefont {H.}~\bibnamefont
  {Johnston}} \emph {et~al.},\ }\href
  {https://doi.org/10.1051/0004-6361/201834714} {\bibfield  {journal} {\bibinfo
   {journal} {Astron. Astrophys.}\ }\textbf {\bibinfo {volume} {624}},\
  \bibinfo {pages} {A30} (\bibinfo {year} {2019})},\ \Eprint
  {https://arxiv.org/abs/1811.09598} {arXiv:1811.09598 [astro-ph.CO]}
  \BibitemShut {NoStop}%
\bibitem [{\citenamefont {Yao}\ \emph {et~al.}(2020)\citenamefont {Yao},
  \citenamefont {Shan}, \citenamefont {Zhang}, \citenamefont {Kneib},\ and\
  \citenamefont {Jullo}}]{Yao:2020jpj}%
  \BibitemOpen
  \bibfield  {author} {\bibinfo {author} {\bibfnamefont {J.}~\bibnamefont
  {Yao}}, \bibinfo {author} {\bibfnamefont {H.}~\bibnamefont {Shan}}, \bibinfo
  {author} {\bibfnamefont {P.}~\bibnamefont {Zhang}}, \bibinfo {author}
  {\bibfnamefont {J.-P.}\ \bibnamefont {Kneib}},\ and\ \bibinfo {author}
  {\bibfnamefont {E.}~\bibnamefont {Jullo}},\ }\href@noop {} {\  (\bibinfo
  {year} {2020})},\ \Eprint {https://arxiv.org/abs/2002.09826}
  {arXiv:2002.09826 [astro-ph.CO]} \BibitemShut {NoStop}%
\bibitem [{\citenamefont {Mandelbaum}\ \emph {et~al.}(2019)\citenamefont
  {Mandelbaum} \emph {et~al.}}]{Mandelbaum:2019zej}%
  \BibitemOpen
  \bibfield  {author} {\bibinfo {author} {\bibfnamefont {R.}~\bibnamefont
  {Mandelbaum}} \emph {et~al.} (\bibinfo {collaboration} {LSST Dark Energy
  Science}),\ }\href@noop {} {\  (\bibinfo {year} {2019})},\ \Eprint
  {https://arxiv.org/abs/1903.09323} {arXiv:1903.09323 [astro-ph.CO]}
  \BibitemShut {NoStop}%
\bibitem [{\citenamefont {Vlah}\ \emph {et~al.}(2020)\citenamefont {Vlah},
  \citenamefont {Chisari},\ and\ \citenamefont {Schmidt}}]{Vlah:2019byq}%
  \BibitemOpen
  \bibfield  {author} {\bibinfo {author} {\bibfnamefont {Z.}~\bibnamefont
  {Vlah}}, \bibinfo {author} {\bibfnamefont {N.~E.}\ \bibnamefont {Chisari}},\
  and\ \bibinfo {author} {\bibfnamefont {F.}~\bibnamefont {Schmidt}},\ }\href
  {https://doi.org/10.1088/1475-7516/2020/01/025} {\bibfield  {journal}
  {\bibinfo  {journal} {JCAP}\ }\textbf {\bibinfo {volume} {01}},\ \bibinfo
  {pages} {025}},\ \Eprint {https://arxiv.org/abs/1910.08085} {arXiv:1910.08085
  [astro-ph.CO]} \BibitemShut {NoStop}%
\bibitem [{\citenamefont {Fortuna}\ \emph {et~al.}(2020)\citenamefont
  {Fortuna}, \citenamefont {Hoekstra}, \citenamefont {Joachimi}, \citenamefont
  {Johnston}, \citenamefont {Chisari}, \citenamefont {Georgiou},\ and\
  \citenamefont {Mahony}}]{Fortuna:2020vsz}%
  \BibitemOpen
  \bibfield  {author} {\bibinfo {author} {\bibfnamefont {M.~C.}\ \bibnamefont
  {Fortuna}}, \bibinfo {author} {\bibfnamefont {H.}~\bibnamefont {Hoekstra}},
  \bibinfo {author} {\bibfnamefont {B.}~\bibnamefont {Joachimi}}, \bibinfo
  {author} {\bibfnamefont {H.}~\bibnamefont {Johnston}}, \bibinfo {author}
  {\bibfnamefont {N.~E.}\ \bibnamefont {Chisari}}, \bibinfo {author}
  {\bibfnamefont {C.}~\bibnamefont {Georgiou}},\ and\ \bibinfo {author}
  {\bibfnamefont {C.}~\bibnamefont {Mahony}},\ }\href@noop {} {\  (\bibinfo
  {year} {2020})},\ \Eprint {https://arxiv.org/abs/2003.02700}
  {arXiv:2003.02700 [astro-ph.CO]} \BibitemShut {NoStop}%
\bibitem [{\citenamefont {Lewandowski}\ \emph {et~al.}(2015)\citenamefont
  {Lewandowski}, \citenamefont {Perko},\ and\ \citenamefont
  {Senatore}}]{Lewandowski:2014rca}%
  \BibitemOpen
  \bibfield  {author} {\bibinfo {author} {\bibfnamefont {M.}~\bibnamefont
  {Lewandowski}}, \bibinfo {author} {\bibfnamefont {A.}~\bibnamefont {Perko}},\
  and\ \bibinfo {author} {\bibfnamefont {L.}~\bibnamefont {Senatore}},\ }\href
  {https://doi.org/10.1088/1475-7516/2015/05/019} {\bibfield  {journal}
  {\bibinfo  {journal} {JCAP}\ }\textbf {\bibinfo {volume} {1505}}\bibfield
  {number} {\bibinfo  {number} { (05)},\ \bibinfo {pages} {019}},\ }\Eprint
  {https://arxiv.org/abs/1412.5049} {arXiv:1412.5049 [astro-ph.CO]}
  \BibitemShut {NoStop}%
\bibitem [{\citenamefont {Chen}\ \emph {et~al.}(2019)\citenamefont {Chen},
  \citenamefont {Castorina},\ and\ \citenamefont {White}}]{Chen:2019cfu}%
  \BibitemOpen
  \bibfield  {author} {\bibinfo {author} {\bibfnamefont {S.-F.}\ \bibnamefont
  {Chen}}, \bibinfo {author} {\bibfnamefont {E.}~\bibnamefont {Castorina}},\
  and\ \bibinfo {author} {\bibfnamefont {M.}~\bibnamefont {White}},\ }\href
  {https://doi.org/10.1088/1475-7516/2019/06/006} {\bibfield  {journal}
  {\bibinfo  {journal} {JCAP}\ }\textbf {\bibinfo {volume} {1906}}\bibfield
  {number} {\bibinfo  {number} { (06)},\ \bibinfo {pages} {006}},\ }\Eprint
  {https://arxiv.org/abs/1903.00437} {arXiv:1903.00437 [astro-ph.CO]}
  \BibitemShut {NoStop}%
\bibitem [{\citenamefont {Semboloni}\ \emph {et~al.}(2013)\citenamefont
  {Semboloni}, \citenamefont {Hoekstra},\ and\ \citenamefont
  {Schaye}}]{Semboloni:2012yh}%
  \BibitemOpen
  \bibfield  {author} {\bibinfo {author} {\bibfnamefont {E.}~\bibnamefont
  {Semboloni}}, \bibinfo {author} {\bibfnamefont {H.}~\bibnamefont
  {Hoekstra}},\ and\ \bibinfo {author} {\bibfnamefont {J.}~\bibnamefont
  {Schaye}},\ }\href {https://doi.org/10.1093/mnras/stt1013} {\bibfield
  {journal} {\bibinfo  {journal} {Mon. Not. Roy. Astron. Soc.}\ }\textbf
  {\bibinfo {volume} {434}},\ \bibinfo {pages} {148} (\bibinfo {year}
  {2013})},\ \Eprint {https://arxiv.org/abs/1210.7303} {arXiv:1210.7303
  [astro-ph.CO]} \BibitemShut {NoStop}%
\bibitem [{\citenamefont {Mead}\ \emph
  {et~al.}(2020{\natexlab{a}})\citenamefont {Mead}, \citenamefont {Tr\"oster},
  \citenamefont {Heymans}, \citenamefont {Van~Waerbeke},\ and\ \citenamefont
  {McCarthy}}]{Mead:2020qgo}%
  \BibitemOpen
  \bibfield  {author} {\bibinfo {author} {\bibfnamefont {A.}~\bibnamefont
  {Mead}}, \bibinfo {author} {\bibfnamefont {T.}~\bibnamefont {Tr\"oster}},
  \bibinfo {author} {\bibfnamefont {C.}~\bibnamefont {Heymans}}, \bibinfo
  {author} {\bibfnamefont {L.}~\bibnamefont {Van~Waerbeke}},\ and\ \bibinfo
  {author} {\bibfnamefont {I.}~\bibnamefont {McCarthy}},\ }\href
  {https://doi.org/10.1051/0004-6361/202038308} {\bibfield  {journal} {\bibinfo
   {journal} {Astron. Astrophys.}\ }\textbf {\bibinfo {volume} {641}},\
  \bibinfo {pages} {A130} (\bibinfo {year} {2020}{\natexlab{a}})},\ \Eprint
  {https://arxiv.org/abs/2005.00009} {arXiv:2005.00009 [astro-ph.CO]}
  \BibitemShut {NoStop}%
\bibitem [{\citenamefont {Mead}\ \emph
  {et~al.}(2020{\natexlab{b}})\citenamefont {Mead}, \citenamefont {Brieden},
  \citenamefont {Tr\"oster},\ and\ \citenamefont {Heymans}}]{Mead:2020vgs}%
  \BibitemOpen
  \bibfield  {author} {\bibinfo {author} {\bibfnamefont {A.}~\bibnamefont
  {Mead}}, \bibinfo {author} {\bibfnamefont {S.}~\bibnamefont {Brieden}},
  \bibinfo {author} {\bibfnamefont {T.}~\bibnamefont {Tr\"oster}},\ and\
  \bibinfo {author} {\bibfnamefont {C.}~\bibnamefont {Heymans}},\ }\href@noop
  {} {\  (\bibinfo {year} {2020}{\natexlab{b}})},\ \Eprint
  {https://arxiv.org/abs/2009.01858} {arXiv:2009.01858 [astro-ph.CO]}
  \BibitemShut {NoStop}%
\bibitem [{\citenamefont {Debackere}\ \emph {et~al.}(2020)\citenamefont
  {Debackere}, \citenamefont {Schaye},\ and\ \citenamefont
  {Hoekstra}}]{Debackere:2019cec}%
  \BibitemOpen
  \bibfield  {author} {\bibinfo {author} {\bibfnamefont {S.~N.}\ \bibnamefont
  {Debackere}}, \bibinfo {author} {\bibfnamefont {J.}~\bibnamefont {Schaye}},\
  and\ \bibinfo {author} {\bibfnamefont {H.}~\bibnamefont {Hoekstra}},\ }\href
  {https://doi.org/10.1093/mnras/stz3446} {\bibfield  {journal} {\bibinfo
  {journal} {Mon. Not. Roy. Astron. Soc.}\ }\textbf {\bibinfo {volume} {492}},\
  \bibinfo {pages} {2285} (\bibinfo {year} {2020})},\ \Eprint
  {https://arxiv.org/abs/1908.05765} {arXiv:1908.05765 [astro-ph.CO]}
  \BibitemShut {NoStop}%
\bibitem [{\citenamefont {Eifler}\ \emph {et~al.}(2015)\citenamefont {Eifler},
  \citenamefont {Krause}, \citenamefont {Dodelson}, \citenamefont {Zentner},
  \citenamefont {Hearin},\ and\ \citenamefont {Gnedin}}]{Eifler:2014iva}%
  \BibitemOpen
  \bibfield  {author} {\bibinfo {author} {\bibfnamefont {T.}~\bibnamefont
  {Eifler}}, \bibinfo {author} {\bibfnamefont {E.}~\bibnamefont {Krause}},
  \bibinfo {author} {\bibfnamefont {S.}~\bibnamefont {Dodelson}}, \bibinfo
  {author} {\bibfnamefont {A.}~\bibnamefont {Zentner}}, \bibinfo {author}
  {\bibfnamefont {A.}~\bibnamefont {Hearin}},\ and\ \bibinfo {author}
  {\bibfnamefont {N.}~\bibnamefont {Gnedin}},\ }\href
  {https://doi.org/10.1093/mnras/stv2000} {\bibfield  {journal} {\bibinfo
  {journal} {Mon. Not. Roy. Astron. Soc.}\ }\textbf {\bibinfo {volume} {454}},\
  \bibinfo {pages} {2451} (\bibinfo {year} {2015})},\ \Eprint
  {https://arxiv.org/abs/1405.7423} {arXiv:1405.7423 [astro-ph.CO]}
  \BibitemShut {NoStop}%
\bibitem [{\citenamefont {Mohammed}\ and\ \citenamefont
  {Gnedin}(2018)}]{Mohammed:2017nei}%
  \BibitemOpen
  \bibfield  {author} {\bibinfo {author} {\bibfnamefont {I.}~\bibnamefont
  {Mohammed}}\ and\ \bibinfo {author} {\bibfnamefont {N.~Y.}\ \bibnamefont
  {Gnedin}},\ }\href {https://doi.org/10.3847/1538-4357/aad3b1} {\bibfield
  {journal} {\bibinfo  {journal} {Astrophys. J.}\ }\textbf {\bibinfo {volume}
  {863}},\ \bibinfo {pages} {173} (\bibinfo {year} {2018})},\ \Eprint
  {https://arxiv.org/abs/1707.02332} {arXiv:1707.02332 [astro-ph.CO]}
  \BibitemShut {NoStop}%
\bibitem [{\citenamefont {Huang}\ \emph {et~al.}(2019)\citenamefont {Huang},
  \citenamefont {Eifler}, \citenamefont {Mandelbaum},\ and\ \citenamefont
  {Dodelson}}]{Huang:2018wpy}%
  \BibitemOpen
  \bibfield  {author} {\bibinfo {author} {\bibfnamefont {H.-J.}\ \bibnamefont
  {Huang}}, \bibinfo {author} {\bibfnamefont {T.}~\bibnamefont {Eifler}},
  \bibinfo {author} {\bibfnamefont {R.}~\bibnamefont {Mandelbaum}},\ and\
  \bibinfo {author} {\bibfnamefont {S.}~\bibnamefont {Dodelson}},\ }\href
  {https://doi.org/10.1093/mnras/stz1714} {\bibfield  {journal} {\bibinfo
  {journal} {Mon. Not. Roy. Astron. Soc.}\ }\textbf {\bibinfo {volume} {488}},\
  \bibinfo {pages} {1652} (\bibinfo {year} {2019})},\ \Eprint
  {https://arxiv.org/abs/1809.01146} {arXiv:1809.01146 [astro-ph.CO]}
  \BibitemShut {NoStop}%
\bibitem [{\citenamefont {Schneider}\ and\ \citenamefont
  {Teyssier}(2015)}]{Schneider:2015wta}%
  \BibitemOpen
  \bibfield  {author} {\bibinfo {author} {\bibfnamefont {A.}~\bibnamefont
  {Schneider}}\ and\ \bibinfo {author} {\bibfnamefont {R.}~\bibnamefont
  {Teyssier}},\ }\href {https://doi.org/10.1088/1475-7516/2015/12/049}
  {\bibfield  {journal} {\bibinfo  {journal} {JCAP}\ }\textbf {\bibinfo
  {volume} {12}},\ \bibinfo {pages} {049}},\ \Eprint
  {https://arxiv.org/abs/1510.06034} {arXiv:1510.06034 [astro-ph.CO]}
  \BibitemShut {NoStop}%
\bibitem [{\citenamefont {Schneider}\ \emph {et~al.}(2019)\citenamefont
  {Schneider}, \citenamefont {Teyssier}, \citenamefont {Stadel}, \citenamefont
  {Chisari}, \citenamefont {Le~Brun}, \citenamefont {Amara},\ and\
  \citenamefont {Refregier}}]{Schneider:2018pfw}%
  \BibitemOpen
  \bibfield  {author} {\bibinfo {author} {\bibfnamefont {A.}~\bibnamefont
  {Schneider}}, \bibinfo {author} {\bibfnamefont {R.}~\bibnamefont {Teyssier}},
  \bibinfo {author} {\bibfnamefont {J.}~\bibnamefont {Stadel}}, \bibinfo
  {author} {\bibfnamefont {N.~E.}\ \bibnamefont {Chisari}}, \bibinfo {author}
  {\bibfnamefont {A.~M.~C.}\ \bibnamefont {Le~Brun}}, \bibinfo {author}
  {\bibfnamefont {A.}~\bibnamefont {Amara}},\ and\ \bibinfo {author}
  {\bibfnamefont {A.}~\bibnamefont {Refregier}},\ }\href
  {https://doi.org/10.1088/1475-7516/2019/03/020} {\bibfield  {journal}
  {\bibinfo  {journal} {JCAP}\ }\textbf {\bibinfo {volume} {03}},\ \bibinfo
  {pages} {020}},\ \Eprint {https://arxiv.org/abs/1810.08629} {arXiv:1810.08629
  [astro-ph.CO]} \BibitemShut {NoStop}%
\bibitem [{\citenamefont {Aric\`o}\ \emph
  {et~al.}(2020{\natexlab{a}})\citenamefont {Aric\`o}, \citenamefont {Angulo},
  \citenamefont {Hern\'andez-Monteagudo}, \citenamefont {Contreras},
  \citenamefont {Zennaro}, \citenamefont {Pellejero-Iba\~nez},\ and\
  \citenamefont {Rosas-Guevara}}]{Arico:2019ykw}%
  \BibitemOpen
  \bibfield  {author} {\bibinfo {author} {\bibfnamefont {G.}~\bibnamefont
  {Aric\`o}}, \bibinfo {author} {\bibfnamefont {R.~E.}\ \bibnamefont {Angulo}},
  \bibinfo {author} {\bibfnamefont {C.}~\bibnamefont {Hern\'andez-Monteagudo}},
  \bibinfo {author} {\bibfnamefont {S.}~\bibnamefont {Contreras}}, \bibinfo
  {author} {\bibfnamefont {M.}~\bibnamefont {Zennaro}}, \bibinfo {author}
  {\bibfnamefont {M.}~\bibnamefont {Pellejero-Iba\~nez}},\ and\ \bibinfo
  {author} {\bibfnamefont {Y.}~\bibnamefont {Rosas-Guevara}},\ }\href
  {https://doi.org/10.1093/mnras/staa1478} {\bibfield  {journal} {\bibinfo
  {journal} {Mon. Not. Roy. Astron. Soc.}\ }\textbf {\bibinfo {volume} {495}},\
  \bibinfo {pages} {4800} (\bibinfo {year} {2020}{\natexlab{a}})},\ \Eprint
  {https://arxiv.org/abs/1911.08471} {arXiv:1911.08471 [astro-ph.CO]}
  \BibitemShut {NoStop}%
\bibitem [{\citenamefont {Aric\`o}\ \emph
  {et~al.}(2020{\natexlab{b}})\citenamefont {Aric\`o}, \citenamefont {Angulo},
  \citenamefont {Hern\'andez-Monteagudo}, \citenamefont {Contreras},\ and\
  \citenamefont {Zennaro}}]{Arico:2020yyf}%
  \BibitemOpen
  \bibfield  {author} {\bibinfo {author} {\bibfnamefont {G.}~\bibnamefont
  {Aric\`o}}, \bibinfo {author} {\bibfnamefont {R.~E.}\ \bibnamefont {Angulo}},
  \bibinfo {author} {\bibfnamefont {C.}~\bibnamefont {Hern\'andez-Monteagudo}},
  \bibinfo {author} {\bibfnamefont {S.}~\bibnamefont {Contreras}},\ and\
  \bibinfo {author} {\bibfnamefont {M.}~\bibnamefont {Zennaro}},\ }\href@noop
  {} {\  (\bibinfo {year} {2020}{\natexlab{b}})},\ \Eprint
  {https://arxiv.org/abs/2009.14225} {arXiv:2009.14225 [astro-ph.CO]}
  \BibitemShut {NoStop}%
\bibitem [{\citenamefont {Schneider}\ \emph
  {et~al.}(2020{\natexlab{b}})\citenamefont {Schneider}, \citenamefont
  {Stoira}, \citenamefont {Refregier}, \citenamefont {Weiss}, \citenamefont
  {Knabenhans}, \citenamefont {Stadel},\ and\ \citenamefont
  {Teyssier}}]{Schneider:2019snl}%
  \BibitemOpen
  \bibfield  {author} {\bibinfo {author} {\bibfnamefont {A.}~\bibnamefont
  {Schneider}}, \bibinfo {author} {\bibfnamefont {N.}~\bibnamefont {Stoira}},
  \bibinfo {author} {\bibfnamefont {A.}~\bibnamefont {Refregier}}, \bibinfo
  {author} {\bibfnamefont {A.~J.}\ \bibnamefont {Weiss}}, \bibinfo {author}
  {\bibfnamefont {M.}~\bibnamefont {Knabenhans}}, \bibinfo {author}
  {\bibfnamefont {J.}~\bibnamefont {Stadel}},\ and\ \bibinfo {author}
  {\bibfnamefont {R.}~\bibnamefont {Teyssier}},\ }\href
  {https://doi.org/10.1088/1475-7516/2020/04/019} {\bibfield  {journal}
  {\bibinfo  {journal} {JCAP}\ }\textbf {\bibinfo {volume} {04}},\ \bibinfo
  {pages} {019}},\ \Eprint {https://arxiv.org/abs/1910.11357} {arXiv:1910.11357
  [astro-ph.CO]} \BibitemShut {NoStop}%
\bibitem [{\citenamefont {Tr\"oster}\ \emph {et~al.}(2019)\citenamefont
  {Tr\"oster}, \citenamefont {Ferguson}, \citenamefont {Harnois-D\'eraps},\
  and\ \citenamefont {McCarthy}}]{Troster:2019mys}%
  \BibitemOpen
  \bibfield  {author} {\bibinfo {author} {\bibfnamefont {T.}~\bibnamefont
  {Tr\"oster}}, \bibinfo {author} {\bibfnamefont {C.}~\bibnamefont {Ferguson}},
  \bibinfo {author} {\bibfnamefont {J.}~\bibnamefont {Harnois-D\'eraps}},\ and\
  \bibinfo {author} {\bibfnamefont {I.~G.}\ \bibnamefont {McCarthy}},\ }\href
  {https://doi.org/10.1093/mnrasl/slz075} {\bibfield  {journal} {\bibinfo
  {journal} {Mon. Not. Roy. Astron. Soc.}\ }\textbf {\bibinfo {volume} {487}},\
  \bibinfo {pages} {L24} (\bibinfo {year} {2019})},\ \Eprint
  {https://arxiv.org/abs/1903.12173} {arXiv:1903.12173 [astro-ph.CO]}
  \BibitemShut {NoStop}%
\bibitem [{\citenamefont {Villaescusa-Navarro}\ \emph
  {et~al.}(2020)\citenamefont {Villaescusa-Navarro} \emph
  {et~al.}}]{Villaescusa-Navarro:2020rxg}%
  \BibitemOpen
  \bibfield  {author} {\bibinfo {author} {\bibfnamefont {F.}~\bibnamefont
  {Villaescusa-Navarro}} \emph {et~al.},\ }\href@noop {} {\  (\bibinfo {year}
  {2020})},\ \Eprint {https://arxiv.org/abs/2010.00619} {arXiv:2010.00619
  [astro-ph.CO]} \BibitemShut {NoStop}%
\bibitem [{\citenamefont {Cooray}\ and\ \citenamefont
  {Sheth}(2002)}]{Cooray:2002dia}%
  \BibitemOpen
  \bibfield  {author} {\bibinfo {author} {\bibfnamefont {A.}~\bibnamefont
  {Cooray}}\ and\ \bibinfo {author} {\bibfnamefont {R.~K.}\ \bibnamefont
  {Sheth}},\ }\href {https://doi.org/10.1016/S0370-1573(02)00276-4} {\bibfield
  {journal} {\bibinfo  {journal} {Phys. Rept.}\ }\textbf {\bibinfo {volume}
  {372}},\ \bibinfo {pages} {1} (\bibinfo {year} {2002})},\ \Eprint
  {https://arxiv.org/abs/astro-ph/0206508} {arXiv:astro-ph/0206508 [astro-ph]}
  \BibitemShut {NoStop}%
\bibitem [{\citenamefont {Parimbelli}\ \emph {et~al.}(2019)\citenamefont
  {Parimbelli}, \citenamefont {Viel},\ and\ \citenamefont
  {Sefusatti}}]{Parimbelli:2018yzv}%
  \BibitemOpen
  \bibfield  {author} {\bibinfo {author} {\bibfnamefont {G.}~\bibnamefont
  {Parimbelli}}, \bibinfo {author} {\bibfnamefont {M.}~\bibnamefont {Viel}},\
  and\ \bibinfo {author} {\bibfnamefont {E.}~\bibnamefont {Sefusatti}},\ }\href
  {https://doi.org/10.1088/1475-7516/2019/01/010} {\bibfield  {journal}
  {\bibinfo  {journal} {JCAP}\ }\textbf {\bibinfo {volume} {01}},\ \bibinfo
  {pages} {010}},\ \Eprint {https://arxiv.org/abs/1809.06634} {arXiv:1809.06634
  [astro-ph.CO]} \BibitemShut {NoStop}%
\bibitem [{\citenamefont {Gonzalez}\ \emph {et~al.}(2013)\citenamefont
  {Gonzalez}, \citenamefont {Sivanandam}, \citenamefont {Zabludoff},\ and\
  \citenamefont {Zaritsky}}]{Gonzalez:2013awy}%
  \BibitemOpen
  \bibfield  {author} {\bibinfo {author} {\bibfnamefont {A.~H.}\ \bibnamefont
  {Gonzalez}}, \bibinfo {author} {\bibfnamefont {S.}~\bibnamefont
  {Sivanandam}}, \bibinfo {author} {\bibfnamefont {A.~I.}\ \bibnamefont
  {Zabludoff}},\ and\ \bibinfo {author} {\bibfnamefont {D.}~\bibnamefont
  {Zaritsky}},\ }\href {https://doi.org/10.1088/0004-637X/778/1/14} {\bibfield
  {journal} {\bibinfo  {journal} {Astrophys. J.}\ }\textbf {\bibinfo {volume}
  {778}},\ \bibinfo {pages} {14} (\bibinfo {year} {2013})},\ \Eprint
  {https://arxiv.org/abs/1309.3565} {arXiv:1309.3565 [astro-ph.CO]}
  \BibitemShut {NoStop}%
\bibitem [{\citenamefont {Grandis}\ \emph {et~al.}(2019)\citenamefont
  {Grandis}, \citenamefont {Mohr}, \citenamefont {Dietrich}, \citenamefont
  {Bocquet}, \citenamefont {Saro}, \citenamefont {Klein}, \citenamefont
  {Paulus},\ and\ \citenamefont {Capasso}}]{Grandis:2018mle}%
  \BibitemOpen
  \bibfield  {author} {\bibinfo {author} {\bibfnamefont {S.}~\bibnamefont
  {Grandis}}, \bibinfo {author} {\bibfnamefont {J.~J.}\ \bibnamefont {Mohr}},
  \bibinfo {author} {\bibfnamefont {J.~P.}\ \bibnamefont {Dietrich}}, \bibinfo
  {author} {\bibfnamefont {S.}~\bibnamefont {Bocquet}}, \bibinfo {author}
  {\bibfnamefont {A.}~\bibnamefont {Saro}}, \bibinfo {author} {\bibfnamefont
  {M.}~\bibnamefont {Klein}}, \bibinfo {author} {\bibfnamefont
  {M.}~\bibnamefont {Paulus}},\ and\ \bibinfo {author} {\bibfnamefont
  {R.}~\bibnamefont {Capasso}},\ }\href {https://doi.org/10.1093/mnras/stz1778}
  {\bibfield  {journal} {\bibinfo  {journal} {Mon. Not. Roy. Astron. Soc.}\
  }\textbf {\bibinfo {volume} {488}},\ \bibinfo {pages} {2041} (\bibinfo {year}
  {2019})},\ \Eprint {https://arxiv.org/abs/1810.10553} {arXiv:1810.10553
  [astro-ph.CO]} \BibitemShut {NoStop}%
\bibitem [{\citenamefont {Soergel}\ \emph {et~al.}(2016)\citenamefont {Soergel}
  \emph {et~al.}}]{Soergel:2016mce}%
  \BibitemOpen
  \bibfield  {author} {\bibinfo {author} {\bibfnamefont {B.}~\bibnamefont
  {Soergel}} \emph {et~al.} (\bibinfo {collaboration} {DES, SPT}),\ }\href
  {https://doi.org/10.1093/mnras/stw1455} {\bibfield  {journal} {\bibinfo
  {journal} {Mon. Not. Roy. Astron. Soc.}\ }\textbf {\bibinfo {volume} {461}},\
  \bibinfo {pages} {3172} (\bibinfo {year} {2016})},\ \Eprint
  {https://arxiv.org/abs/1603.03904} {arXiv:1603.03904 [astro-ph.CO]}
  \BibitemShut {NoStop}%
\bibitem [{\citenamefont {Vikram}\ \emph {et~al.}(2017)\citenamefont {Vikram},
  \citenamefont {Lidz},\ and\ \citenamefont {Jain}}]{Vikram:2016dpo}%
  \BibitemOpen
  \bibfield  {author} {\bibinfo {author} {\bibfnamefont {V.}~\bibnamefont
  {Vikram}}, \bibinfo {author} {\bibfnamefont {A.}~\bibnamefont {Lidz}},\ and\
  \bibinfo {author} {\bibfnamefont {B.}~\bibnamefont {Jain}},\ }\href
  {https://doi.org/10.1093/mnras/stw3311} {\bibfield  {journal} {\bibinfo
  {journal} {Mon. Not. Roy. Astron. Soc.}\ }\textbf {\bibinfo {volume} {467}},\
  \bibinfo {pages} {2315} (\bibinfo {year} {2017})},\ \Eprint
  {https://arxiv.org/abs/1608.04160} {arXiv:1608.04160 [astro-ph.CO]}
  \BibitemShut {NoStop}%
\bibitem [{\citenamefont {Tanimura}\ \emph
  {et~al.}(2020{\natexlab{a}})\citenamefont {Tanimura}, \citenamefont
  {Zaroubi},\ and\ \citenamefont {Aghanim}}]{Tanimura:2020une}%
  \BibitemOpen
  \bibfield  {author} {\bibinfo {author} {\bibfnamefont {H.}~\bibnamefont
  {Tanimura}}, \bibinfo {author} {\bibfnamefont {S.}~\bibnamefont {Zaroubi}},\
  and\ \bibinfo {author} {\bibfnamefont {N.}~\bibnamefont {Aghanim}},\
  }\href@noop {} {\  (\bibinfo {year} {2020}{\natexlab{a}})},\ \Eprint
  {https://arxiv.org/abs/2007.02952} {arXiv:2007.02952 [astro-ph.CO]}
  \BibitemShut {NoStop}%
\bibitem [{\citenamefont {Lim}\ \emph {et~al.}(2020)\citenamefont {Lim},
  \citenamefont {Mo}, \citenamefont {Wang},\ and\ \citenamefont
  {Yang}}]{Lim:2019jfn}%
  \BibitemOpen
  \bibfield  {author} {\bibinfo {author} {\bibfnamefont {S.}~\bibnamefont
  {Lim}}, \bibinfo {author} {\bibfnamefont {H.}~\bibnamefont {Mo}}, \bibinfo
  {author} {\bibfnamefont {H.}~\bibnamefont {Wang}},\ and\ \bibinfo {author}
  {\bibfnamefont {X.}~\bibnamefont {Yang}},\ }\href
  {https://doi.org/10.3847/1538-4357/ab63df} {\bibfield  {journal} {\bibinfo
  {journal} {Astrophys. J.}\ }\textbf {\bibinfo {volume} {889}},\ \bibinfo
  {pages} {48} (\bibinfo {year} {2020})},\ \Eprint
  {https://arxiv.org/abs/1912.10152} {arXiv:1912.10152 [astro-ph.GA]}
  \BibitemShut {NoStop}%
\bibitem [{\citenamefont {Battaglia}\ \emph {et~al.}(2017)\citenamefont
  {Battaglia}, \citenamefont {Ferraro}, \citenamefont {Schaan},\ and\
  \citenamefont {Spergel}}]{Battaglia:2017neq}%
  \BibitemOpen
  \bibfield  {author} {\bibinfo {author} {\bibfnamefont {N.}~\bibnamefont
  {Battaglia}}, \bibinfo {author} {\bibfnamefont {S.}~\bibnamefont {Ferraro}},
  \bibinfo {author} {\bibfnamefont {E.}~\bibnamefont {Schaan}},\ and\ \bibinfo
  {author} {\bibfnamefont {D.}~\bibnamefont {Spergel}},\ }\href
  {https://doi.org/10.1088/1475-7516/2017/11/040} {\bibfield  {journal}
  {\bibinfo  {journal} {JCAP}\ }\textbf {\bibinfo {volume} {11}},\ \bibinfo
  {pages} {040}},\ \Eprint {https://arxiv.org/abs/1705.05881} {arXiv:1705.05881
  [astro-ph.CO]} \BibitemShut {NoStop}%
\bibitem [{\citenamefont {Tanimura}\ \emph
  {et~al.}(2020{\natexlab{b}})\citenamefont {Tanimura}, \citenamefont
  {Hinshaw}, \citenamefont {McCarthy}, \citenamefont {Van~Waerbeke},
  \citenamefont {Aghanim}, \citenamefont {Ma}, \citenamefont {Mead},
  \citenamefont {Tr\"oster}, \citenamefont {Hojjati},\ and\ \citenamefont
  {Moraes}}]{Tanimura:2019whh}%
  \BibitemOpen
  \bibfield  {author} {\bibinfo {author} {\bibfnamefont {H.}~\bibnamefont
  {Tanimura}}, \bibinfo {author} {\bibfnamefont {G.}~\bibnamefont {Hinshaw}},
  \bibinfo {author} {\bibfnamefont {I.~G.}\ \bibnamefont {McCarthy}}, \bibinfo
  {author} {\bibfnamefont {L.}~\bibnamefont {Van~Waerbeke}}, \bibinfo {author}
  {\bibfnamefont {N.}~\bibnamefont {Aghanim}}, \bibinfo {author} {\bibfnamefont
  {Y.-Z.}\ \bibnamefont {Ma}}, \bibinfo {author} {\bibfnamefont
  {A.}~\bibnamefont {Mead}}, \bibinfo {author} {\bibfnamefont {T.}~\bibnamefont
  {Tr\"oster}}, \bibinfo {author} {\bibfnamefont {A.}~\bibnamefont {Hojjati}},\
  and\ \bibinfo {author} {\bibfnamefont {B.}~\bibnamefont {Moraes}},\ }\href
  {https://doi.org/10.1093/mnras/stz3130} {\bibfield  {journal} {\bibinfo
  {journal} {Mon. Not. Roy. Astron. Soc.}\ }\textbf {\bibinfo {volume} {491}},\
  \bibinfo {pages} {2318} (\bibinfo {year} {2020}{\natexlab{b}})},\ \Eprint
  {https://arxiv.org/abs/1903.06654} {arXiv:1903.06654 [astro-ph.CO]}
  \BibitemShut {NoStop}%
\bibitem [{\citenamefont {Amodeo}\ \emph {et~al.}(2020)\citenamefont {Amodeo}
  \emph {et~al.}}]{Amodeo:2020mmu}%
  \BibitemOpen
  \bibfield  {author} {\bibinfo {author} {\bibfnamefont {S.}~\bibnamefont
  {Amodeo}} \emph {et~al.},\ }\href@noop {} {\  (\bibinfo {year} {2020})},\
  \Eprint {https://arxiv.org/abs/2009.05558} {arXiv:2009.05558 [astro-ph.CO]}
  \BibitemShut {NoStop}%
\bibitem [{\citenamefont {Ma}\ \emph {et~al.}(2020)\citenamefont {Ma},
  \citenamefont {Gong}, \citenamefont {Troster},\ and\ \citenamefont
  {Van~Waerbeke}}]{Ma:2020cir}%
  \BibitemOpen
  \bibfield  {author} {\bibinfo {author} {\bibfnamefont {Y.-Z.}\ \bibnamefont
  {Ma}}, \bibinfo {author} {\bibfnamefont {Y.}~\bibnamefont {Gong}}, \bibinfo
  {author} {\bibfnamefont {T.}~\bibnamefont {Troster}},\ and\ \bibinfo {author}
  {\bibfnamefont {L.}~\bibnamefont {Van~Waerbeke}},\ }\href@noop {} {\
  (\bibinfo {year} {2020})},\ \Eprint {https://arxiv.org/abs/2010.15064}
  {arXiv:2010.15064 [astro-ph.CO]} \BibitemShut {NoStop}%
\bibitem [{\citenamefont {Pandey}\ \emph {et~al.}(2020)\citenamefont {Pandey},
  \citenamefont {Baxter},\ and\ \citenamefont {Hill}}]{Pandey:2019uxy}%
  \BibitemOpen
  \bibfield  {author} {\bibinfo {author} {\bibfnamefont {S.}~\bibnamefont
  {Pandey}}, \bibinfo {author} {\bibfnamefont {E.}~\bibnamefont {Baxter}},\
  and\ \bibinfo {author} {\bibfnamefont {J.}~\bibnamefont {Hill}},\ }\href
  {https://doi.org/10.1103/PhysRevD.101.043525} {\bibfield  {journal} {\bibinfo
   {journal} {Phys. Rev. D}\ }\textbf {\bibinfo {volume} {101}},\ \bibinfo
  {pages} {043525} (\bibinfo {year} {2020})},\ \Eprint
  {https://arxiv.org/abs/1909.00405} {arXiv:1909.00405 [astro-ph.CO]}
  \BibitemShut {NoStop}%
\bibitem [{\citenamefont {Battaglia}\ \emph {et~al.}(2019)\citenamefont
  {Battaglia} \emph {et~al.}}]{Battaglia:2019dew}%
  \BibitemOpen
  \bibfield  {author} {\bibinfo {author} {\bibfnamefont {N.}~\bibnamefont
  {Battaglia}} \emph {et~al.},\ }\href@noop {} {\  (\bibinfo {year} {2019})},\
  \Eprint {https://arxiv.org/abs/1903.04647} {arXiv:1903.04647 [astro-ph.CO]}
  \BibitemShut {NoStop}%
\bibitem [{\citenamefont {Hojjati}\ \emph {et~al.}(2015)\citenamefont
  {Hojjati}, \citenamefont {McCarthy}, \citenamefont {Harnois-Deraps},
  \citenamefont {Ma}, \citenamefont {Van~Waerbeke}, \citenamefont {Hinshaw},\
  and\ \citenamefont {Le~Brun}}]{Hojjati:2014usa}%
  \BibitemOpen
  \bibfield  {author} {\bibinfo {author} {\bibfnamefont {A.}~\bibnamefont
  {Hojjati}}, \bibinfo {author} {\bibfnamefont {I.~G.}\ \bibnamefont
  {McCarthy}}, \bibinfo {author} {\bibfnamefont {J.}~\bibnamefont
  {Harnois-Deraps}}, \bibinfo {author} {\bibfnamefont {Y.-Z.}\ \bibnamefont
  {Ma}}, \bibinfo {author} {\bibfnamefont {L.}~\bibnamefont {Van~Waerbeke}},
  \bibinfo {author} {\bibfnamefont {G.}~\bibnamefont {Hinshaw}},\ and\ \bibinfo
  {author} {\bibfnamefont {A.~M.}\ \bibnamefont {Le~Brun}},\ }\href
  {https://doi.org/10.1088/1475-7516/2015/10/047} {\bibfield  {journal}
  {\bibinfo  {journal} {JCAP}\ }\textbf {\bibinfo {volume} {10}},\ \bibinfo
  {pages} {047}},\ \Eprint {https://arxiv.org/abs/1412.6051} {arXiv:1412.6051
  [astro-ph.CO]} \BibitemShut {NoStop}%
\bibitem [{\citenamefont {Hojjati}\ \emph {et~al.}(2017)\citenamefont {Hojjati}
  \emph {et~al.}}]{Hojjati:2016nbx}%
  \BibitemOpen
  \bibfield  {author} {\bibinfo {author} {\bibfnamefont {A.}~\bibnamefont
  {Hojjati}} \emph {et~al.},\ }\href {https://doi.org/10.1093/mnras/stx1659}
  {\bibfield  {journal} {\bibinfo  {journal} {Mon. Not. Roy. Astron. Soc.}\
  }\textbf {\bibinfo {volume} {471}},\ \bibinfo {pages} {1565} (\bibinfo {year}
  {2017})},\ \Eprint {https://arxiv.org/abs/1608.07581} {arXiv:1608.07581
  [astro-ph.CO]} \BibitemShut {NoStop}%
\bibitem [{\citenamefont {Hill}\ and\ \citenamefont
  {Spergel}(2014)}]{Hill:2013dxa}%
  \BibitemOpen
  \bibfield  {author} {\bibinfo {author} {\bibfnamefont {J.}~\bibnamefont
  {Hill}}\ and\ \bibinfo {author} {\bibfnamefont {D.~N.}\ \bibnamefont
  {Spergel}},\ }\href {https://doi.org/10.1088/1475-7516/2014/02/030}
  {\bibfield  {journal} {\bibinfo  {journal} {JCAP}\ }\textbf {\bibinfo
  {volume} {02}},\ \bibinfo {pages} {030}},\ \Eprint
  {https://arxiv.org/abs/1312.4525} {arXiv:1312.4525 [astro-ph.CO]}
  \BibitemShut {NoStop}%
\end{thebibliography}%


\appendix

\section{Comparison of simulations and parametric model for baryonic effects}\label{App:model_fits}

In Sec.~\ref{sec:marginalisation_mead}, we investigated whether including a model for baryonic effects in the prediction for the matter power spectrum can reduce the bias on $M_\nu$ after marginalization over the model's parameters. For this to be a valid method, we must be reasonably confident that the model can capture a realistic range of baryonic effects. In this appendix, we check this for the hydrodynamical simulations we use in this work, assuming that this set of simulations itself spans a realistic range of effects. More detailed comparisons between these simulations and observations will likely be required to fully justify this assumption, but this is beyond the scope of this work.

We compare model predictions from Ref.~\cite{Mead:2016zqy} with the power spectrum ratios $\hat{R}(k,z)$ measured from simulations, using the following statistic:
\be
\Delta(A,\eta_0) \equiv \sum_{i,j}\lb \hat{R}(k_i,z_j)
	- \frac{P_m(k_i,z_j; A, \eta_0)}{P_{\rm DMO}(k_i,z_j)} \rb^2,
	\label{chi2_mead}
\ee
where $P_{\rm DMO}$ is evaluated with the fiducial values of $A$ and $\eta_0$ from Sec.~\ref{sec:marginalisation_mead}. We sum over $k$ and $z$ points at which simulation measurements are available over $0\leq z \leq 2$ and $1\,\invMpcnoh \leq k \leq 10 \,\invMpcnoh$, chosen to correspond roughly to the ranges in which baryonic effects significantly affect $C_L^{\kappa\kappa}$ for $L\lesssim 3000$~\cite{Chung:2019bsk}. Eq.~\eqref{chi2_mead} is equivalent to a~$\chi^2$ statistic that weights all points equally, motivated by other work that has found sample-variance uncertainties on~$\hat{R}$ to be roughly scale-independent~\cite{Chisari:2018prw,vanDaalen:2019pst,Foreman:2019ahr}. Our goal is only to examine the best-fit predictions of the model for each simulation, rather than fully quantify the goodness of fit (which we cannot do without better knowledge of the uncertainties on $\hat{R}$), so we simply use unit weights in Eq.~\eqref{chi2_mead}.
We evaluate $P_m$ and $P_{\rm DMO}$ at our fiducial cosmology, because our goal is to check how well the model from Ref.~\cite{Mead:2016zqy} can reproduce our range of $\hat{R}$ curves with cosmology held fixed.

\begin{figure*}
\begin{center}
\includegraphics[width=0.75\textwidth]{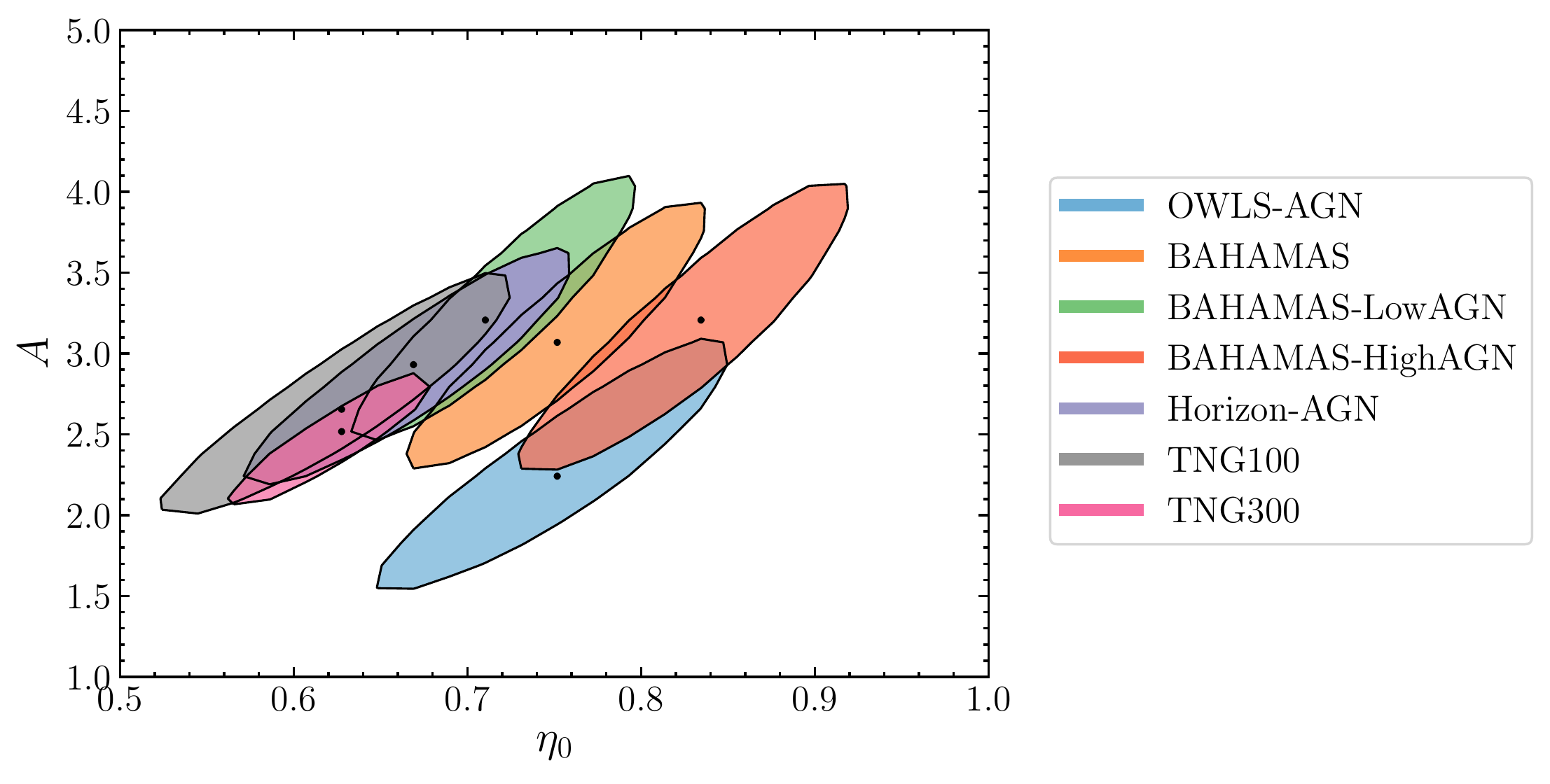}
\end{center}
\caption{Contour plots in the $A,\eta_0$ plane. We show contours of  $\Delta(A,\eta_0)$, where $\Delta(A,\eta_0)$ is defined in Eq.~\eqref{chi2_mead}. The points which minimise $\Delta(A,\eta_0)$ are shown, and the contours are filled in for values of $\Delta(A,\eta_0)$ which  $\Delta(A,\eta_0) < \Delta_{\rm min} + 5\times 10^{-5}$, corresponding to the separate $\Delta_{\rm min}$ for each simulation. 
 } \label{fig:sims_loglikelihoods}
\end{figure*}

\begin{figure*}
\begin{center}
\includegraphics[width=0.3\textwidth]{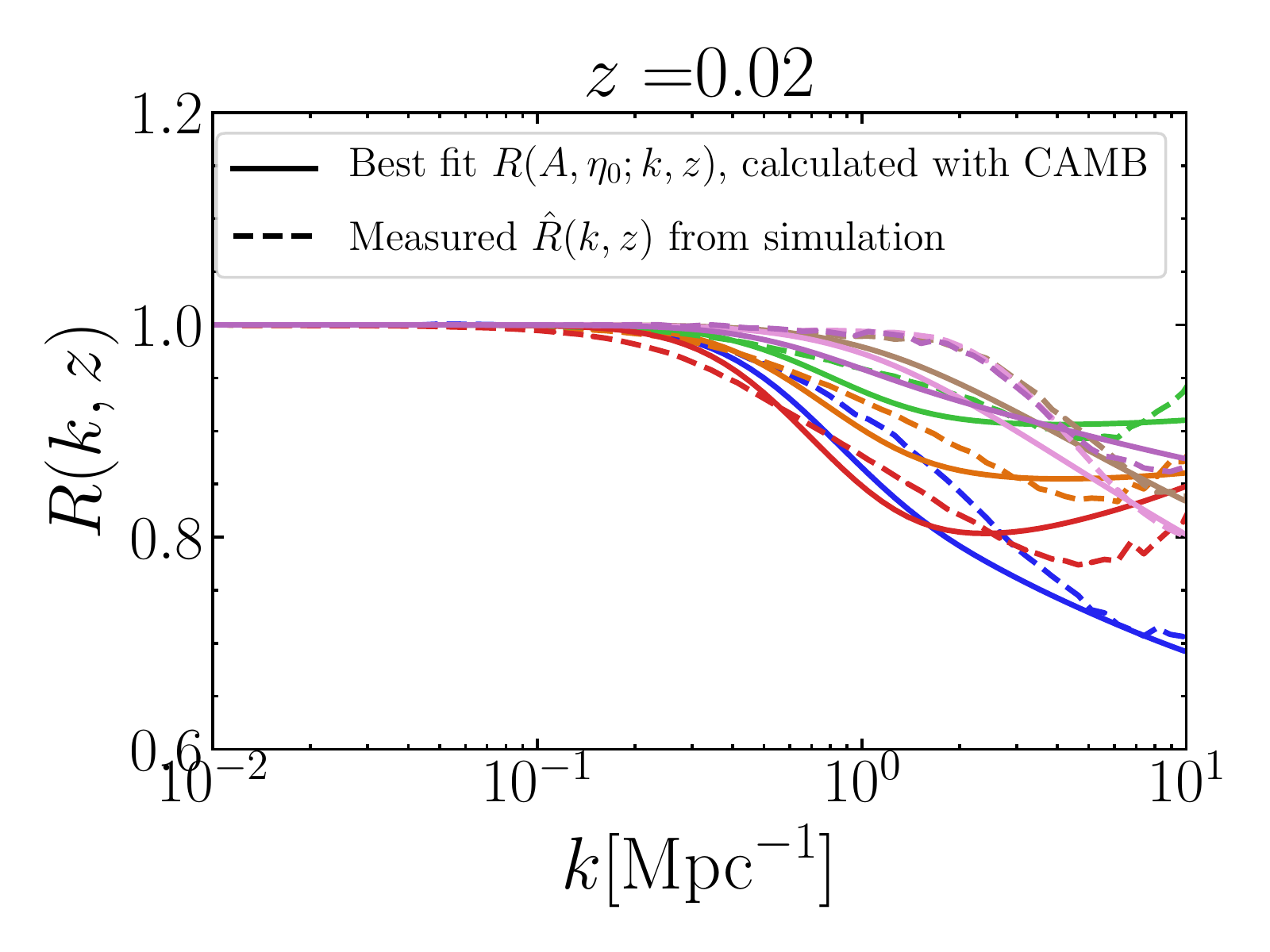}
\includegraphics[width=0.3\textwidth]{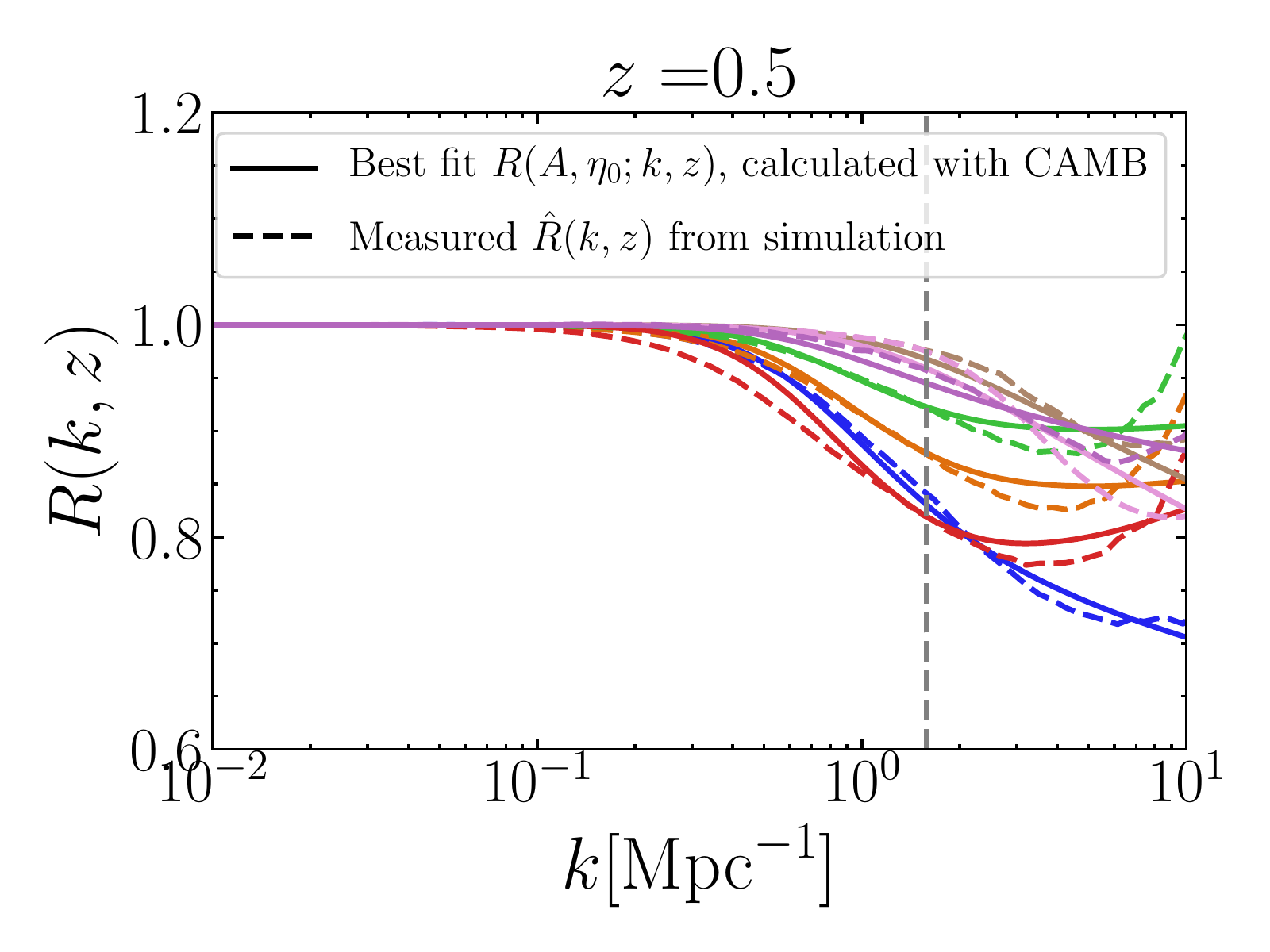}
\includegraphics[width=0.3\textwidth]{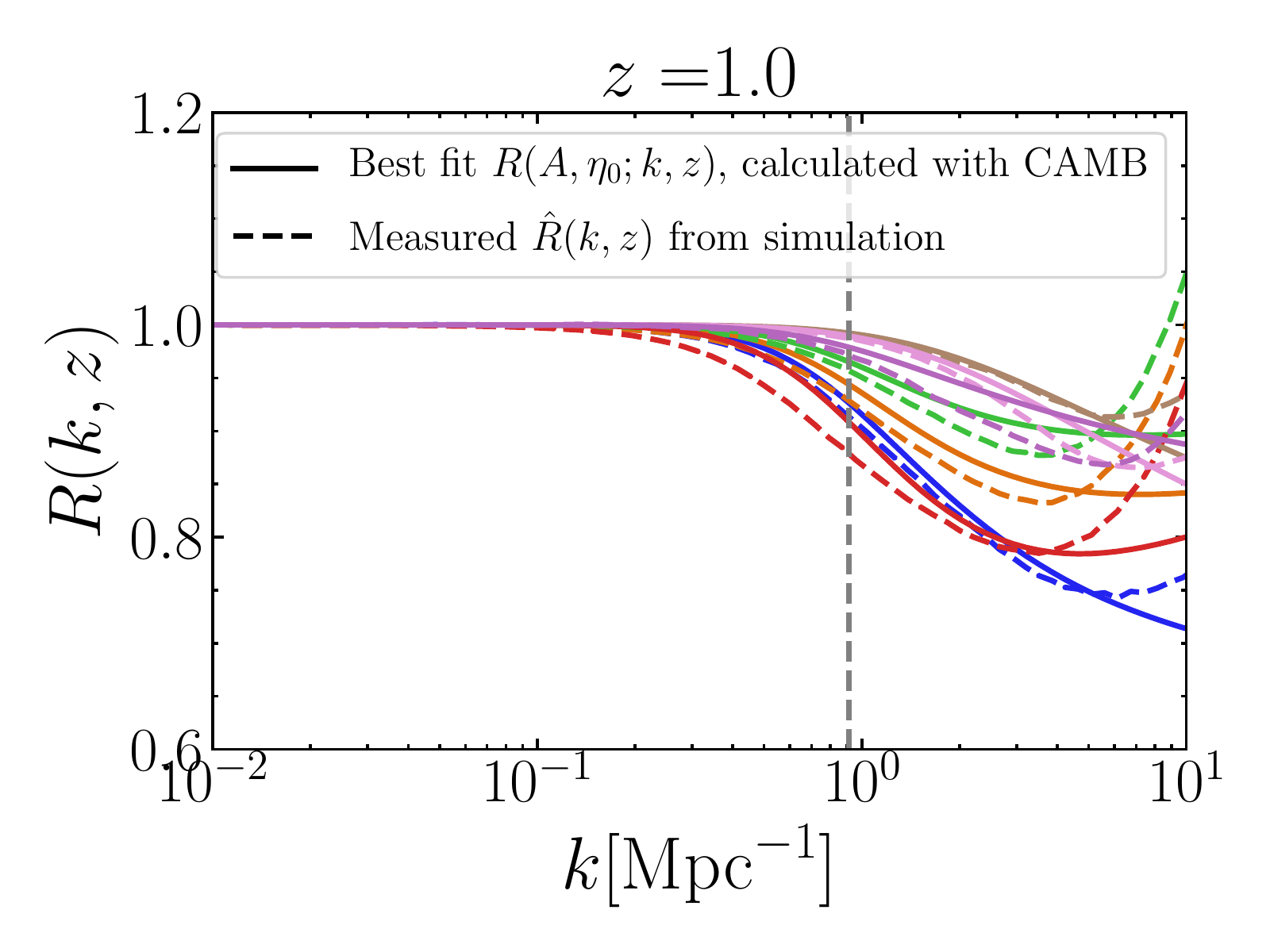}\\
\includegraphics[width=0.3\textwidth]{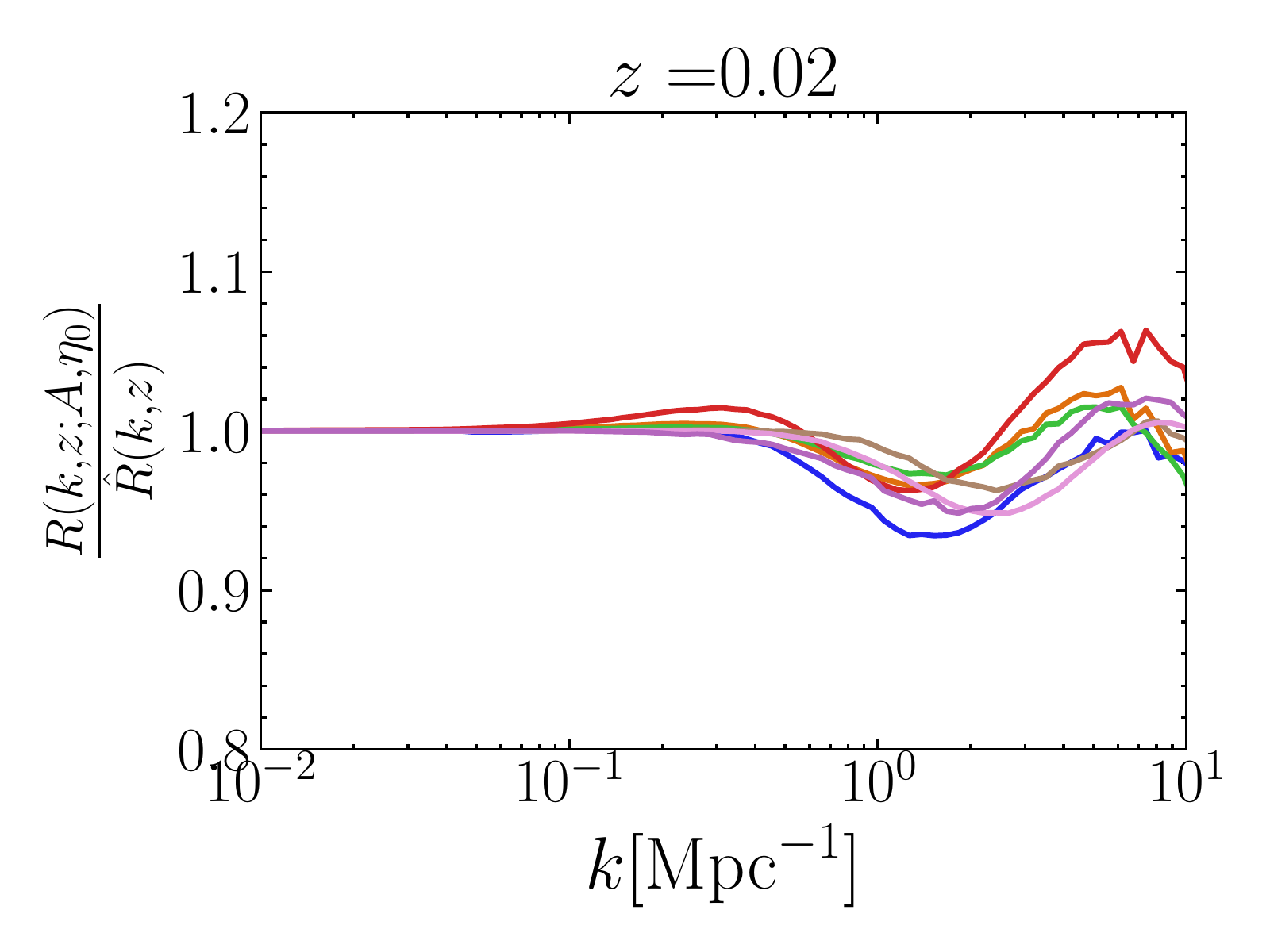}
\includegraphics[width=0.3\textwidth]{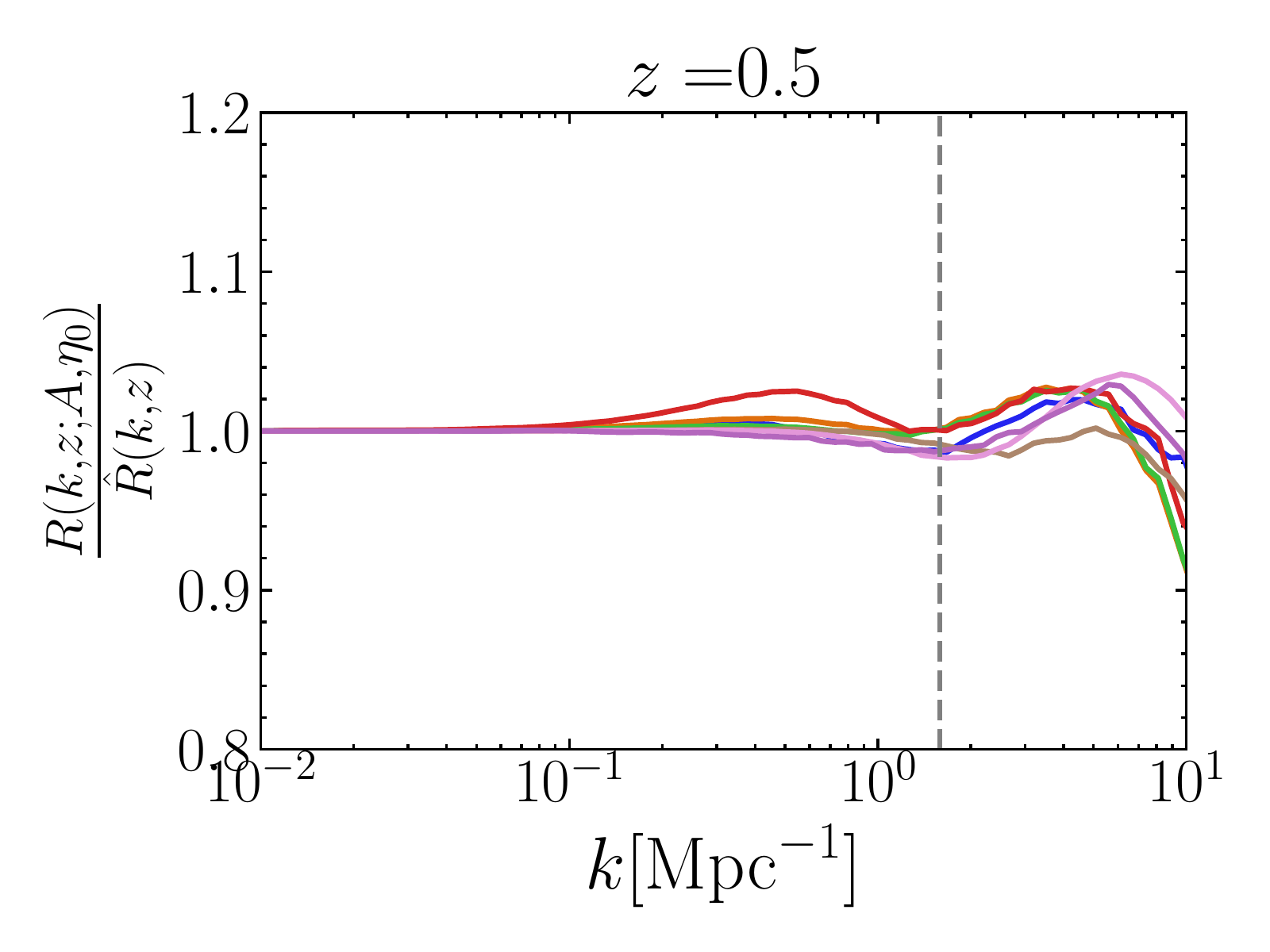}
\includegraphics[width=0.3\textwidth]{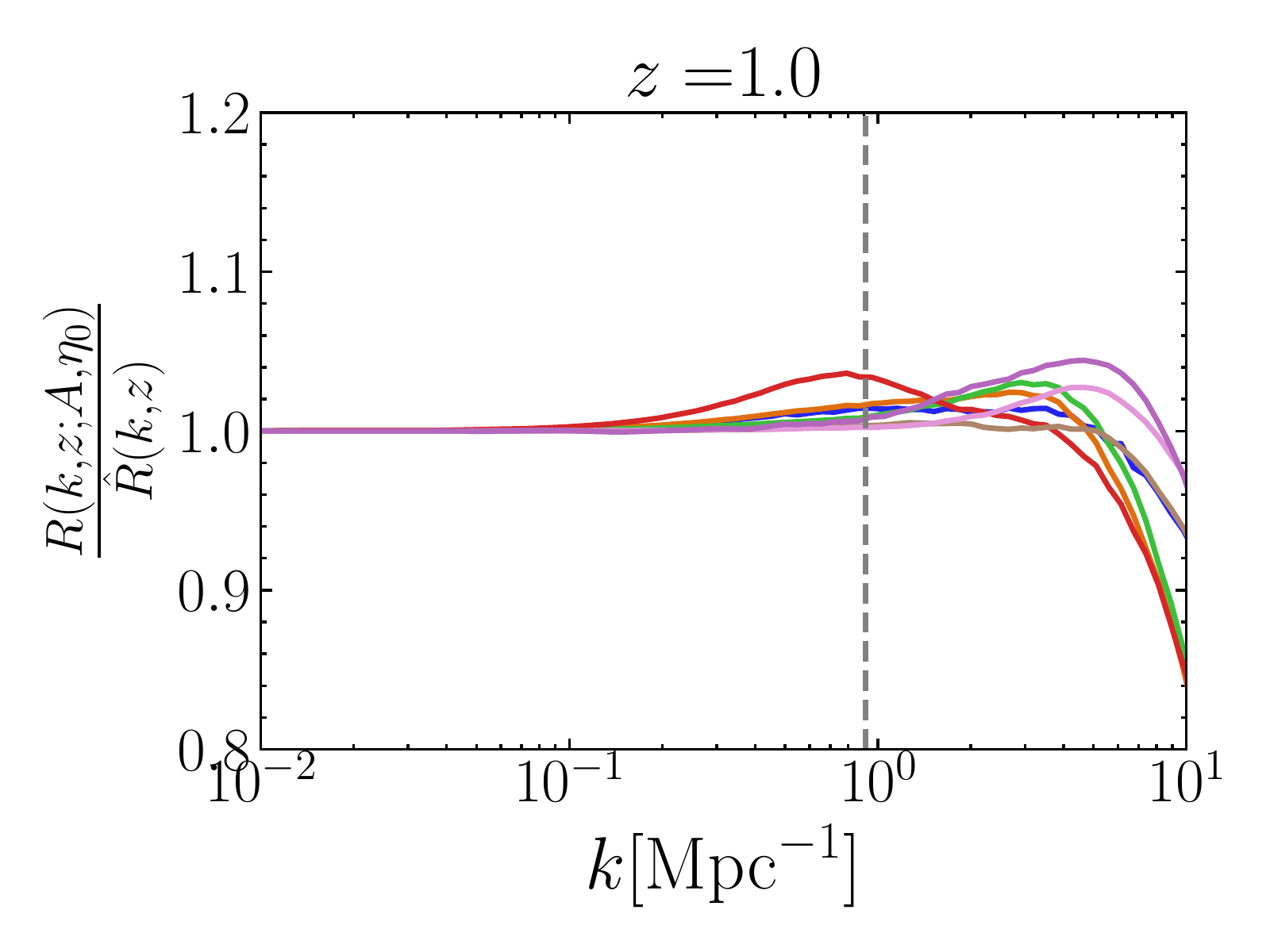}\\
\includegraphics[scale=0.5]{Images/legend_summary_method.pdf}

\end{center}
\caption{{\textit{ Top:}} The response functions $\hat{R}(k,z)$ measured from the simulations (solid lines) and the $R(A,\eta_0;k,z)$ (from the baryonic model) that minimize $\Delta(A,\eta_0)$, at various redshifts. {\textit Bottom:} Ratios of best-fit and measured $R$ functions. A dashed vertical line is shown at the $k$ that is equivalent to $3100/\chi(z)$ at each redshift, i.e.\ the maximum $k$ used to calculate $C_L^{\kappa\kappa}$ in the Limber approximation at each $z$.
}
\label{fig:bestfits}
\end{figure*}

\begin{figure*}
\includegraphics[width=0.4\textwidth]{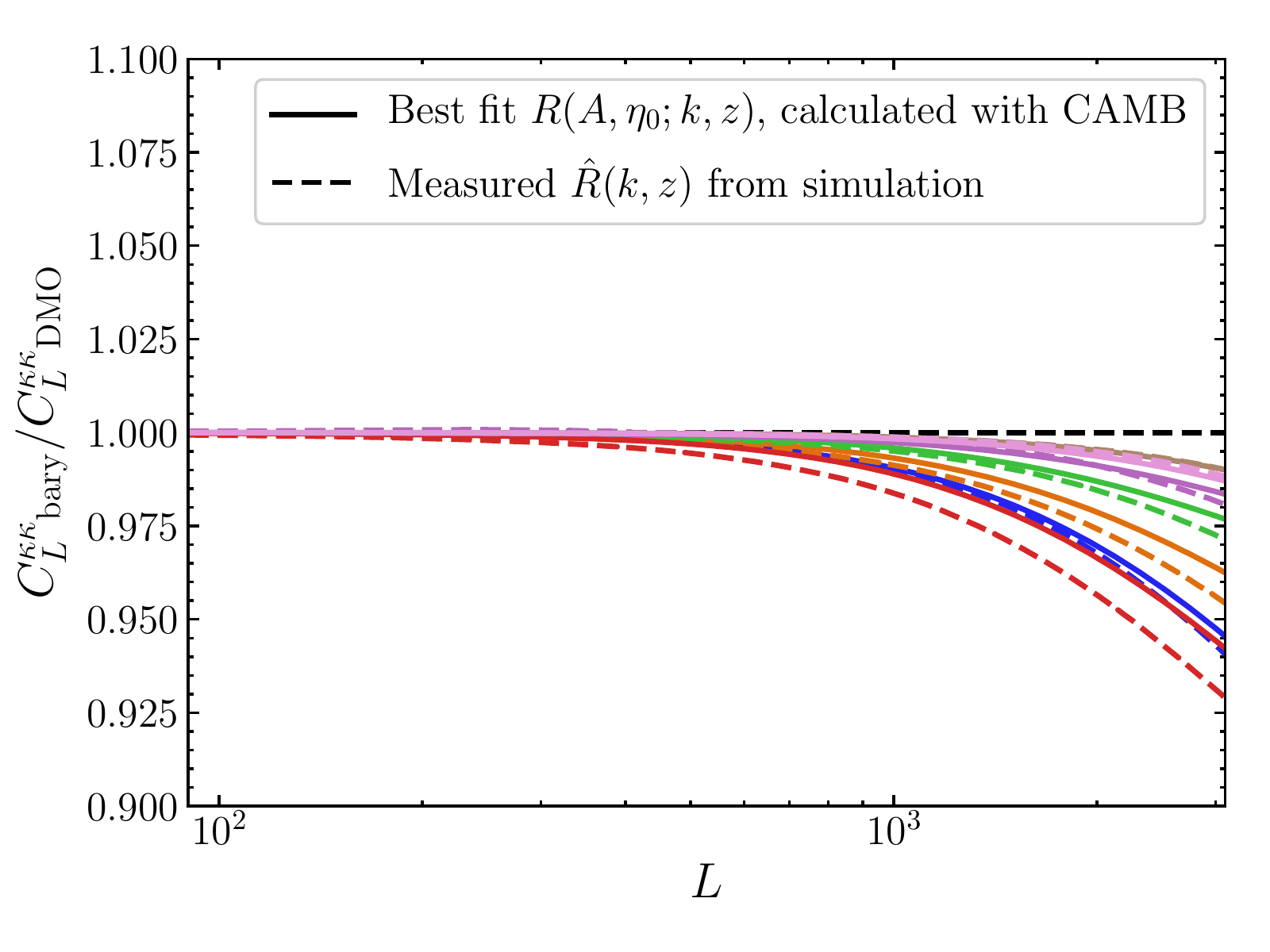}
\includegraphics[width=0.4\textwidth]{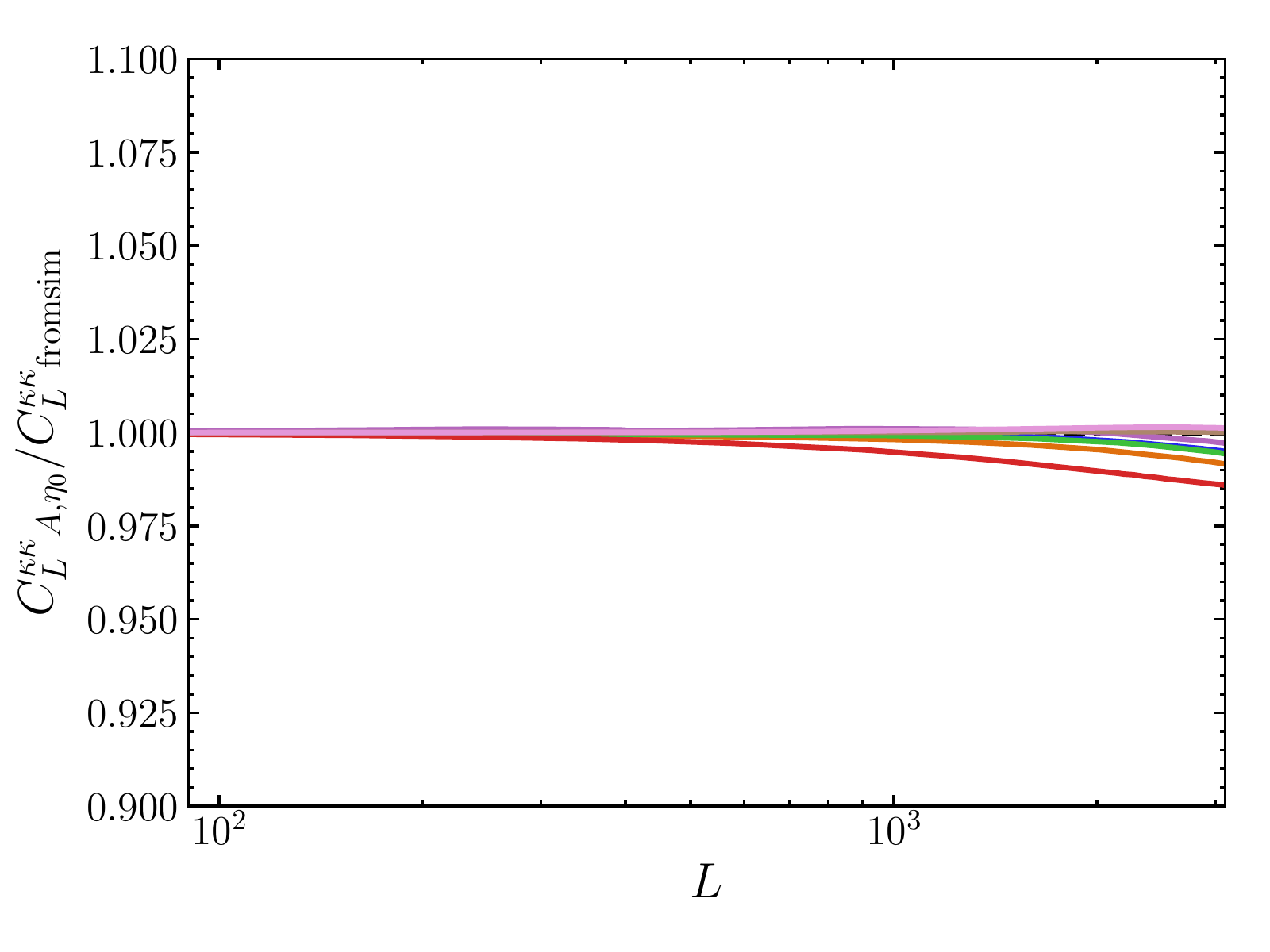}
\includegraphics[scale=0.55]{Images/legend_summary_method.pdf}

\caption{{\ti Left:} The ratio of the baryonic lensing power spectra to the DMO power spectra, computed with the $P(k,z;A,\eta_0)$ at the ``best fit'' values found for each simulation (in dashed lines). Also shown (in solid lines) is the ratio of the power spectra with the response function $\hat R(k,z)$ measured directly from the simulations. {\ti Right:} The ratio of the lensing power spectra computed at the best-fit $A$,$\eta_0$ and that computed with $\hat R(k,z)$ from the simulations. We find that the model is capable of reproducing the simulation measurements at better than 2\% accuracy for $L<3100$.}\label{fig:clkk_mead}
\end{figure*}

In Fig.~\ref{fig:sims_loglikelihoods}, we show contour plots of $ \Delta(A,\eta_0)$ for each simulation. These plots clearly imply a degeneracy between $A$ and $\eta_0$ for each simulation, roughly consistent with the degeneracy directions seen in the fits in Ref.~\cite{Mead:2015yca} (see their Fig.~6). Fig.~\ref{fig:bestfits} shows the best-fitting predictions for $\hat{R}(k,z)$ from minimizing $\Delta(A,\eta_0)$ with respect to the two parameters, at a few representative  redshifts. We find that for all simulations, the model can describe the power spectrum suppression to better than $\sim 5\%$ over the scales of interest, with better fits at lower $z$. While other models have been shown to match a subset of these simulations at higher precision (e.g.~\cite{Mead:2020vgs}), the $\sim 5\%$ precision we find for the model from Ref.~\cite{Mead:2016zqy} is sufficient to use it our proof-of-concept forecast in Sec.~\ref{sec:marginalisation_mead}. To see this, note that 5\% systematic errors in $P_m$ over the scales we fit for translate into $\sim$2\% errors  in $C_L^{\kappa\kappa}$ (see Fig.~\ref{fig:clkk_mead}), and Fig.~\ref{fig:baryons_clkk} shows that the simulation-derived $C_L^{\kappa\kappa}$ curves are still distinguishable from the effect of massive neutrinos even with this level of errors.

\end{document}